\documentclass[a4paper,11pt]{article}
\usepackage{epsfig}
\usepackage{amsmath,dsfont}
\usepackage{amssymb}
\usepackage{caption}
\usepackage{mathtools}
\usepackage{graphicx}
\usepackage{hyperref}
\usepackage{subcaption}
\usepackage{algorithm2e}
\usepackage{float}
\usepackage{bbold}
\usepackage[nameinlink]{cleveref}
\usepackage[toc,page]{appendix}
\usepackage{float}
\usepackage{subcaption}
\usepackage{multirow}
\restylefloat{table}
\usepackage{booktabs}
\usepackage[dvipsnames]{xcolor}
\usepackage{tikz}
\usepackage{pdflscape}
\usepackage{lscape}
\usepackage{soul,xcolor,cancel}
\date{\today}
 \normalsize

\newcommand{\bmat}{\left(\begin{array}}
\newcommand{\emat}{\end{array}\right)}
\newcommand{\be}{\begin{equation}}
\newcommand{\ee}{\end{equation}}
\newcommand{\bea}{\begin{eqnarray}}
\newcommand{\eea}{\end{eqnarray}}

\def\gtwid{\mathrel{\raise.3ex\hbox{$>$\kern-.75em\lower1ex\hbox{$\sim$}}}}
\def\ltwid{\mathrel{\raise.3ex\hbox{$<$\kern-.75em\lower1ex\hbox{$\sim$}}}}
\def\gev{{\rm \, Ge\kern-0.125em V}}
\def\tev{{\rm \, Te\kern-0.125em V}}

\def    \be            {\begin{equation}}
\def    \ee            {\end{equation}}
\def    \bea           {\begin{eqnarray}}
\def    \eea           {\end{eqnarray}}
\def\eps{\epsilon}
\def\a{\alpha}

\def\d{\delta}
\def\n{\nu}

\def\sig{\sigma}

\def\lam{\lambda}
\def\th{\theta}

\def\nn{\nonumber}
\def\d{\delta}
\def\D{\Delta}
\def\s{\sigma}
\def\r{\rho}
\def\t{\theta}

\def\me{m_{e}}
\def\mee{m_{ee}}
\def\cx{c_{12}}
\def\sx{s_{12}}
\def\cy{c_{23}}
\def\sy{s_{23}}
\def\cz{c_{13}}
\def\sz{s_{13}}

\begin{document}
\renewcommand{\thefootnote}{\fnsymbol{footnote}}

\vspace{.3cm}

\title{\Large\bf Texture of Single Vanishing Subtrace in Neutrino Mass Matrix}

\author
{ \hspace{-3.cm} \it \bf  A. Ismael$^{1,2}$\thanks{ahmedEhusien@sci.asu.edu.eg}, M. AlKhateeb$^{3,4}$\thanks{mohkha88@gmail.com},  N.
Chamoun$^{5}$\thanks{nidal.chamoun@hiast.edu.sy, nchamoun@th.physik.uni-bonn.de}  and E. I. Lashin$^{1,2,6}$\thanks{slashin@zewailcity.edu.eg, elashin@ictp.it}
 \\\hspace{-3.cm}
 \footnotesize$^1$ Department of Physics, Faculty of Science, Ain Shams University, Cairo 11566,  Egypt.\\\hspace{-3.cm}
 \footnotesize$^2$ Centre for Fundamental Physics, Zewail City of Science and
Technology,\\\hspace{-3.cm}
 \footnotesize  6 October City, Giza 12578, Egypt.\\\hspace{-3.cm}
\footnotesize$^3$ Department of Physics, Faculty of Science, Damascus University,  Syria.\\\hspace{-3.cm}
\footnotesize$^4$ D$\acute{e}$partement de Physique, University$\acute{e}$ de Cergy-Pontoise, F-95302 Cergy-Pontoise, France.\\\hspace{-3.cm}
\footnotesize$^5$  Physics Department, HIAST, P.O.Box 31983, Damascus,
Syria.\\\hspace{-3.cm}
\footnotesize$^6$ The Abdus Salam, ICTP, P. O. Box 586, 34100 Trieste,  Italy.
}

\date{\today}

\maketitle
\begin{abstract}
We consider a texture for the neutrino mass matrix characterized by one vanishing $2\times2$ subtrace. We analyze phenomenologically and analytically all the six possible patterns, and show that all non-singular ones are able to accommodate the experimental bounds, whereas singular patterns allow only for four inverted-hierarchy type textures. We then present some possible realizations of this texture, within seesaw scenarios, either directly or indirectly by relating it to zero-textures.
\end{abstract}

\maketitle

\begin{minipage}[h]{14.0cm}
\end{minipage}

\vskip 0.3cm \hrule \vskip 0.5cm

\section{Introduction}

The fact that neutrinos are massive was the first firm sign of physics beyond standard model \cite{neutrinoMass}. Many flavor models for neutrino mass matrix were conceived, motivated by phenomenological data on neutrino oscillations \cite{oscillation}. Zero textures were studied extensively \cite{xing,M0texture,0texture}, but other forms of textures were equally studied, such as zero minors \cite{minor}, and partial $\mu-\tau$ symmetry textures \cite{pmt}.

The objective of this work is to study the texture characterized by one vanishing subtrace{ , motivated by many factors. First, a particular texture of vanishing two subtraces was studied in \cite{Lashin}, where analytical expressions for the measurable neutrino observables were derived, and numerical analysis was done to show that $8$ patterns,  out of the $15$ independent ones, can  accommodate data. Second, the so-called $\mu-\tau$ antisymmetry texture was studied in, say, \cite{mtas}. This texture, in addition to the condition ($M_{e \mu}=-M_{e \tau}$) which can be implemented also in $\mu$-$\tau$ symmetry textures \cite{pmt}, included also a certain vanishing subtrace condition ($M_{\mu \mu}+M_{\tau \tau}=0$). Third, the one vanishing $2 \times 2$ subtrace texture can be seen actually as a generalization of the zero texture when the latter is regarded as a vanishing $1 \times 1$ subtrace texture. Fourth, since the equal entries texture has been of interest recently, either in Hybrid texture (e.g. \cite{Liu}) or simply in assuming two such equalities (e.g. \cite{Gautam}), and since such an equality condition is closely related to a condition of vanishing sum of the corresponding two entries, so studying textures of vanishing subtrace adds further analysis into these studies}.  We implement the new experimental bounds of \cite{Lisi}, with the new updates on the
 non-vanishing value of the smallest mixing matrix \cite{daya}, and carry out a complete numerical analysis, where all the free parameters are scanned within their experimentally accepted ranges. We discuss non-singular patterns having all the neutrino masses non vanishing and singular patterns where one of the masses is zero. We find all six non-singular textures able to accommodate the experimental data. As to singular textures, only four textures can accommodate data of inverted hierarchy type. We then address the question of how to realize such a vanishing subtrace texture. First, we relate the symmetry imposing the vanishing subtrace pattern to another symmetry    which forces zeros at specific locations. The former pattern  arises upon rotating the zero-texture form. Although the texture form is imposed at high scale, however many arguments \cite{cherif} were presented in favor of keeping the form when running into low scale. The method we suggest for realizing the vanishing subtrace texture applies only to four out of six possible patterns. However, it is generic enough to be applicable to any specific texture related by unitary transformation to zero textures, and we apply it successfully  within  type I and type II seesaw scenarios. Second, we present direct realizations of the textures  within  type I and type II seesaw scenarios, without relating them to zero textures, based on discrete symmetries.

 The plan of the paper is as follows. In section 2,
  we present the notation and adopted conventions. In section 3, we explain the method to follow for the phenomenology. In section 4, we present the analysis of all the non-singular six patterns of one vanishing subtrace supplemented  with one single table summarizing all the predictions of the various patterns.  Subsections therein correspond to these different  patterns where for each one we report the relevant defining quantities, correlation plots and one representative point for each type of hierarchy. We repeat the analysis in section 5 for the singular patterns. Section 6 outlines the problem of how to build models implementing the vanishing subtrace texture whether it be directly or indirectly. In section 7, we develop the generic method relating the symmetry of vanishing subtraces to that of zero textures and use it as an indirect method for getting the vanishing subtrace texture.  In section 8, we clarify the notion of flavor basis which is of paramount relevance in our discussion. The aforementioned indirect method is applied within  type I  seesaw scenarios in order to realize invertible (singular) vanishing subtrace textures  in section 9 (10). We reapply this indirect method in section 11 but within type II seesaw scenario. In sections 12 and 13, we present respectively a direct way to impose the textures within type I and type II seesaw scenarios.  Summary and conclusion are presented in section 14.


\section{Notation}
In the `flavor' basis, where the charged lepton mass matrix is diagonal and thus the observed neutrino mixing matrix comes solely from the neutrino sector,
 we have
\be
M_\nu = V_{\mbox{\tiny PMNS}} \left( \begin{matrix}m_1 & 0 & 0 \cr 0 & m_2 & 0 \cr 0 & 0
& m_3\end{matrix} \right) (V_{\mbox{\tiny PMNS}})^T,
 \label{Mnu}
\ee
\be
\left.
\begin{array}{lll}
P = \mbox{diag}\left(e^{i\rho},e^{i\sigma},1\right)\,, U \; = R_{23}\left(\t_{23}\right)\; R_{13}\left(\t_{13}\right)\; \mbox{diag}\left(1,e^{-i\d},1\right)\; R_{12}\left(\t_{12}\right),\\\\
V_{\mbox{\tiny PMNS}} = U\;P  {\footnotesize = \left ( \begin{array}{ccc} c_{12}\, c_{13} e^{i\rho} & s_{12}\, c_{13} e^{i\sigma}& s_{13} \\ (- c_{12}\, s_{23}
\,s_{13} - s_{12}\, c_{23}\, e^{-i\delta}) e^{i\rho} & (- s_{12}\, s_{23}\, s_{13} + c_{12}\, c_{23}\, e^{-i\delta})e^{i\sigma}
& s_{23}\, c_{13}\, \\ (- c_{12}\, c_{23}\, s_{13} + s_{12}\, s_{23}\, e^{-i\delta})e^{i\rho} & (- s_{12}\, c_{23}\, s_{13}
- c_{12}\, s_{23}\, e^{-i\delta})e^{i\sigma} & c_{23}\, c_{13} \end{array} \right ),}
\end{array}
\right\}
\label{defv}
\ee
where $R_{ij}\left(\t_{ij}\right)$ is the rotation matrix in the $(i,j)$-plane by angle $\t_{ij}$, and $s_{12} \equiv \sin\theta_{12} \ldots$. Note that in this adopted parametrization, the
third column of $V_{\mbox{\tiny PMNS}}$ is real.

The mass spectrum is classified into
two categories:
\begin{itemize}
\item Normal hierarchy: characterized by $m_1< m_2 < m_3$ and
is denoted by ${\bf N}$. \item Inverted hierarchy: characterized
by $m_3 < m_1< m_2$ and is denoted by ${\bf I}$.
\end{itemize}

The  neutrino mass-squared differences, characterizing respectively solar and
atmospheric neutrino mass-squared differences together with their ratio $R_\nu$, are defined as,
\be
\delta m^2 \; \equiv \; m_2^2-m_1^2 \; , \; \Delta m^2  \;\equiv \; \left|m_3^2-{1\over 2}\left(m_1^2+m_2^2\right)\right|,\;\;  R_\nu = \delta m^2 /  \Delta m^2.
\label{sqm}
\ee

Two parameters which put bounds on the neutrino mass scales, through studying  beta-decay kinematics and neutrinoless double-beta decay, are the
effective electron-neutrino mass:
\be
\me \; = \; \sqrt{\sum_{i=1}^{3} \displaystyle \left (
|V_{e i}|^2 m^2_i \right )} \;\; ,
\label{medef}
\ee
and the effective Majorana mass term
$\mee$:
\be
\mee \; = \; \left | m_1
V^2_{e1} + m_2 V^2_{e2} + m_3 V^2_{e3} \right | \; = \; \left | M_{\n 11} \right |.
\label{mee}
\ee
Cosmological observations put bounds on the  `sum'
parameter $\Sigma$:
\be
\Sigma = \sum_{i=1}^{3} m_i.
\ee
The last measurable quantity we shall consider is the Jarlskog rephasing invariant
defined by:
\be
\label{jg}
J = s_{12}\,c_{12}\,s_{23}\, c_{23}\, s_{13}\,c_{13}^2 \sin{\delta}.
\ee

The experimental bounds for the oscillation parameters are summarized in Table~(\ref{TableLisi})
{\sf
\begin{table}[h]
\centering
\scalebox{0.8}{
\begin{tabular}{cccccc}
\toprule
Parameter & Hierarchy & Best fit & $1 \sigma$ & $2 \sigma$ & $3 \sigma$ \\
\toprule
$\delta m^2$ $(10^{-5} \text{eV}^2)$ & NH, IH & 7.37 & [7.21,7.54] & [7.07,7.73] & [6.93,7.96] \\
\midrule
 \multirow{2}{*}{$\Delta m^2$ $(10^{-3} \text{eV}^2)$} & NH & 2.53 & [2.50,2.57] & [2.45,2.61] & [2.41,2.65] \\
 \cmidrule{2-6}
           & IH & 2.51 & [2.47,2.54] & [2.43,2.58] & [2.39,2.62]\\
\midrule
\multirow{2}{*}{$R_\nu $} & NH & 0.029 & [0.028,0.030] & [0.027,0.031] & [0.026,0.033] \\
\cmidrule{2-6}
      & IH & 0.029 & [0.028,0.030] & [0.027,0.032] & [0.026,0.033]\\
\midrule
$\theta_{12}$ ($^{\circ}$) & NH, IH & 33.02 & [32.02,34.09] & [30.98,35.30] & [30.00,36.51] \\
\midrule
\multirow{2}{*}{$\theta_{13}$ ($^{\circ}$)}  & NH & 8.43 & [8.30,8.55] & [8.11,8.74] & [7.92,8.90]] \\
\cmidrule{2-6}
    & IH & 8.45 & [8.27,8.59] & [8.08,8.78]& [7.92,8.94]]\\
\midrule
\multirow{2}{*}{$\theta_{23}$ ($^{\circ}$)}  & NH & 40.69 & [39.82,41.89]& [38.93,43.29] & [38.10,51.66]] \\
\cmidrule{2-6}
      & \multirow{2}{*}{IH} & \multirow{2}{*}{42.42} & $[40.23,42.02] \cup$ & $[39.18,44.02] \cup $ & [38.29,52.90]\\
  & & &  $ [48.86,51.06]$ & $[46.89,52.01]$ &   \\
\midrule
\multirow{4}{*}{$\delta$ ($^{\circ}$)}  & \multirow{2}{*}{NH} & \multirow{2}{*}{248.40} & [212.40,289.80] & [180.00,342.00] & $[0.00,30.60] \cup $ \\
& & & &  & [136.80,360]\\
\cmidrule{2-6}
 & \multirow{2}{*}{IH} & \multirow{2}{*}{235.80} & [201.60,291.60] & [165.60,338.40] & $[0,27.00] \cup $  \\
& & & &  & [124.20,360]\\
\bottomrule
\end{tabular}}
\caption{\footnotesize Allowed $1$-$2$-$3\s$-ranges for the neutrino oscillation parameters: mixing angles, Dirac phase $\delta$, mass-square differences together with the $R_\nu$ parameter, taken from the global fit to neutrino oscillation data \cite{Lisi}.  The quantities $\d m^2$, $\Delta m^2$ and $R_\nu$ are respectively defined as  $m_2^2-m_1^2$,   $\left|m_3^2 -\left(m_1^2 + m_2^2\right)/2\right|$ and $\d m^2/\Delta m^2$. Normal and Inverted Hierarchies are
respectively denoted by NH and IH.}
\label{TableLisi}
\end{table}
}

For the non oscillation parameters  $\Sigma$ , $\mee$ and $\me$, we adopt the ranges reported in the recent reference \cite{heinrich}
for the first two, while for $\me$ we use more stringent values found in the earlier reference  \cite{Cuoricino}:
\be
\left.
\begin{array}{lll}
\Sigma &<& 0.7 \;\mbox{eV}, \\
\mee & < & 0.3\; \mbox{eV}, \\
\me &<& 1.8\; \mbox{eV}.
\end{array}
\right\}
\label{nosdata}
\ee

\section{Texture of one-vanishing subtrace}
We denote by $\mathbf C_{ij}$ the texture where the subtrace corresponding to the
$ij^{th}$ element (i.e. the trace of the sub-matrix obtained
by deleting the $i^{th}$ row and the $j^{th}$ column of $M_\nu$) is equal to zero.
We have six possibilities of having one subtrace vanishing. Let the diagonal
elements of the trace-free submatrix corresponding
to $\mathbf C_{ij}$  be the elements at the ($a,b$) and ($c,d$) entries of $M_\n$, then the
vanishing subtrace condition is written as:
\be
 M_{\nu\;ab} + M_{\nu\;cd} = 0,
\label{det1}
\ee
then we have
\be
\sum_{\ell=1}^{3}\left(
U_{a\ell}U_{b\ell}+ U_{c\ell}U_{d\ell} \right)
\lambda_l = 0.
\label{U's}
\ee
with $\lambda_1 = m_1 e^{2i\r}, \lambda_2=m_2 e^{2i\s}$ and $\lambda_3=m_3$.
This leads to:
\be
\left.
\begin{array}{lll}
\displaystyle \frac{m_1}{m_3} &=&
\displaystyle \frac{Re(A_3) Im(A_2 e^{2i\s})-Re(A_2
e^{2i\s}) Im(A_3) }
{Im(A_1 e^{2i\r} ) Re(A_2 e^{2i\sigma})-Re(A_1 e^{2i\r}) Im(A_2 e^{2i\s})} ,\\\\
\displaystyle \frac{m_2}{m_3} &=& \displaystyle \frac{Im(A_3)Re(A_1 e^{2i\r}) -Re(A_3) Im(A_1 e^{2i\r})}
{Im(A_1 e^{2i\r} ) Re(A_2 e^{2i\sigma})-Re(A_1 e^{2i\r}) Im(A_2 e^{2i\s})}
\end{array}
\right\}
\label{massratio}
\ee
where $A_\a$ is defined as,
\be
A_\a =  \left(
U_{a\a}U_{b\a}+ U_{c\a}U_{d\a} \right), \;\;\;  \a=1,2,3.
\label{Ah}
\ee
We see that knowing the mixing and phase angles we can get mass ratios. Considering now
\be
\label{computemass}
m_3=\sqrt{\frac{\d m^2}{(\frac{m_2}{m_3})^2-(\frac{m_1}{m_3})^2}} ,\;\;\;\; m_1=m_3 \times \frac{m_1}{m_3}, m_2=m_3 \times \frac{m_2}{m_3}
\ee
 we see that knowing $\d m^2$ will allow us now to compute the mass spectrum  and all the neutrino observables. Thus our input parameters will be the seven parameters (three mixing angles + three phase angles + solar mass squared difference) which for the texture imposing one complex condition (two real conditions) allow us to determine the $9$ degrees of freedom of the neutrino mass matrix. We then can compute all the observable quantities and test the experimental bounds in Table~(\ref{TableLisi}) of $\Delta m^2$   and in  Eqs. (\ref{nosdata}) of the remaining mass bounds, and draw correlation plots of the accepted points.

Also, one should investigate the
possibility, for each pattern, to have singular (non-invertible) mass matrix. The viable singular mass matrix is
characterized by one of the two masses ($m_1$ for N hierarchy, and $m_3$ for I hierarchy) being equal to zero, as compatibility with the data prevents the
simultaneous vanishing of two masses:
\begin{itemize}
\item The vanishing of $m_1$ together with Eqs.(\ref{sqm},\ref{U's},\ref{Ah})  imply that the mass
spectrum of $m_2$ and $m_3$ takes the values $\sqrt{\d
m^2}$ and $\sqrt{\Delta m^2+\d m^2/2}$   respectively, and we get
\be
\left.
\begin{array}{lll}
\displaystyle \Delta m^2 &=& \displaystyle \d m^2 \left(\left| \frac{A_2}{A_3}\right|^2-\frac{1}{2}\right),\\
\displaystyle e^{2\,i\,\s} & = &  -\displaystyle {A_3 \, m_3 \over A_2\,m_2}.
\end{array}
\right\}
\label{m1=0Delta}
\ee
\item The vanishing
of $m_3$ together with Eqs.(\ref{sqm},\ref{U's},\ref{Ah}) imply that the mass spectrum of $m_2$
and $m_1$  takes the values $\sqrt{\Delta m^2+  \d m^2/2}$ and
$\sqrt{\Delta m^2- \d m^2/2}$
respectively, and we get
\be
\label{m3=0Delta}
\left.
\begin{array}{lll}
\displaystyle \Delta m^2 &=&\displaystyle  \frac{1}{2}\, \d m^2 \left( \frac{\left|\frac{A_1}{A_2}\right|^2+1}{\left|\frac{A_1}{A_2}\right|^2-1}\right),\\
\displaystyle e^{2\,i\,\left(\r - \s\right)} & = &  -\displaystyle  {A_2 \, m_2 \over A_1\,m_1}.
\end{array}
\right\}
\ee
\end{itemize}


\section{Phenomenological analysis for non-singular textures}
The parameter space is seven-dimensional representing the parameters ($\theta_{12},\theta_{13},\theta_{23}, \d,\r,\s, \d m^2$) within their allowed experimental ranges, where we throw $N$ points uniformly in the corresponding parameter space and test using the Eqs.(\ref{massratio}, \ref{computemass}) first to check the hierarchy type, then to see whether or not the bounds of $\D m^2$ with those of Eq.(\ref{nosdata}) are satisfied. Since the experimental bounds stated in Table~(\ref{TableLisi}) are not identical for the two types of hierarchy, then the parameter spaces in both cases are different, and one is obliged to repeat the sampling in the two cases, imposing the desired type of hierarchy with the other experimental bounds on the accepted points.
The number of points $N$ needed for a statistically significant sampling is found be at least of the order $10^7-10^{10}$.

In each of the following subsections, labeled by the textures $\mathbf C_{ij}$, and for each corresponding pattern we provide the analytic expressions of the quantities $A_{\a}$, defined in Eq.(\ref{Ah}), which characterize the pattern.  We find  that all the textures accommodate data for all types of  hierarchy and at all statistical
levels. All various predictions concerning the ranges spanned  by mixing angles, phase angles, neutrino masses, $\me$, $\mee$ and $J$ are summarized in Table~(\ref{tab_pred_non_sing}).
No signature is apparent in the case of normal ordering for the spanned ranges of neutrino masses presented  in Table~(\ref{tab_pred_non_sing}).  However, in the case of inverted ordering of the neutrino masses, we see  that $m_3$ can reach a vanishing value for the  textures $\mathbf C_{12}$, $\mathbf C_{13}$ at all $\s$-levels, and only at $2$-$3$-$\s$-levels for the textures  $\mathbf C_{22}$ and $\mathbf C_{33}$. In contrast, $m_3$ is never vanishing for the  textures $\mathbf C_{11}$ and $\mathbf C_{23}$. Thus, the textures $\mathbf C_{12}$, $\mathbf C_{13}$, $\mathbf C_{22}$ and $\mathbf C_{33}$ are predicted to allow for singular
mass matrix, as will be shown later to be the case. The ranges spanned by the parameter $J$,  in  Table~(\ref{tab_pred_non_sing}),  show that $J$ at the $1$-$2$-$\s$-levels for normal ordering and $1$-$\s$-level for inverted ordering  is negative in all textures, which puts the Dirac phase $\d$ in the third and fourth quarters. Also from Table~(\ref{tab_pred_non_sing}), the ranges spanned by the phase angles ($\r$, $\s$) indicate that for the texture $\mathbf C_{12}$ in case of normal hierarchy and at the $2\s$-level there are gaps ($\s \not \in [90^0,150^0]$ and $\r \not \in [34^0,101^0]$), and a similar gap ($\r \not \in [0^0,18^0]$) for the texture $\mathbf C_{13}$ which becomes ($\r \not \in [0^0,5^0]$) at the $3\s$-level. However, in the case of inverted ordering, we see at all levels that the phase $\r$ for the texture $\mathbf C_{23}$ is bound to be in the interval ($[60^0,120^0]$).

We present for each texture with either hierarchy type the neutrino mass matrix obtained at one representative point chosen from the points accepted out of those generated randomly in the corresponding parameter space at the $3$-$\s$-level. The choice of the  representative point is made in such a way to be close as possible to the best fit values for mixing and Dirac phase angles.

Finally, we plot all the possible correlations at the $2$-$\s$-level. We show for each texture with either ordering twenty correlations. All correlations for each texture  are organized in a single figure divided into left and right panels. The left panel of the figure consists of two columns where the first (second) column is devoted for normal (inverted) hierarchy and shows three correlations amidst the mixing angles, three correlations amidst the phase angles and three correlations of $\d$ with $(J, \mee, \text{LNM})$ (LNM=least neutrino mass) and finally the correlation ($m_3,m_{23}\equiv \frac{m_2}{m_3}$). On the other hand, we follow for  the right panel of the figure  the same division strategy as in the left one, but each column includes all the nine inter-correlations between the phase angles and the mixing angles, and the correlation ($m_3,m_{21}\equiv \frac{m_2}{m_1}$). For the sake of convenience and easy referencing, each subfigure is labeled by three letters which indicate the vertical positioning $(a,b,c,\cdots)$, the type of ordering ($N$=Normal, $I$ = Inverted) and the paneling ($L$=Left, $R$ = Right). The last row in the figure thus gives information on the severity of the mass hierarchy.

Irrespective of the ordering  (normal or  inverted),  we find in all the textures a sinusoidal correlation between ($\d, J$) which is a direct consequence of Eq.(\ref{jg}) where $J$ depends on mixing angles and Dirac phase $\d$. The variations due to the mixing angles in this relation are tiny because of  the tight range allowed for the mixing angles, and thus $J \propto \sin{\d}$. The appearing sinusoidal curve is not a full sine curve which would have covered a complete cycle, rather it is a portion depending on the admissible range for $\d$.  Another generic feature that we find is the quasi degeneracy of the first two neutrino masses characterized by $m_1\approx m_2$.

In the case of normal ordering, we see, for the textures $\mathbf C_{12}$ and $\mathbf C_{13}$, sizable forbidden bands for both $\s$ and $\rho$ that tend to diminish as the statistical level increases, and a quasi degenerate spectrum for all neutrino masses with $(0.7 \le m_{23} < 1)$.
As to the textures $\mathbf C_{11}$ an $\mathbf C_{22}$, we see that there remain persistent forbidden bands for $(\s,\r)$ at all statistical levels, and that we can have a mild or  moderate mass hierarchy  characterized by $(0.4 \le  m_{23} < 1)$ in texture $\mathbf C_{22}$, whereas we have  a quasi degenerate spectrum for all masses with  $ m_{23} \approx  1$  in texture $\mathbf C_{11}$. Moreover, for the latter texture $\mathbf C_{11}$, we find  two ribbons for the correlation ($\d,\s$). Regarding  the texture $\mathbf C_{23}$, we see that there are forbidden bands for $\r$, and that the mass hierarchy can be mild or moderate $(0.4 \le  m_{23} \le 0.9)$. This situation repeats itself for the texture $C_{33}$ where we have $(0.35 \le  m_{23} \le 0.9)$.

In the case of inverted ordering, we find  for the texture $\mathbf C_{11}$ two ribbons for the correlation ($\d,\s$)  and  a mild or  moderate mass hierarchy  characterized by $(1 <  m_{23} \le 3)$. For  the textures $\mathbf C_{12}$,  $\mathbf C_{13}$ and $\mathbf C_{22}$, we may get an acute hierarchy  reaching a strength $m_{23} \approx  10^4$ for $\mathbf C_{12}$ and $m_{23} \approx  10^3$ for both $\mathbf C_{13}$ and $\mathbf C_{22}$.  We get for the mass spectrum a mild hierarchy characterized by
 $(1.2 \le m_{23} \le 3)$ in the texture ($\mathbf C_{23}$), where, in addition, we find  forbidden bands for $(\s, \r)$. Finally for the texture $\mathbf C_{33}$, we have again  forbidden bands for $(\s, \r)$, but the hierarchy can be severe reaching a strength $m_{23} \approx  10^4$ .
\begin{landscape}
\begin{table}[H]
 \begin{center}
\scalebox{0.7}{
{\tiny
 \begin{tabular}{c|c|c|c|c|c|c|c|c|c|c|c|c}
 \hline
 \hline
\multicolumn{13}{c}{\mbox{Model} $\mathbf {C_{11}}\equiv M_{\n\,22} + M_{\n\,33} = 0$} \\
\hline
\hline
  \mbox{quantity} & $\th_{12}^{\circ}$ & $\th_{23}^{\circ}$& $\th_{13}^{\circ}$ & $m_1$ $(10^{-1} \text{eV})$ & $m_2$ $(10^{-1} \text{eV})$ & $m_3$ $(10^{-1} \text{eV})$ & $\d^{\circ}$ & $\r^{\circ}$ & $\s^{\circ}$ & $\me$ $(10^{-1} \text{eV})$
 & $\mee$ $(10^{-1} \text{eV})$ & $J$ $(10^{-1})$\\
 \hline
 \multicolumn{13}{c}{\mbox{Normal  Hierarchy}} \\
 \cline{1-13}
 $1\, \sig$ &$32.49 - 33.97 $& $40.58 - 41.76$ &$8.33 - 8.51$ &$2.11 - 2.29$ &$2.11 - 2.29$ &
  $2.17 - 2.35$& $212.79 - 222.37$ &$119.07 - 130.46$ & $120.43 - 132.71$ &$2.11 - 2.29$ &
  $2.05 - 2.23$ & $-0.22 -  -0.18$  \\
 \hline
 $2\, \sig$ &$31.02 - 35.26 $& $38.97 - 43.28$ &$8.12 - 8.74$ &$1.67 - 2.31$ &$1.67 - 2.31$ &
  $1.74 - 2.37$& $180.07 - 220.9 \cup 319.2 - 341.98$ &$49.31 - 132.06$ & $
  47.83 - 72.82 \cup 86.54 - 129.47$ &$1.67 - 2.31$ &
  $1.59 - 2.23$ & $-0.23 -  0.00$  \\
 \hline
 $3\, \sig$ &$30.02 - 36.51 $& $38.13 - 51.66$ &$7.92 - 8.90$ &$1.57 - 2.31$ &$1.58 - 2.31$ &
  $1.65 - 2.37$& $0.27 - 30.33 \cup 138.2 - 224.8 \cup 316.5 - 359.51$ &$46.41 - 131.49$ & $47.41 - 135.33$ &$1.58 -  2.32$ &
  $1.50  -  2.24$ & $-0.24 - 0.22$  \\
 \hline
 \multicolumn{13}{c}{\mbox{Inverted  Hierarchy}} \\
 \cline{1-13}
 $1\, \sig$ &$32.02 - 34.09$ & $40.23 - 42.02 \cup  48.86 - 51.06$ & $8.27 - 8.59$ & $0.54 -  2.25$ &
 $0.55 - 2.25$ & $0.21 - 2.20$ & $201.70 - 291.57$ & $0.14 - 179.82$ & $0.09 - 33.77 \cup 98.85 - 179.97$ &
  $0.54 -  2.25$& $ 0.19 - 2.10$ & $ -0.34 -  -0.12 $ \\
 \hline
 $2\, \sig$ &$30.98 - 35.30$ & $39.18 -44.01 \cup  46.89 - 52.01$ & $8.08 - 8.78$ & $0.52 - 1.99$ &
 $0.53 - 2.00$ & $0.19 -  1.93$ & $165.76 - 338.29$ & $0.02 - 179.91$ & $0.09 - 179.94$ &$ 0.52 -   1.99$& $ 0.17 - 1.92$ & $ -0.35  - 0.08$  \\
 \hline
 $3\, \sig$ & $30.01 - 36.51$ & $38.29 - 52.90$ & $7.92 - 8.94$ & $0.52 - 2.35$ &
 $ 0.52 -  2.35$ & $0.17  - 2.30$ & $0.12 - 26.96 \cup 124.30 - 359.88$ & $0.01 - 179.96$ & $0.17 -  179.98$ &
  $0.51 - 2.35$ & $ 0.15 -  2.24$& $ -0.36 - 0.29$  \\
 \hline
 \hline
\multicolumn{13}{c}{\mbox{Model} $\mathbf{C_{12}}\equiv M_{\n\,12} + M_{\n\,33} = 0$} \\
\hline
  \mbox{quantity} & $\th_{12}^{\circ}$ & $\th_{23}^{\circ}$& $\th_{13}^{\circ}$ & $m_1$ $(10^{-1} \text{eV})$ & $m_2$ $(10^{-1} \text{eV})$ & $m_3$ $(10^{-1} \text{eV})$ & $\d^{\circ}$ & $\r^{\circ}$ & $\s^{\circ}$ & $\me$ $(10^{-1} \text{eV})$
 & $\mee$ $(10^{-1} \text{eV})$ & $J$ $(10^{-1})$\\
 \hline
 \multicolumn{13}{c}{\mbox{Normal  Hierarchy}} \\
 \cline{1-13}
 $1\, \sig$ &$32.02 - 34.09$& $39.82 - 41.89$ &$8.30 - 8.55$ &$0.41 - 2.24$ &$0.42 - 2.24
$ &
  $0.65 - 2.30$& $212.47 - 289.79$ &$0.28 - 6.26 \cup 103.7 - 179.70$ & $0.00 - 56.69 \cup 157 - 179.99$ &$ 0.42 - 2.24$ &
  $0.18 - 2.16$ & $-0.34 -  -0.17$  \\
 \hline
 $2\, \sig$ & $30.98 - 35.30$ &$38.93 - 43.29$ & $8.11 - 8.74$ &$0.37 -  1.97$ &$0.38 - 1.97$
 &$0.62 - 2.03$ & $200.90 - 341.98$ & $ 0.05 - 34.62 \cup 102.4 - 179.97$ & $0.05 - 91.38 \cup 152.1  - 179.66$& $0.38 - 1.97$ &
 $0.14 - 1.83$ & $-0.35 -  -0.10$  \\
 \hline
 $3\, \sig$ &$ 30.00 - 36.51$ & $38.10 - 51.66$ & $7.92 - 8.90$& $0.26 - 2.08$ & $0.28 -    2.09$ &
  $0.56 - 2.15$& $0.01 - 30.59 \cup 136.9 - 168.5 \cup 198.3 -  359.77$ & $0.21 -83.27 \cup 96.99 - 179.97$ &$0.08 - 179.79$ & $ 0.28 - 2.09$ &
  $0.10 - 2.03$ & $-0.35 - 0.23$ \\
 \hline
 \multicolumn{13}{c}{\mbox{Inverted  Hierarchy}} \\
 \cline{1-13}
 $1\, \sig$ & $32.02 - 34.09$  &$40.23 - 42.02 \cup 48.86 - 51.06$ &$8.27 - 8.59$ &$0.49 - 1.87$ &$0.50 - 1.87$&
 $0.00 - 1.80$ &$201.88 -  291.51$ &$0.01 - 179.79$ &$0.10 - 179.94$ &$0.49 -  1.87$
 &$0.03 - 1.78$ &$-0.33 -   -0.12$  \\
 \hline
 $2\, \sig$ &$30.98 - 35.30$ & $39.18 - 44.02 \cup 46.9  - 51.99$ & $8.08 - 8.78$ & $0.49 - 2.28$ &
 $0.50  -  2.28$ & $0.00 -  2.22$ & $165.60 -  337.76$ & $0.06 - 179.97$ &$0.24 - 179.99$ &
  $0.49 - 2.28$& $ 0.29 - 2.17$ & $-0.35 -   0.01$  \\
 \hline
 $3\, \sig$ & $30.00 - 36.51$ & $38.30 - 52.90$ & $7.92 - 8.94$ & $0.49 -   2.11$ &
 $0.49 - 2.11$ & $0.00 - 2.05$ & $0.14 -26.23 \cup 124.2 - 358.22$ & $0.08 -  179.88$ &$0.01 - 179.95$ &
  $0.48 - 2.11$ & $0.29 -  2.08$& $-0.36 -  0.29$  \\
 \hline
 \hline
\multicolumn{13}{c}{\mbox{Model} $\mathbf{C_{13}}\equiv M_{\n\,12} + M_{\n\,23} = 0$} \\
\hline
  \mbox{quantity} & $\th_{12}^{\circ}$ & $\th_{23}^{\circ}$& $\th_{13}^{\circ}$ & $m_1$ $(10^{-1} \text{eV})$ & $m_2$ $(10^{-1} \text{eV})$ & $m_3$ $(10^{-1} \text{eV})$ & $\d^{\circ}$ & $\r^{\circ}$ & $\s^{\circ}$ & $\me$ $(10^{-1} \text{eV})$
 & $\mee$ $(10^{-1} \text{eV})$ & $J$ $(10^{-1})$\\
 \hline
 \multicolumn{13}{c}{\mbox{Normal  Hierarchy}} \\
 \cline{1-13}
 $1\, \sig$ &$32.02 - 34.09$& $39.82 - 41.89$ &$8.30 - 8.55$ &$0.46 - 2.24$ &$0.47 - 2.24$ &
  $0.68 - 2.30$& $212.40 - 289.78$ &$42.35 - 155.60$ & $17.14 - 102.78$ &$0.47 - 2.24$ &
  $0.17 - 2.17$ & $-0.33   - -0.17$  \\
 \hline
 $2\, \sig$ &$30.98 - 35.30$& $38.93 - 43.29$ &$8.11 - 8.74$ &$0.45 - 2.24$ &$0.46 - 2.24$ &
  $0.68 -  2.30$& $180.13 - 341.49$ &$18.38 - 172.91$ & $0.21 - 151.7 \cup 175.7 - 179.95$ &$0.46 -  2.24$ &
  $0.14 - 2.08$ & $-0.35  - -0.00$  \\
 \hline
 $3\, \sig$ &$30.00 - 36.50$& $38.10 - 51.64$ &$7.92 - 8.90$ &$0.43 - 2.30$ &$0.44 - 2.30$ &
  $0.66 - 2.35$& $13.07 - 30.30 \cup 136.8  - 351.91$ &$5.89 - 174.88$ & $0.01 -  179.99$ &$0.44 - 2.30$ &
  $0.13  -  2.20$ & $-0.36 -   0.23$  \\
 \hline
 \multicolumn{13}{c}{\mbox{Inverted  Hierarchy}} \\
 \cline{1-13}
 $1\, \sig$ &$32.02 - 34.09$ & $40.23 - 42.02 \cup 48.89 - 51.06$ & $8.27 - 8.59$ & $0.49 - 2.24$ &
 $0.50 - 2.24$ & $ 0.00 - 2.18$ & $ 201.96 -  291.57$ & $0.00 -  179.94$ & $0.57 - 177.09$ &
  $0.49 -  2.23$& $ 0.35 - 2.08$ & $-0.34 -   -0.12$  \\
 \hline
 $2\, \sig$ &$30.98 - 35.30$ & $ 39.18 - 44.02 \cup 46.9  -  52.01$ & $8.08 -  8.78$ & $0.49 -  2.34$ &
 $0.50 -  2.34$ & $ 0.00 - 2.29$ & $ 204.64 - 338.03$ & $0.19 - 179.99$ & $0.03 -  179.80$ &
  $0.49 -  2.34$& $  0.31 -  2.16$ & $-0.35 -  -0.12$  \\
 \hline
 $3\, \sig$ & $30.00 - 36.51$ & $38.29 - 52.90$ & $7.92 - 8.94$ & $0.49  - 2.22$ &
 $0.49 - 2.22$ & $0.00 - 2.16$ & $0.69 -26.44 \cup 124.2 - 150.7 \cup 208.9 - 359.07$ & $ 0.04 - 179.95$ & $0.04 - 179.68$ &
  $0.48 - 2.21$ & $0.30 - 2.09$& $-0.36 -  0.27$  \\
 \hline
 \hline
\multicolumn{13}{c}{\mbox{Model} $\mathbf{C_{22}}\equiv M_{\n\,11} + M_{\n\,33} = 0$} \\
\hline
 \mbox{quantity} & $\th_{12}^{\circ}$ & $\th_{23}^{\circ}$& $\th_{13}^{\circ}$ & $m_1$ $(10^{-1} \text{eV})$ & $m_2$ $(10^{-1} \text{eV})$ & $m_3$ $(10^{-1} \text{eV})$ & $\d^{\circ}$ & $\r^{\circ}$ & $\s^{\circ}$ & $\me$ $(10^{-1} \text{eV})$
 & $\mee$ $(10^{-1} \text{eV})$ & $J$ $(10^{-1})$\\
 \hline
 \multicolumn{13}{c}{\mbox{Normal  Hierarchy}} \\
 \cline{1-13}
 $1\, \sig$ &$32.02 -  34.09$& $39.82 - 41.89$ &$8.30 - 8.55$ &$0.25 - 2.10$ &$0.26 - 2.11$ &
  $0.56 - 2.16$& $212.40 - 289.79$ &$76.46 - 112.48$ & $1.96 - 176.32$ &$0.26 - 2.11$ &
  $0.22 - 1.74$ & $-0.33 - -0.17$  \\
 \hline
 $2\, \sig$ &$30.98 - 35.30$& $38.93 - 43.29$ &$8.11 - 8.74$ &$0.20 - 2.26$ &$0.22 - 2.26$ &
  $0.54 - 2.32$& $180.03 -261.4 \cup 273.4 - 341.94$ &$58.82 - 115.13$ & $5.26 - 177.74$ &$0.22 - 2.26$ &
  $0.18 - 1.70$ & $-0.33 -  0.00$  \\
 \hline
 $3\, \sig$ &$30.01 - 36.51$& $38.12 - 51.66$ &$7.92 - 8.90$ &$0.12 - 2.04$ &$0.15 - 2.04$ &
  $0.51 - 2.10$& $0.14 - 30.57 \cup 136.9 - 359.99$ &$52.46 -  128.42$ & $2.15 - 176.90$ &$0.15 - 2.04$ &
  $0.11 -  1.44$ & $-0.34 -  0.24$  \\
 \hline
 \multicolumn{13}{c}{\mbox{Inverted  Hierarchy}} \\
 \cline{1-13}
 $1\, \sig$ &$32.02 - 34.09$ & $40.24 - 42.02 \cup 48.86 - 51.06$ & $8.27 - 8.59$ & $0.56 - 2.173$ &
 $0.57 -  2.18$ & $ 0.27 -  2.12$ & $ 201.66 - 254.8 \cup 268 - 291.59$ & $67.13 - 112.45$ & $0.01 -  66.87 \cup 85.25 - 179.92$ &
  $0.56 - 2.17$& $ 0.23 -  1.92$ & $-0.33 -   -0.12$  \\
 \hline
 $2\, \sig$ &$30.99 - 35.30$ & $ 39.18 - 44.01 \cup 46.89 - 52.01$ & $8.08 - 8.78$ & $0.49 - 2.24$ &
 $0.50 - 2.24$ & $ 0.00 - 2.18$ & $ 165.61 - 251.5 \cup 270.9  - 338.38$ & $23.07 - 27.31 \cup 37.98 - 143.13$ & $0.08 - 63.81 \cup 87.39 - 179.86$ &
  $0.49  - 2.23$& $  0.16 - 1.79$ & $-0.35 - 0.08$  \\
 \hline
 $3\, \sig$ & $30.00 - 36.51$ & $38.29 - 52.90$ & $7.92 - 8.94$ & $0.49 - 2.23$ &
 $0.49 -  2.23$ & $0.00 -  2.17$ & $0.02 - 26.84 \cup 124.9 - 240.2 \cup 272.3 - 359.99$ & $4.58 - 172.25$ & $0.03 - 179.85$ &
  $0.48 - 2.23$ & $0.14 - 1.28$& $-0.34 -   0.28$  \\
 \hline
 \hline
\multicolumn{13}{c}{\mbox{Model} $\mathbf{C_{23}}\equiv M_{\n\,11} + M_{\n\,23} = 0$} \\
\hline
 \mbox{quantity} & $\th_{12}^{\circ}$ & $\th_{23}^{\circ}$& $\th_{13}^{\circ}$ & $m_1$ $(10^{-1} \text{eV})$ & $m_2$ $(10^{-1} \text{eV})$ & $m_3$ $(10^{-1} \text{eV})$ & $\d^{\circ}$ & $\r^{\circ}$ & $\s^{\circ}$ & $\me$ $(10^{-1} \text{eV})$
 & $\mee$ $(10^{-1} \text{eV})$ & $J$ $(10^{-1})$\\
 \hline
 \multicolumn{13}{c}{\mbox{Normal  Hierarchy}} \\
 \cline{1-13}
 $1\, \sig$ &$32.02 - 34.09$& $39.82 - 41.89$ &$8.30 - 8.55$ &$0.17 - 2.05$ &$0.19 - 2.05$ &
  $0.53 - 2.11$& $212.45 - 289.78$ &$61.39 - 117.58$ & $17.76 - 143.84$ &$0.20 - 2.05$ &
  $0.16 - 1.28$ & $-0.33 -  -0.17$  \\
 \hline
 $2\, \sig$ &$30.98 - 35.30$& $38.93 - 43.29$ &$8.11 - 8.74$ &$0.17 - 1.56$ &$0.19 - 1.57$ &
  $0.53 - 1.64$& $180.92  - 341.37$ &$60.64 - 119.29$ & $12.48 - 177.66$ &$0.19 - 1.57$ &
  $0.16 - 1.40$ & $-0.34 -   -0.01$  \\
 \hline
 $3\, \sig$ &$30.00 - 36.51$& $38.11 - 51.66$ &$7.92 -8.90$ &$0.17 - 2.14$ &$0.19 - 2.14$ &
  $0.53 - 2.20$& $1.56 - 30.52 \cup 136.9 - 352.44$ &$59.80 - 120.73$ & $4.88 - 179.03$ &$0.19 - 2.14$ &
  $0.16 - 1.56$ & $-0.36 -  0.24$  \\
 \hline
 \multicolumn{13}{c}{\mbox{Inverted  Hierarchy}} \\
 \cline{1-13}
 $1\, \sig$ &$32.02 - 34.09$ & $40.23 - 42.02 \cup 48.86 - 51.06$ & $8.27 - 8.59$ & $0.53 - 2.23$ &
 $0.54 - 2.23$ & $ 0.19 - 2.17$ & $ 201.96 - 291.56$ & $61.76 - 118.35$ & $0.07 - 52.85 \cup 108 - 179.93$ &
  $0.52 - 2.23$& $ 0.02 -  1.13$ & $-0.34 -   -0.12$  \\
 \hline
 $2\, \sig$ &$30.98 - 35.30$ & $ 39.20 - 44.02 \cup 46.89 - 52.00$ & $8.08 - 8.78$ & $0.52 - 2.23$ &
 $0.53 - 2.24$ & $ 0.17 - 2.18$ & $ 167.35 - 172.5 \cup 186.6 - 338.37$ & $60.24 - 120.08$ & $0.17 - 75.97 \cup 98.74 - 179.99$ &
  $0.52 - 2.23$& $  0.17 - 1.91$ & $-0.35 -  - 0.04 \cup 0.04  -  0.07$  \\
 \hline
 $3\, \sig$ & $30.00 - 36.50$ & $38.29 - 52.89$ & $7.92 - 8.94$ & $0.51 - 2.24$ &
 $0.52 - 2.24$ & $0.14 - 2.18$ & $3.85 -26.92 \cup 124.3 - 174.2 \cup 184.5 - 355.95$ & $ 58.70 -  121.76$ & $0.00 - 78.15 \cup 102.9 - 179.88$ &
  $0.51 - 2.24$ & $0.14 - 1.78$& $-0.36 - -0.02 \cup 0.02 - 0.30$  \\
 \hline
 \hline
\multicolumn{13}{c}{\mbox{Model} $\mathbf{C_{33}}\equiv M_{\n\,11} + M_{\n\,22} = 0$} \\
\hline
 \mbox{quantity} & $\th_{12}^{\circ}$ & $\th_{23}^{\circ}$& $\th_{13}^{\circ}$ & $m_1$ $(10^{-1} \text{eV})$ & $m_2$ $(10^{-1} \text{eV})$ & $m_3$ $(10^{-1} \text{eV})$ & $\d^{\circ}$ & $\r^{\circ}$ & $\s^{\circ}$ & $\me$ $(10^{-1} \text{eV})$
 & $\mee$ $(10^{-1} \text{eV})$ & $J$ $(10^{-1})$\\
 \hline
 \multicolumn{13}{c}{\mbox{Normal  Hierarchy}} \\
 \cline{1-13}
 $1\, \sig$ &$32.02 - 34.09$& $39.82 - 41.89$ &$8.30 - 8.55$ &$0.16 - 1.92$ &$0.18 - 1.92$ &
  $0.53 - 1.98$& $212.41 - 289.68$ &$63.95 - 126.18$ & $0.05 - 179.89$ &$0.18 - 1.92$ &
  $0.13 - 1.40$ & $-0.33 -  -0.17$  \\
 \hline
 $2\, \sig$ &$30.98 -  35.30$& $38.93 - 43.29$ &$8.11 - 8.74$ &$0.12 - 2.19$ &$0.15 - 2.19$ &
  $0.52  -  2.25$& $180.00 - 341.94$ &$55.80 - 128.67$ & $0.82 - 179.18$ &$0.15 -  2.19$ &
  $0.11 - 2.09$ & $-0.34 -  0.00$  \\
 \hline
 $3\, \sig$ &$30.00 - 36.50$& $38.10 - 51.64$ &$7.92 - 8.90$ &$0.12 - 1.97
$ &$0.15 -   1.97$ &
  $0.51 -  2.03$& $0.12 - 30.38 \cup 137.7 -  359.89$ &$51.70 - 129.52$ & $0.47 - 177.96$ &$0.14 - 1.97$ &
  $0.01 - 1.02$ & $-0.33 - 0.23$  \\
 \hline
 \multicolumn{13}{c}{\mbox{Inverted  Hierarchy}} \\
 \cline{1-13}
 $1\, \sig$ &$32.01 - 34.09$ & $40.23 - 42.02 \cup 48.86 -  51.06$ & $8.27 - 8.59$ & $0.50 - 2.16$ &
 $0.50 - 2.16$ & $ 0.027 -   2.10$ & $ 201.61 -  291.1$ & $56.68 - 131.75$ & $0.007 - 94.1 \cup   141.7 - 179.73$ &
  $0.49 -  2.16$& $ 0.19  - 1.93$ & $ -0.33 -  -0.12$  \\
 \hline
 $2\, \sig$ &$30.98 -  35.30$ & $ 39.18  - 44.02 \cup 46.92 - 52.01$ & $8.08 - 8.78$ & $0.50 - 2.20$ &
 $0.50 -  2.20$ & $ 0.00 - 2.14$ & $ 165.60 - 267.5 \cup  286.3 - 338.21$ & $14.89 -  169.88$ & $0.01 - 86.3 \cup  106.4 - 179.75$ &
  $0.49 -  2.20$& $  0.16  -  1.51$ & $-0.34 -  0.08$  \\
 \hline
 $3\, \sig$ & $30.00 - 36.51$ & $38.29 - 52.90$ & $7.92 - 8.94
$ & $0.49 - 2.13$ & $0.49 -  2.13$ & $0.00  -  2.07$ & $1.18 - 26.53 \cup 124.20 - 359.84$ & $ 1.46 -  173.20$ & $0.01 - 83.32 \cup 99.9 - 179.98$ &
  $0.48 - 2.13$ & $0.14 -  1.51$& $-0.33 -  0.29$  \\
 \hline
 \end{tabular}
 }}
 \end{center}
  \caption{\footnotesize  The various prediction for the patterns of
  one vanishing subtrace textures designated by $\mathbf{C_{11}, C_{12}, C_{13}, C_{22}, C_{23}}$ and
  $\mathbf C_{33}$.}
\label{tab_pred_non_sing}
\end{table}
 \end{landscape}


\subsection{Pattern $\mathbf C_{11}$: Vanishing of $M_{\n\,22} + M_{\n\, 33}$}
The relevant expressions for $A_1$, $A_2$ and $A_3$ , as defined in Eq.(\ref{Ah}) for this pattern, are
\bea
A_1 &=& \left(\cx\, \sy\, \sz + \sx\, \cy\, e^{-i\,\d}\right)^2 + \left(\cx\, \cy\, \sz - \sx\, \sy\, e^{-i\,\d}\right)^2,\nn \\
A_2 & = & \left(\sx \, \sy\, \sz - \cx\, \cy\, e^{-i\,\d}\right)^2 + \left(\sx\, \cy\, \sz + \cx\, \sy\, e^{-i\,\d}\right)^2,\nn\\
A_3 &=&  \cz^2.
\label{Atrace_2233}
\eea

For a representative point with normal ordering, we take $\theta_{12} = 33.2327^\circ$,    $\theta_{23} = 41.7746^\circ$,    $\theta_{13} = 8.5625^\circ$,    $\delta = 189.5139^\circ$,    $\rho = 91.0971^\circ$,    $\sigma = 102.9376^\circ$,    $ m_1 = 0.2162\, \mbox{eV}$,    $ m_2 = 0.21642\, \mbox{eV}$,   $ m_3 = 0.22212\, \mbox{eV}$,    $ \me = 0.21642\, \mbox{eV}$,    $ \mee = 0.20292\, \mbox{eV}$ with the  corresponding mass matrix (in eV):
\be
M_\n = \begin{pmatrix}
  -0.2001-0.0334i &   0.0400+0.0333i &    0.0495-0.0230i \\
 0.0400+0.0333i &  -0.0234-0.0059i &   0.2112-0.0014i \\
 0.0495-0.0230i  &  0.2112-0.0014i  &  0.0234+0.0059i
\end{pmatrix}.
\ee

For an inverted hierarchy representative point we take $\theta_{12} = 33.8335^\circ$,    $\theta_{23} = 42.3044^\circ$,   $\theta_{13} = 8.7834^\circ$,    $\delta = 246.8983^\circ$,    $\rho = 51.0968^\circ$,    $\sigma = 150.5728^\circ$,    $ m_1 = 0.0553\, \mbox{eV}$,    $ m_2 = 0.0560\, \mbox{eV}$,   $ m_3 = 0.0236\, \mbox{eV}$,    $ \me = 0.0550\, \mbox{eV}$,    $ \mee = 0.0220\, \mbox{eV}$  with the  corresponding mass matrix (in eV):
\be
M_\n = \begin{pmatrix}
  0.0014+0.0219i &  0.0286+0.0239i &   -0.0214-0.0263i \\
 0.0286+0.0239i  & -0.0076-0.0055i &   0.0225+0.0000i \\
-0.0214-0.0263i &   0.0225+0.0000i &   0.0076+0.0055i
\end{pmatrix}.
\ee

We see, from Table(\ref{tab_pred_non_sing}), that $m_3$ does not approach a vanishing value in inverted type which indicates that no corresponding singular matrix exists. We see also that $J$ at $1$-$2\s$-levels for normal ordering and $1\s$ for inverted ordering is negative so the corresponding $\d$ is at third or fourth quarters.  For normal ordering, the allowed ranges  for $\r$ and  $\s$ tend to increases as the statistical level increases reaching $ [46.41^\circ,131.49^\circ]$ ($[47.41^\circ,135.33^\circ]$) for $\r$ ($\s$) at $3\s$-level.

\begin{figure}[H]
\begin{minipage}{.5\linewidth}
           \centering
        \begin{tabular}{c}
             \includegraphics[height=25cm,width=8.5cm]{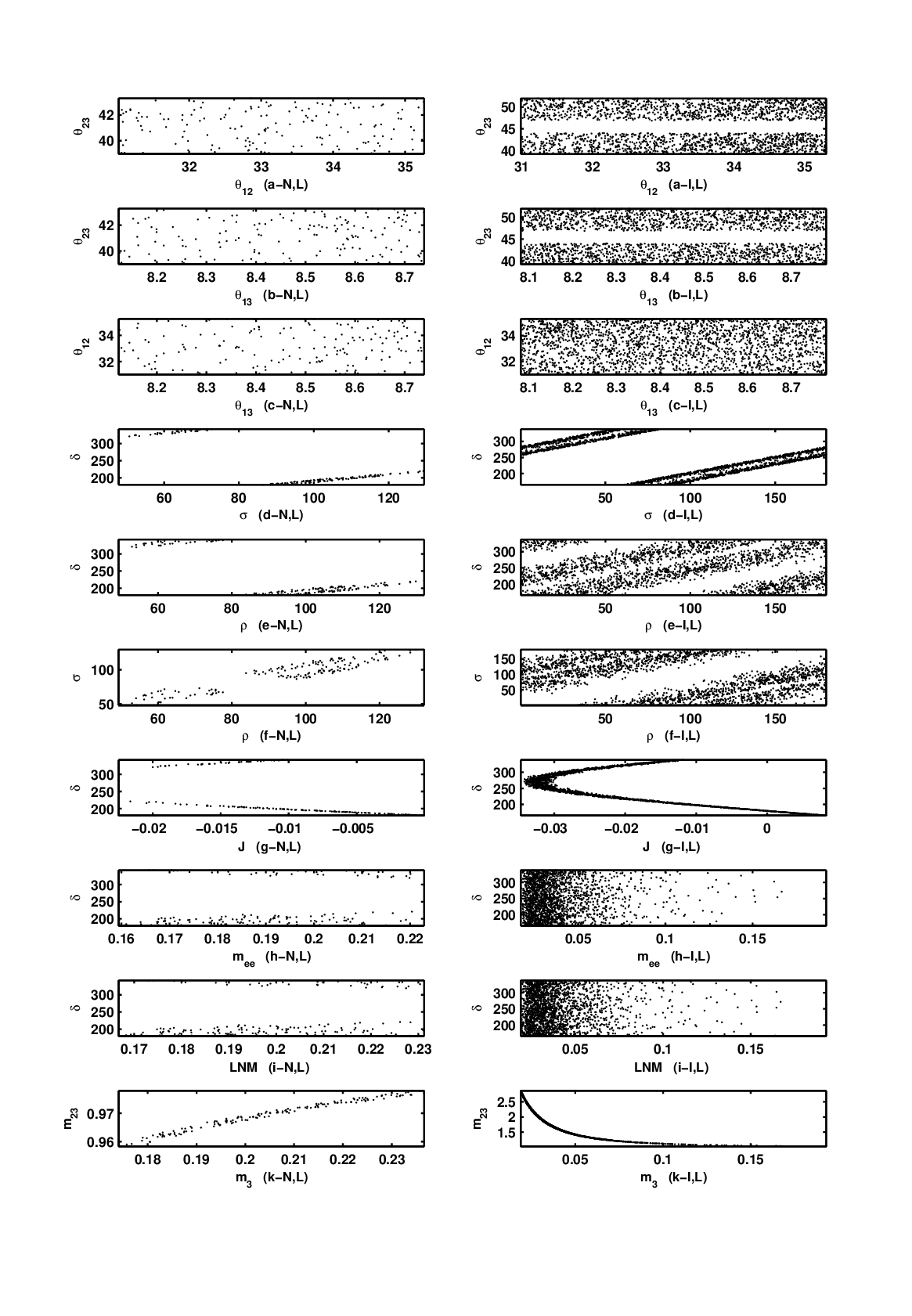}
        \end{tabular}
\end{minipage}
    \begin{minipage}{.5\linewidth}
      \centering
         \begin{tabular}{c}
            \includegraphics[height=25cm,width=8.5cm]{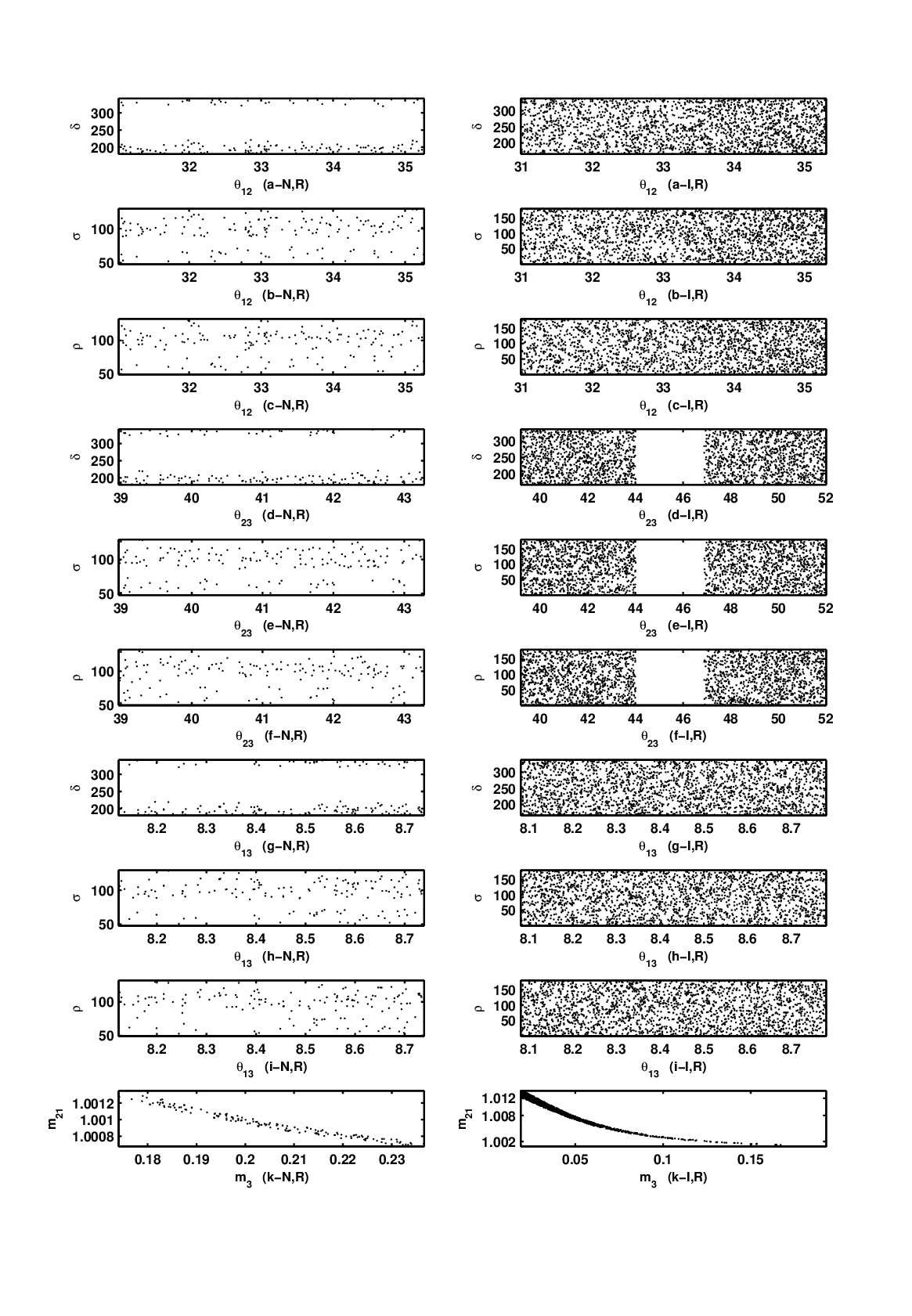}
\end{tabular}
    \end{minipage}
    \vspace{-2.6cm}
     \caption{\footnotesize Pattern $\mathbf{C_{11}} \equiv \mathbf M_{\n\,22} + \mathbf M_{\n\,33}=0$ for non-singular mass matrices: The left panel (the left two columns) shows three correlations amidst the mixing angles, three correlations amidst the phase angles and three correlations of $\d$ with $J, m_{ee}$, LNM (LNM=least neutrino mass) and finally the correlation ($m_3,m_{23}\equiv \frac{m_2}{m_3}$) for normal (N) and Inverted (I) hierarchy. The right panel (the right two columns) shows all the nine inter-correlations between phase angles and mixing angles, and the correlation ($m_3,m_{21}\equiv \frac{m_2}{m_1}$)for normal (N) and Inverted (I) hierarchy. {  Angles (masses) are evaluated in degrees (eV).}}
     \label{fig_Tr2233_2s}
     \end{figure}

For the plots of  Fig.(\ref{fig_Tr2233_2s}),  two ribbons for the correlation ($\d,\s$) exist for both types of hierarchy.  The mass spectrum has a  moderate mass hierarchy  characterized by   $(1 < m_{23}\le 3)$ in the inverted ordering. In contrast, the  mass spectrum is quasi degenerate in the case of normal ordering where $m_1 \approx m_2 \approx m_3$.

\subsection{Pattern $\mathbf C_{12}$: Vanishing of $M_{\n\,21} + M_{\n\, 33}$}
The relevant expressions for $A_1$, $A_2$ and $A_3$ , as defined in Eq.(\ref{Ah}) for this pattern, are
\bea
A_1 &=& -\left(\cx\, \sy\, \sz + \sx \,\cy\, e^{-i\,\d}\right)\,\cx\,\cz + \left(\cx\, \cy\, \sz - \sx\, \sy\, e^{-i\,\d}\right)^2,\nn \\
A_2 & = &  -\left(\sx\, \sy\, \sz - \cx\, \cy\, e^{-i\,\d}\right)\,\sx\,\cz +\left (\sx\, \cy\, \sz + \cx\, \sy\, e^{-i\,\d}\right)^2,\nn\\
A_3 &=&  \sz\,\sy\,\cz + \cy^2\,\cz^2.
\label{Atrace_2133}
\eea

 For a representative point with normal ordering, we take $\theta_{12} = 33.7367^\circ$,   $\theta_{23} = 41.7468^\circ$,   $\theta_{13} = 8.4134^\circ$,    $\delta = 312.5765^\circ$,    $\rho = 151.9557^\circ$,    $\sigma = 63.7910^\circ$,    $ m_1 = 0.0458\, \mbox{eV}$,    $ m_2 = 0.0466\, \mbox{eV}$,   $ m_3 = 0.0684\, \mbox{eV}$,    $ \me = 0.0466\, \mbox{eV}$,    $ \mee = 0.0178\, \mbox{eV}$, with the  corresponding mass matrix (in eV):
\be
M_\n = \begin{pmatrix}
  0.0102-0.0146i &  -0.0255+0.0052i &    0.0343-0.0017i \\
-0.0255+0.0052i  &  0.0288-0.0081i &    0.0404+0.0062i \\
 0.0343-0.0017i  &  0.0404+0.0062i &    0.0255-0.0052i\\
\end{pmatrix}.
\ee

For an inverted hierarchy representative point we take $\theta_{12} = 33.2436^\circ$,    $\theta_{23} = 41.4914^\circ$,    $\theta_{13} = 8.6242^\circ$,    $\delta = 236.7486^\circ$,    $\rho = 149.9638^\circ$,    $\sigma = 123.9278^\circ$,    $ m_1 = 0.0507\, \mbox{eV}$,    $ m_2 = 0.0515\, \mbox{eV}$,   $ m_3 = 0.0115\, \mbox{eV}$,    $ \me = 0.0504\, \mbox{eV}$,    $ \mee = 0.0456\, \mbox{eV}$ with the  corresponding mass matrix (in eV):
\be
M_\n = \begin{pmatrix}
  0.0118-0.0440i &    0.0093-0.0076i &   -0.0083+0.0156i\\
 0.0093-0.0076i &  -0.0194+0.0154i &    0.0255-0.0121i \\
-0.0083+0.0156i &   0.0255-0.0121i &  -0.0093+0.0076i
\end{pmatrix}.
\ee

We see, from  Table(\ref{tab_pred_non_sing}), that $m_3$ can reach zero in inverted type, so we expect a possible singular texture existing. Again $J$ at $1$-$2\s$-levels for normal ordering and $1\s$ for inverted ordering is negative so the corresponding $\d$ is at third or fourth quarters.  For normal ordering,  at $1$-$\s$-level there is a gap $ [6^o,103^o] $ ($[56^o,157^o]$) for $\r$ ($\s$) which becomes at  $2$-$\s$-level $ [34^o,102^o] $ ($[91^o,152^o]$).

\begin{figure}[H]
\begin{minipage}{.5\linewidth}
           \centering
        \begin{tabular}{c}
             \includegraphics[height=25cm,width=8.5cm]{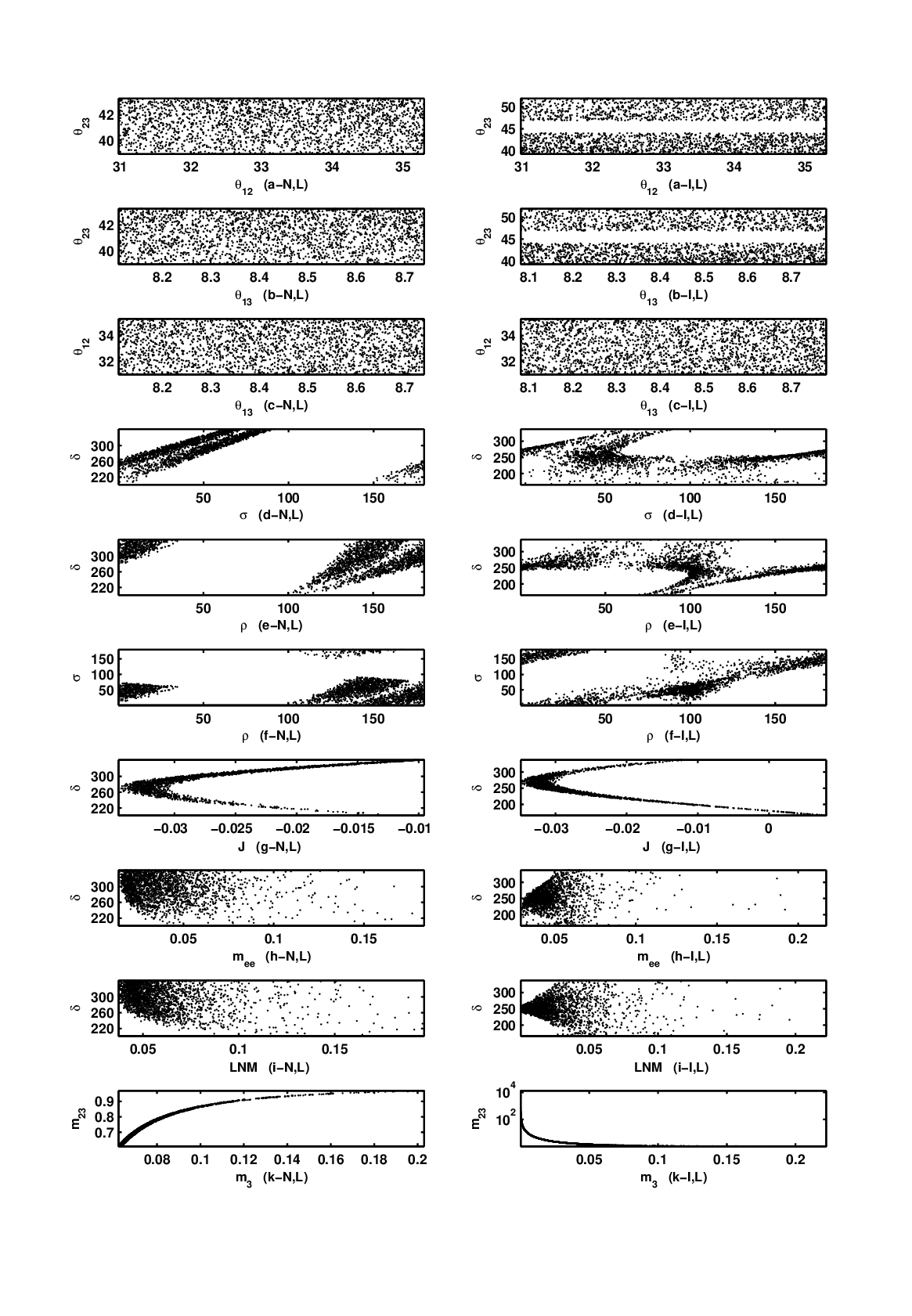}
        \end{tabular}
\end{minipage}
    \begin{minipage}{.5\linewidth}
      \centering
         \begin{tabular}{c}
            \includegraphics[height=25cm,width=8.5cm]{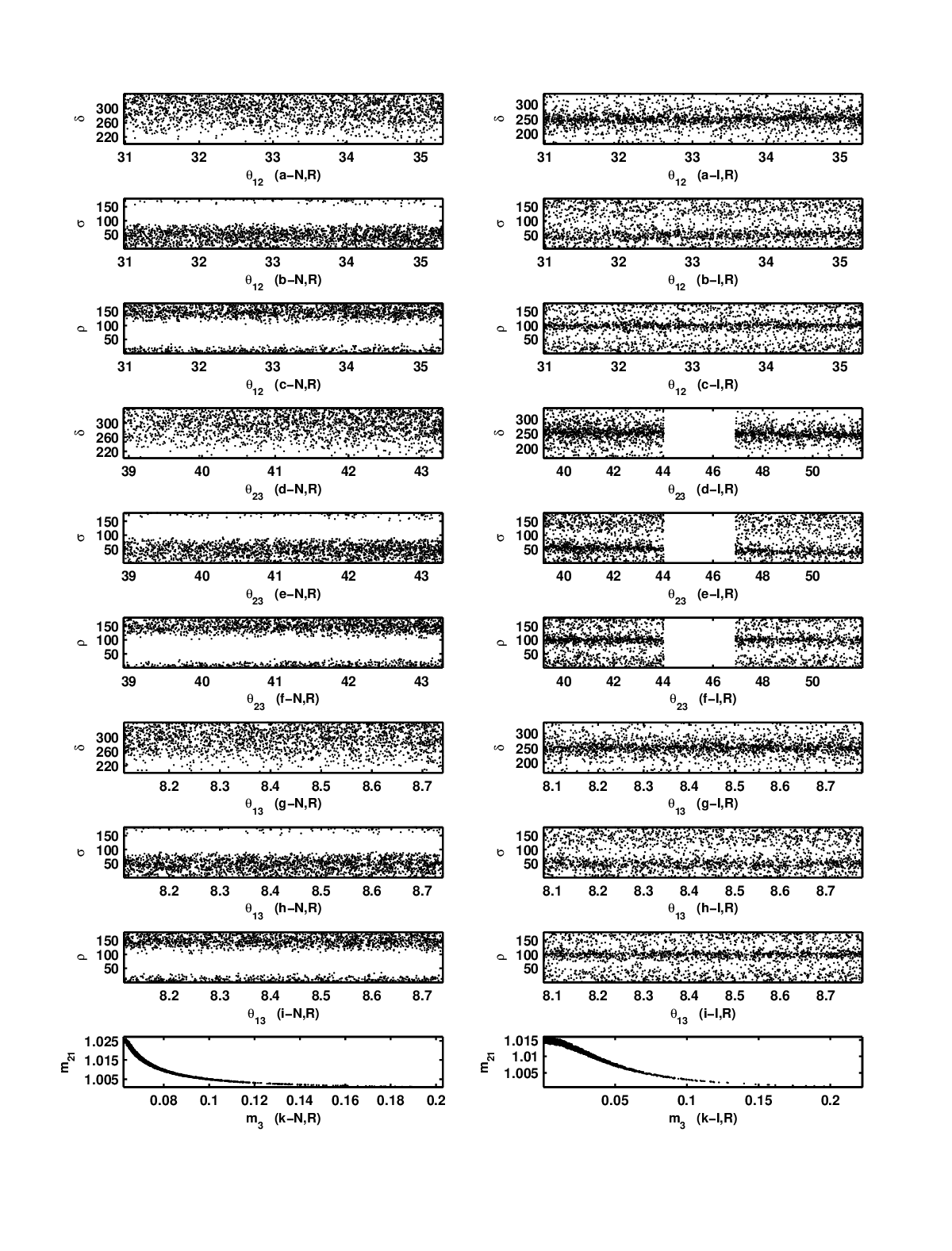}
\end{tabular}
    \end{minipage}
    \vspace{-2.6cm}
     \caption{\footnotesize Pattern $\mathbf{C_{12}} \equiv \mathbf M_{\n\,21} + \mathbf M_{\n\,33}=0$ for non-singular mass matrices: The left panel (the left two columns) shows three correlations amidst the mixing angles, three correlations amidst the phase angles and three correlations of $\d$ with $J, m_{ee}$, LNM (LNM=least neutrino mass) and finally the correlation ($m_3,m_{23}\equiv \frac{m_2}{m_3}$) for normal (N) and Inverted (I) hierarchy. The right panel (the right two columns) shows all the nine inter-correlations between phase angles and mixing angles, and the correlation ($m_3,m_{21}\equiv \frac{m_2}{m_1}$)for normal (N) and Inverted (I) hierarchy. {  Angles (masses) are evaluated in degrees (eV).}}
     \label{fig_Tr2133_2s}
     \end{figure}

For the plots of  Fig.(\ref{fig_Tr2133_2s}) in normal ordering, we find large forbidden gaps for $\r$ and  $\s$ and a quasi degenerate mass spectrum where $(0.6 \le m_{23} \le 0.95)$. As to the plots of  Fig.(\ref{fig_Tr2133_2s}) in inverted type, we see that we may get an acute hierarchy with $m_{23}$  reaching up to $10^4$ which reveals the possibility of vanishing $m_3$.

\subsection{Pattern $\mathbf C_{13}$: Vanishing of $M_{\n\,21} + M_{\n\, 23}$}
The relevant expressions for $A_1$, $A_2$ and $A_3$ , as defined in Eq.(\ref{Ah}) for this pattern, are
\bea
A_1 &=& \left(\cx\, \sy\, \sz + \sx\, \cy\, e^{-i\,\d}\right)\,\left(\cx\, \cy\, \sz  - \cx\,\cz -\sx\, \sy\, e^{-i\,\d}\right),\nn \\
A_2 & = &\left(\sx\, \sy \,\sz - \cx\, \cy\, e^{-i\,\d}\right)\,\left(\sx\, \cy\, \sz  - \sx\,\cz + \cx\, \sy\, e^{-i\,\d}\right),\nn\\
A_3 &=& \cz\,\sy\,\left( \cy\,\cz + \sz \right).
\label{Atrace_1223}
\eea

For a representative point with normal ordering, we take $\theta_{12} = 33.8222^\circ$,    $\theta_{23} = 40.4289^\circ$,    $\theta_{13} = 8.7721^\circ$,    $\delta = 243.7429^\circ$,    $\rho = 148.0834^\circ$,    $\sigma = 34.0333^\circ$,    $ m_1 = 0.1821\, \mbox{eV}$,    $ m_2 = 0.1823\, \mbox{eV}$,   $ m_3 = 0.1888\, \mbox{eV}$,    $ \me = 0.1823\, \mbox{eV}$,    $ \mee = 0.0987\, \mbox{eV}$ with the  corresponding mass matrix (in eV):
\be
M_\n = \begin{pmatrix}
  0.0791-0.0590i &   -0.0909-0.0491i  &  0.0997+0.0538i\\
-0.0909-0.0491i  &  0.1037-0.0459i &   0.0909+0.0491i \\
 0.0997+0.0538i  &   0.0909+0.0491i &    0.0911-0.0527i \\
\end{pmatrix}.
\ee

Equally, for an inverted hierarchy we can take a representative point as follows: $\theta_{12} = 33.8850^\circ$,    $\theta_{23} = 42.2823^\circ$,    $\theta_{13} = 8.5649^\circ$,    $\delta = 244.3791^\circ$,    $\rho = 48.7884^\circ$,    $\sigma = 57.5072^\circ$,    $ m_1 = 0.0644\, \mbox{eV}$,    $ m_2 = 0.0650\, \mbox{eV}$,   $ m_3 = 0.0421\, \mbox{eV}$,    $ \me = 0.0642\, \mbox{eV}$,    $ \mee = 0.0623\, \mbox{eV}$ with the  corresponding mass matrix (in eV):
\be
M_\n = \begin{pmatrix}
 -0.0131+0.0609i  &  0.0099-0.0112i &    0.0023-0.0022i\\
 0.0099-0.0112i   & 0.0508-0.0098i &   -0.0099+0.0112i\\
 0.0023-0.0022i   & -0.0099+0.0112i  &  0.0507-0.0098i\\
\end{pmatrix}.
\ee

We see, from Table(\ref{tab_pred_non_sing}),  that $m_3$ can reach zero in inverted type, so we expect a possible singular texture existing.  Table(\ref{tab_pred_non_sing}) also reveals that $J$, at $1$-$2\s$-levels for both  normal and inverted  ordering, is negative so the corresponding $\d$ is in third or fourth quarters.  For normal ordering, the ranges for $\r$ ($\s$) are restricted to be  $ [42^\circ,155^\circ]$ ($[17^\circ,102^\circ]$) at $1\s$-level, whereas they tend to be wider at $3\s$-level covering $ [6^\circ,175^\circ]$ ($[0.01^\circ,180^\circ]$).

\begin{figure}[H]
\begin{minipage}{.5\linewidth}
           \centering
        \begin{tabular}{c}
             \includegraphics[height=25cm,width=8.5cm]{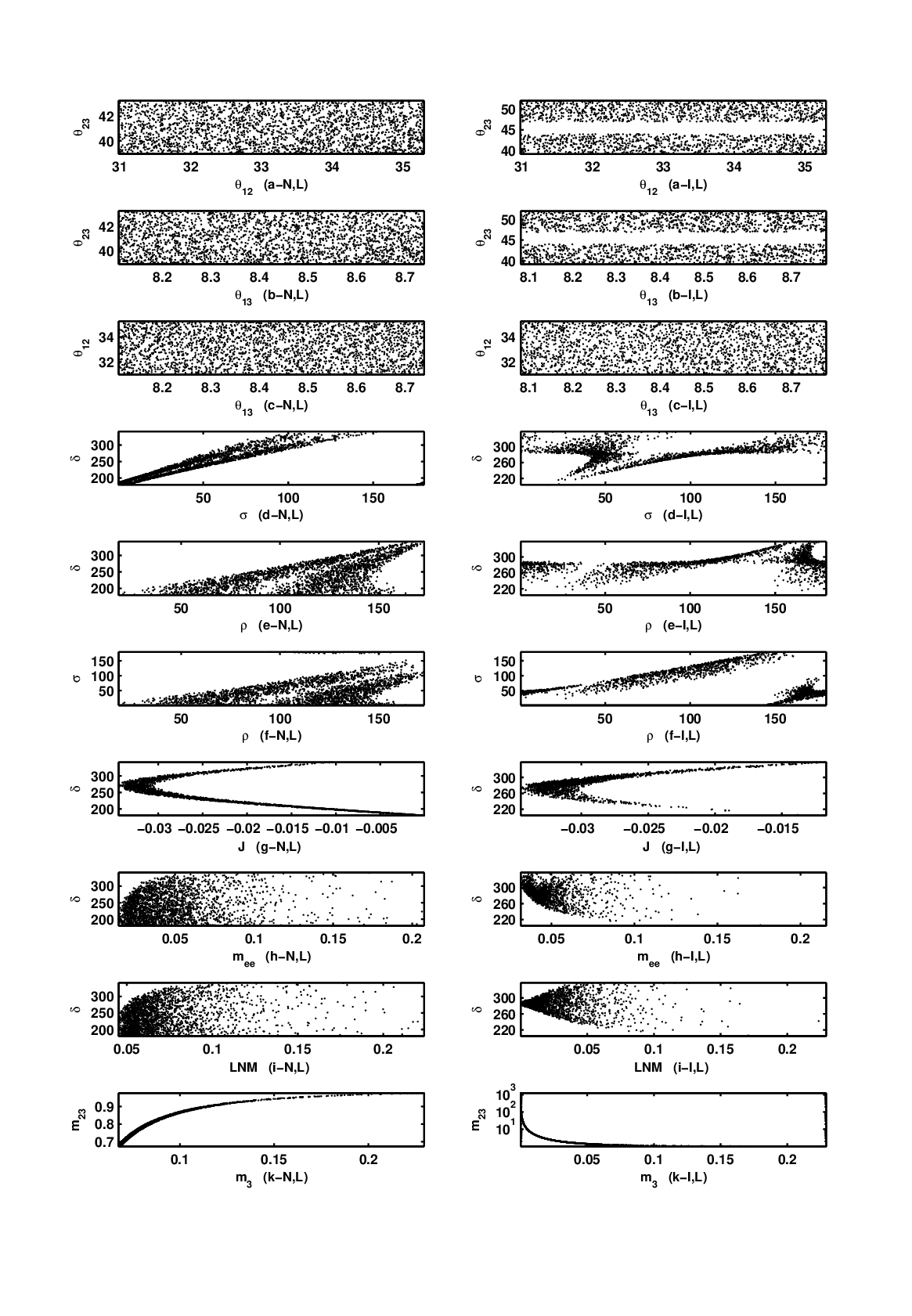}
        \end{tabular}
\end{minipage}
    \begin{minipage}{.5\linewidth}
      \centering
         \begin{tabular}{c}
            \includegraphics[height=25cm,width=8.5cm]{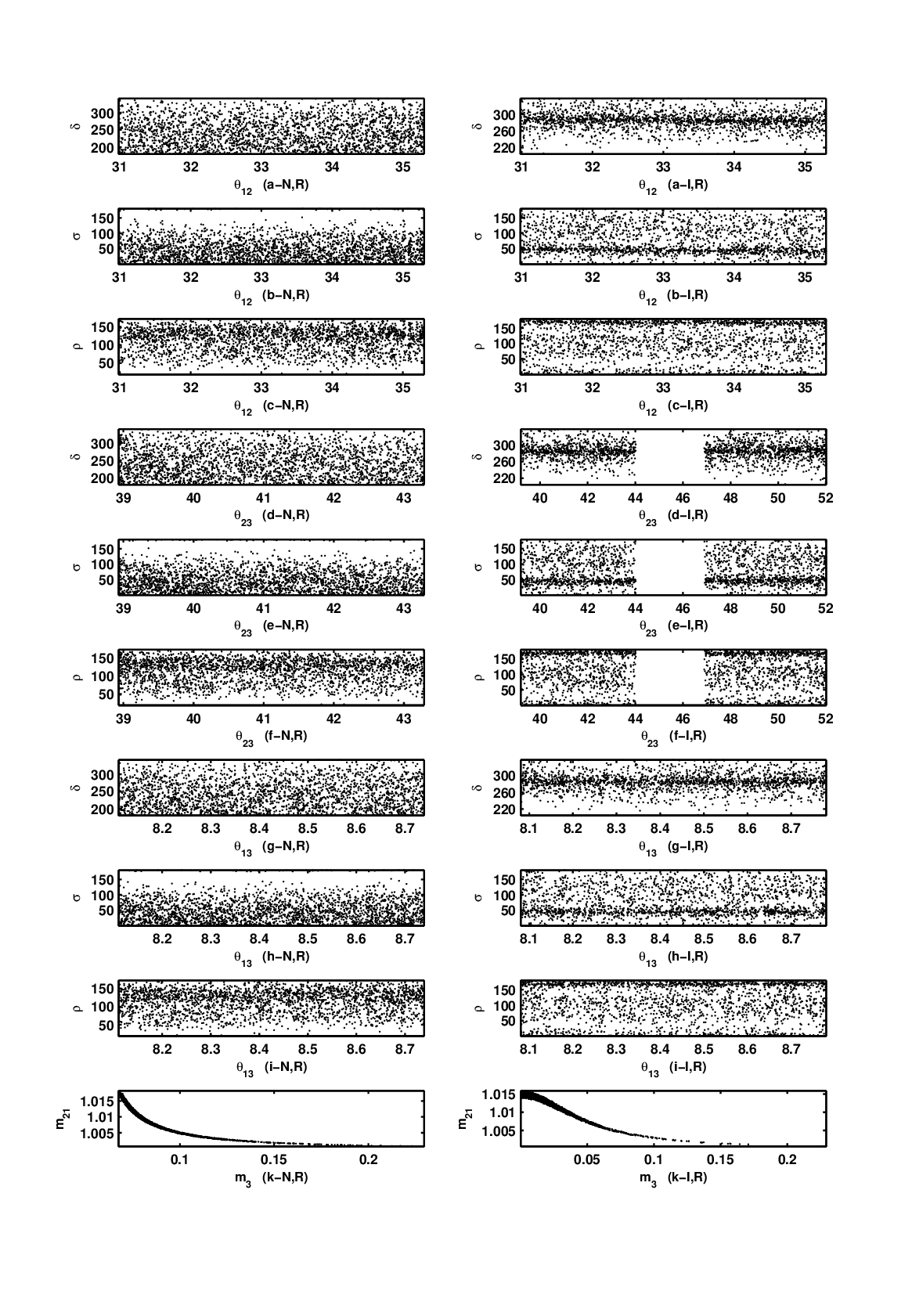}
\end{tabular}
    \end{minipage}
    \vspace{-2.6cm}
     \caption{\footnotesize Pattern $\mathbf{C_{13}} \equiv \mathbf M_{\n\,21} + \mathbf M_{\n\,23}=0$ for non-singular mass matrices: The left panel (the left two columns) shows three correlations amidst the mixing angles, three correlations amidst the phase angles and three correlations of $\d$ with $J, m_{ee}$, LNM (LNM=least neutrino mass) and finally the correlation ($m_3,m_{23}\equiv \frac{m_2}{m_3}$) for normal (N) and Inverted (I) hierarchy. The right panel (the right two columns) shows all the nine inter-correlations between phase angles and mixing angles, and the correlation ($m_3,m_{21}\equiv \frac{m_2}{m_1}$)for normal (N) and Inverted (I) hierarchy. {  Angles (masses) are evaluated in degrees (eV).}}
     \label{fig_Tr1223_2s}
     \end{figure}

For the plots of  Fig.(\ref{fig_Tr1223_2s}) in normal ordering, we find a quasi degenerate mass spectrum where $(0.65 \le m_{23} \le 0.95)$. As to the plots of  Fig.(\ref{fig_Tr1223_2s}) in inverted type, we may get an acute hierarchy with $m_{23}$ reaching up to $10^3$, so a vanishing $m_3$ is possible.


\subsection{Pattern $\mathbf C_{22}$: Vanishing of $M_{\n\,11} + M_{\n\, 33}$}
The relevant expressions for $A_1$, $A_2$ and $A_3$ , as defined in Eq.(\ref{Ah}) for this pattern, are
\bea
A_1 &=& \cx^2\, \cz^2 + \left(\cx\, \cy\, \sz - \sx\, \sy\, e^{-i\,\d}\right)^2,\nn \\
A_2 & = &\sx^2\, \cz^2 + \left(\sx\, \cy\, \sz + \cx\, \sy\, e^{-i\,\d}\right)^2,\nn\\
A_3 &=& \sz^2 + \cy^2\, \cz^2.
\label{Atrace_1133}
\eea

For a representative point with normal ordering, we take $\theta_{12} = 33.8006^\circ$,    $\theta_{23} = 40.7648^\circ$,    $\theta_{13} = 8.4791^\circ$,    $\delta = 300.9481^\circ$,    $\rho = 81.5950^\circ$,    $\sigma = 69.8454^\circ$,    $ m_1 = 0.0398\, \mbox{eV}$,    $ m_2 = 0.0407\, \mbox{eV}$,   $ m_3 = 0.0647\, \mbox{eV}$,    $ \me = 0.0408\, \mbox{eV}$,    $ \mee = 0.0372\, \mbox{eV}$ with the  corresponding mass matrix (in eV):
\be
M_\n = \begin{pmatrix}
 -0.0337+0.0157i &   0.0064+0.0032i &    0.0138-0.0059i\\
 0.0064+0.0032i &   0.0252-0.0235i &    0.0327+0.0196i\\
 0.0138-0.0059i  &  0.0327+0.0196i &    0.0337-0.0157i\\
\end{pmatrix}.
\ee

For an inverted hierarchy representative point, we take  $\theta_{12} = 33.5774^\circ$,    $\theta_{23} = 42.7607^\circ$,    $\theta_{13} = 8.7549^\circ$,    $\delta = 281.0485^\circ$,    $\rho = 99.6048^\circ$,    $\sigma = 167.4130^\circ$,    $ m_1 = 0.0749\, \mbox{eV}$,    $ m_2 = 0.0754\, \mbox{eV}$,   $ m_3 = 0.0571\, \mbox{eV}$,    $ \me = 0.0747\, \mbox{eV}$,    $ \mee = 0.0372\, \mbox{eV}$ with the  corresponding mass matrix (in eV):
\be
M_\n = \begin{pmatrix}
-0.0263-0.0263i &   0.0201+0.0479i &   -0.0011-0.0388i\\
 0.0201+0.0479i  &  0.0162+0.0104i &    0.0336-0.0196i \\
-0.0011-0.0388i  &  0.0336-0.0196i &    0.0263+0.0263i\\
\end{pmatrix}.
\ee

We see,  from Table(\ref{tab_pred_non_sing}), that $m_3$ can reach zero in inverted type, so we expect a possible singular texture existing. Again,  from Table(\ref{tab_pred_non_sing}), $J$ at $1$-$2\s$-levels for normal ordering and $1\s$ for inverted ordering is negative so the corresponding $\d$ is in third or fourth quarters.  For normal ordering, values of $\r$ are restricted to fall in the range  $[52^\circ,128^\circ]$  at the  $3\s$-level.

\begin{figure}[H]
\begin{minipage}{.5\linewidth}
           \centering
        \begin{tabular}{c}
             \includegraphics[height=25cm,width=8.5cm]{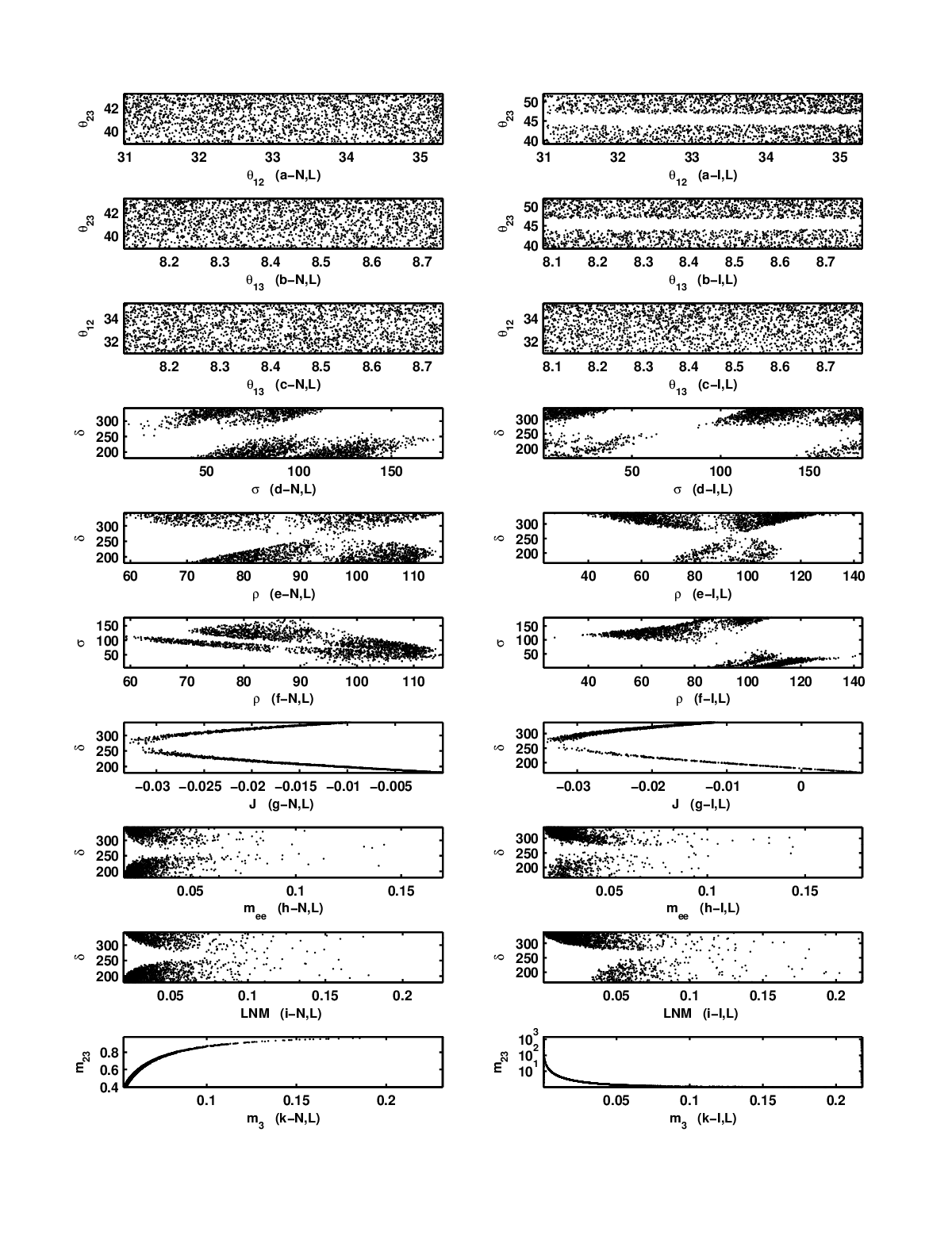}
        \end{tabular}
\end{minipage}
    \begin{minipage}{.5\linewidth}
      \centering
         \begin{tabular}{c}
            \includegraphics[height=25cm,width=8.5cm]{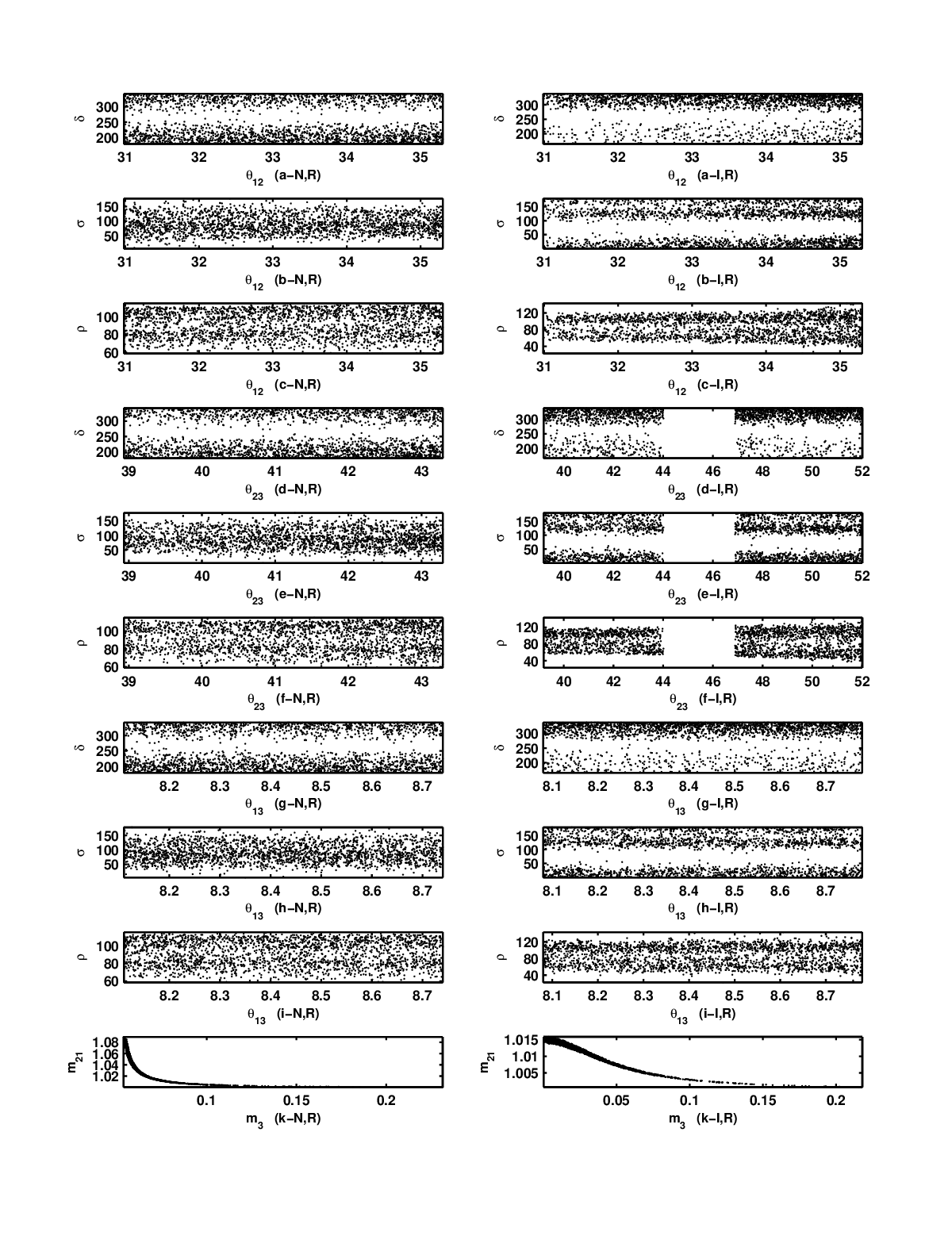}
\end{tabular}
    \end{minipage}
    \vspace{-2.6cm}
     \caption{\footnotesize Pattern $\mathbf{C_{22}} \equiv \mathbf M_{\n\,11} + \mathbf M_{\n\,33}=0$ for non-singular mass matrices: The left panel (the left two columns) shows three correlations amidst the mixing angles, three correlations amidst the phase angles and three correlations of $\d$ with $J, m_{ee}$, LNM (LNM=least neutrino mass) and finally the correlation ($m_3,m_{23}\equiv \frac{m_2}{m_3}$) for normal (N) and Inverted (I) hierarchy. The right panel (the right two columns) shows all the nine inter-correlations between phase angles and mixing angles, and the correlation ($m_3,m_{21}\equiv \frac{m_2}{m_1}$)for normal (N) and Inverted (I) hierarchy. {  Angles (masses) are evaluated in degrees (eV).}}
     \label{fig_Tr1133_2s}
     \end{figure}

For the plots ,  Fig.(\ref{fig_Tr1133_2s}),  in normal ordering, we find  forbidden bands for ($\s,\r$) and a moderate  mass hierarchy where $ ( 0.45 \le m_{23} < 1)$. As to the plots of  Fig.(\ref{fig_Tr1133_2s})  in inverted type, there are  forbidden bands for ($\s,\r$) and the mass hierarchy can become acute with $m_{23}$ reaching up to $10^3$ making a vanishing $m_3$ possible.

\subsection{Pattern $\mathbf C_{23}$: Vanishing of $M_{\n\,11} + M_{\n\, 23}$}
The relevant expressions for $A_1$, $A_2$ and $A_3$ , as defined in Eq.(\ref{Ah}) for this pattern, are
\bea
A_1 &=& \cx^2\, \cz^2 + \left(\cx\, \sy\, \sz + \sx\, \cy\, e^{-i\,\d}\right)\,\left(\cx\, \cy\, \sz - \sx\, \sy\, e^{-i\,\d}\right),\nn \\
A_2 & = & \sx^2\, \cz^2 + \left(\sx\, \sy\, \sz - \cx\, \cy\, e^{-i\,\d}\right)\,\left(\sx\, \cy\, \sz + \cx\, \sy\, e^{-i\,\d}\right),\nn\\
A_3 &=& \sz^2 + \sy \, \cy \, \cz^2.
\label{Atrace_1123}
\eea

For a representative point with normal ordering, we take $\theta_{12} = 33.4546^\circ$,    $\theta_{23} = 42.2981^\circ$,    $\theta_{13} = 8.4653^\circ$,    $\delta = 248.6157^\circ$,    $\rho = 94.3533^\circ$,    $\sigma = 68.6630^\circ$,    $ m_1 = 0.0203\, \mbox{eV}$,    $ m_2 = 0.0221\, \mbox{eV}$,   $ m_3 = 0.0554\, \mbox{eV}$,    $ \me = 0.0222\, \mbox{eV}$,    $ \mee = 0.0175\, \mbox{eV}$ with the  corresponding mass matrix (in eV):
\be
M_\n = \begin{pmatrix}
-0.0173+0.0024i &   0.0012-0.0013i &   0.0136+0.0007i\\
 0.0012-0.0013i &   0.0361+0.0029i  &  0.0173-0.0024i\\
 0.0136+0.0007i  &  0.0173-0.0024i &   0.0369+0.0020i\\
\end{pmatrix}.
\ee

For an inverted hierarchy representative point we take $\theta_{12} = 33.2679^\circ$,    $\theta_{23} = 42.8064^\circ$,    $\theta_{13} = 8.6838^\circ$,    $\delta = 236.7459^\circ$,    $\rho = 98.4416^\circ$,    $\sigma = 2.2660^\circ$,    $ m_1 = 0.0573\, \mbox{eV}$,    $ m_2 = 0.0579\, \mbox{eV}$,   $ m_3 = 0.0288\, \mbox{eV}$,    $ \me = 0.0570\, \mbox{eV}$,    $ \mee = 0.0222\, \mbox{eV}$ with the  corresponding mass matrix (in eV):
\be
M_\n = \begin{pmatrix}
-0.0198-0.0100i &   -0.0214+0.0285i &    0.0299-0.0243i\\
-0.0214+0.0285i  &  0.0122-0.0172i &    0.0198+0.0100i\\
 0.0299-0.0243i  &  0.0198+0.0100i &   0.0042-0.0042i\\
\end{pmatrix}.
\ee

We see,  from Table(\ref{tab_pred_non_sing}), that $m_3$ can not reach zero, so we expect no viable corresponding singular pattern. Again,  from Table(\ref{tab_pred_non_sing}), $J$ at $1$-$2\s$-levels for normal ordering and $1\s$ for inverted ordering is negative so the corresponding $\d$ is in third or fourth quarters. For both  normal and inverted ordering, the phase $\r$ is bound at all $\s$-levels to be nearly  in the interval ($[60^\circ,120^\circ]$).

\begin{figure}[H]
\begin{minipage}{.5\linewidth}
           \centering
        \begin{tabular}{c}
             \includegraphics[height=25cm,width=8.5cm]{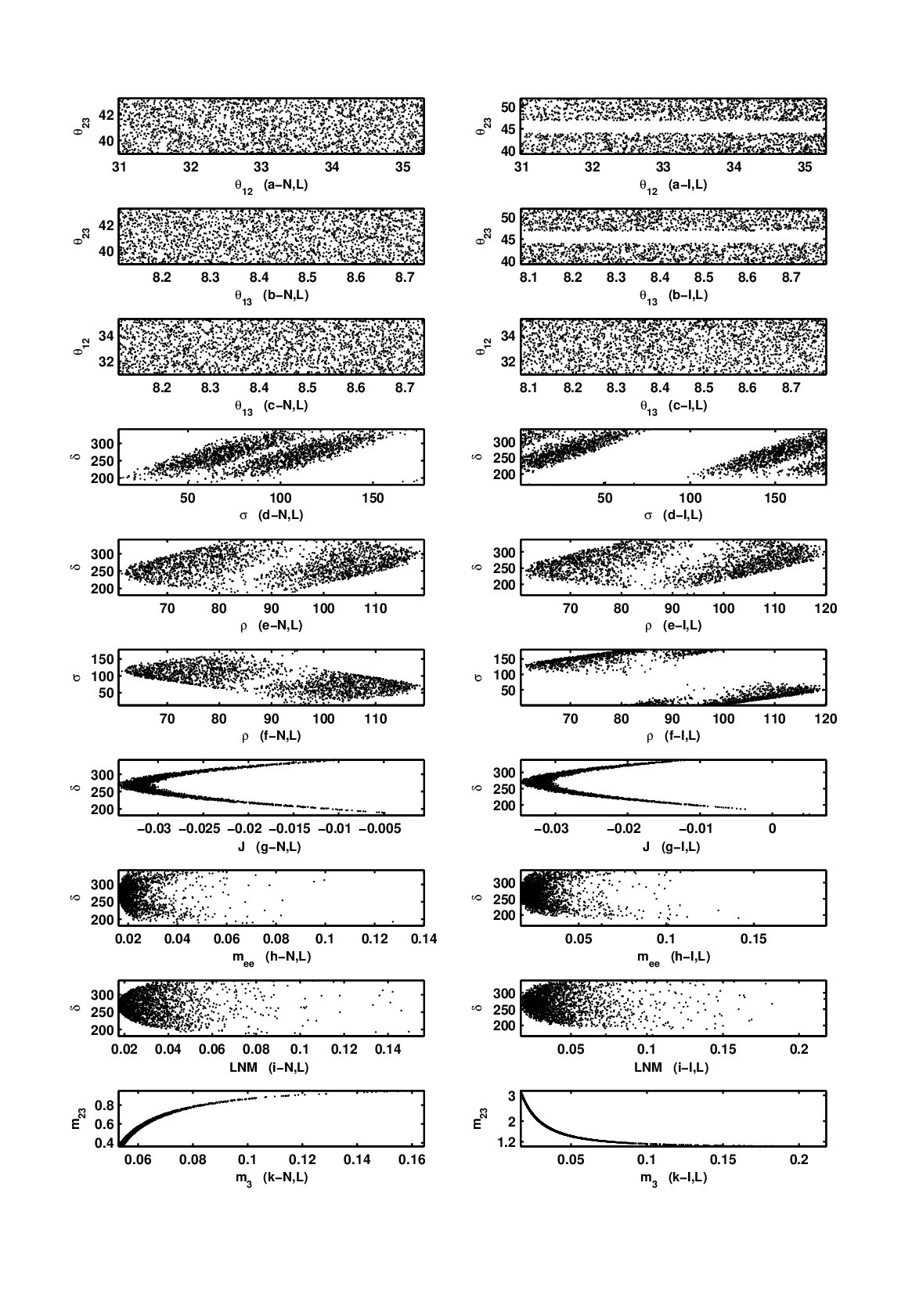}
        \end{tabular}
\end{minipage}
    \begin{minipage}{.5\linewidth}
      \centering
         \begin{tabular}{c}
            \includegraphics[height=25cm,width=8.5cm]{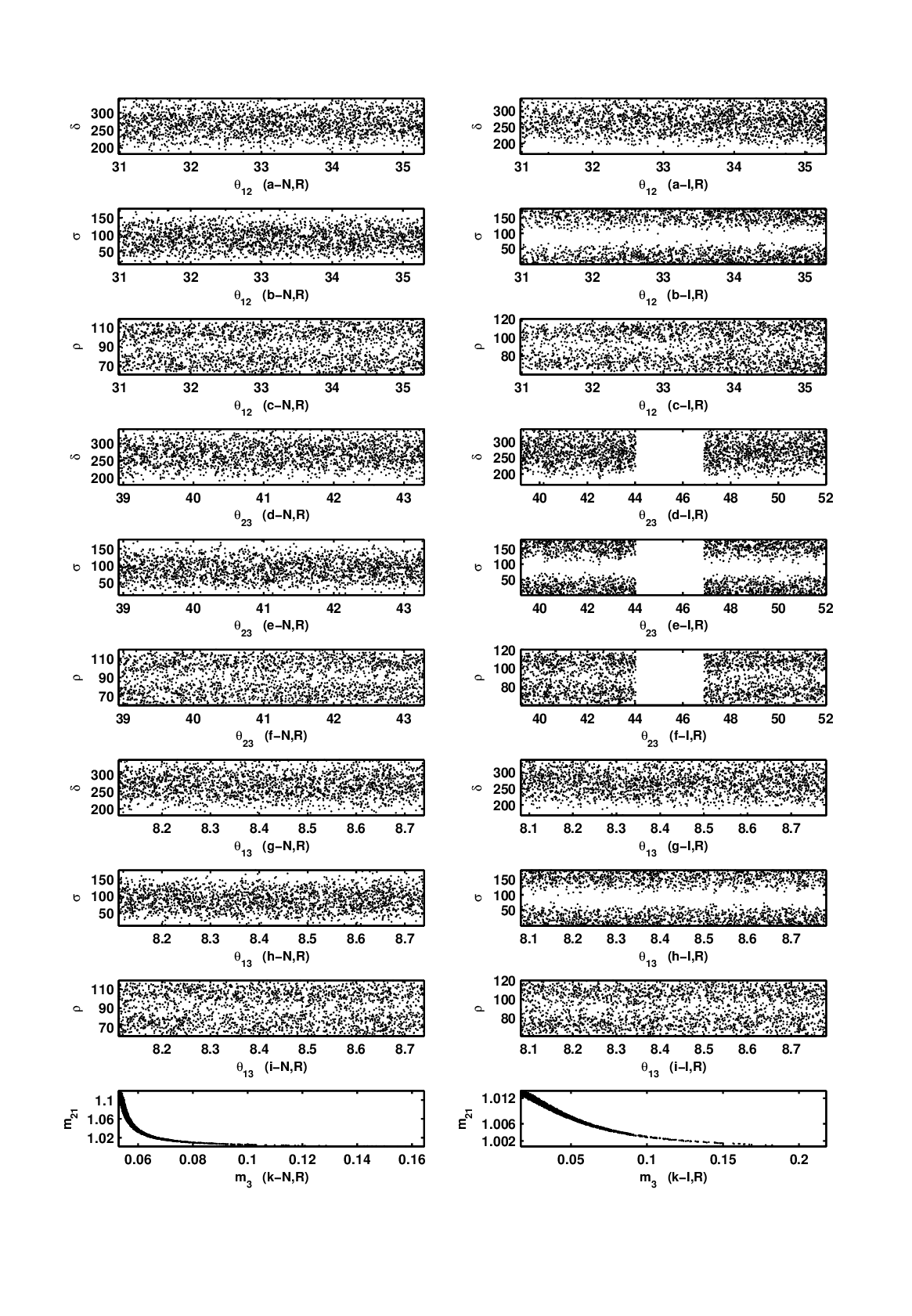}
\end{tabular}
    \end{minipage}
    \vspace{-2.6cm}
     \caption{\footnotesize Pattern $\mathbf{C_{23}} \equiv \mathbf M_{\n\,11} + \mathbf M_{\n\,23}=0$ for non-singular mass matrices: The left panel (the left two columns) shows three correlations amidst the mixing angles, three correlations amidst the phase angles and three correlations of $\d$ with $J, m_{ee}$, LNM (LNM=least neutrino mass) and finally the correlation ($m_3,m_{23}\equiv \frac{m_2}{m_3}$) for normal (N) and Inverted (I) hierarchy. The right panel (the right two columns) shows all the nine inter-correlations between phase angles and mixing angles, and the correlation ($m_3,m_{21}\equiv \frac{m_2}{m_1}$)for normal (N) and Inverted (I) hierarchy. {  Angles (masses) are evaluated in degrees (eV).}}
     \label{fig_Tr1123_2s}
     \end{figure}

For the plots,  Fig.(\ref{fig_Tr1123_2s}),  in both normal and inverted ordering, we get an approximately degenerate spectrum characterized respectively by $ (0.4 \le m_{23} \le 0.9)$ and  $(1.2 \le m_{23} \le 3)$. The plots  in Fig.(\ref{fig_Tr1123_2s}),  also reveal that the phase $\r$ is bound to fall approximately  in the interval ($[60^\circ,120^\circ]$),  while there are forbidden bands for the phase $\s$ for both type of hierarchies.

\subsection{Pattern $\mathbf C_{33}$: Vanishing of $M_{\n\,11} + M_{\n\, 22}$}
The relevant expressions for $A_1$, $A_2$ and $A_3$ , as defined in Eq.(\ref{Ah}) for this pattern, are
\bea
A_1 &=& \cx^2\, \cz^2 + (\cx \,\sy\, \sz + \sx\, \cy\, e^{-i\,\d})^2,\nn \\
A_2 & = & \sx^2\, \cz^2 + (\sx \, \sy\, \sz - \cx \,\cy\, e^{-i\,\d})^2,\nn\\
A_3 &=& \sz^2 + \sy^2 \, \cz^2.
\label{Atrace_1122}
\eea

As for a normal type representative point, we take $\theta_{12} = 33.5935^\circ$,    $\theta_{23} = 40.7528^\circ$,    $\theta_{13} = 8.7162^\circ$,    $\delta = 252.1164^\circ$,    $\rho = 77.1818^\circ$,    $\sigma = 164.3730^\circ$,    $ m_1 = 0.0530\, \mbox{eV}$,    $ m_2 = 0.0537\, \mbox{eV}$,   $ m_3 = 0.0736\, \mbox{eV}$,    $ \me = 0.0537\, \mbox{eV}$,    $ \mee = 0.0184\, \mbox{eV}$ with the  corresponding mass matrix (in eV):
\be
M_\n = \begin{pmatrix}
-0.0170+0.0072i &   0.0158+0.0354i  &  0.0047-0.0320i\\
 0.0158+0.0354i &   0.0170-0.0072i  &  0.0456-0.0010i\\
 0.0047-0.0320i  &  0.0456-0.0010i  &  0.0333+0.0073i\\
\end{pmatrix}.
\ee

For an inverted type representative point, we can take $\theta_{12} = 33.0850^\circ$,    $\theta_{23} = 42.6054^\circ$,    $\theta_{13} = 8.7610^\circ$,    $\delta = 221.0642^\circ$,    $\rho = 123.4419^\circ$,    $\sigma = 60.2387^\circ$,    $ m_1 = 0.0540\, \mbox{eV}$,    $ m_2 = 0.0547\, \mbox{eV}$,   $ m_3 = 0.0231\, \mbox{eV}$,    $ \me = 0.0537\, \mbox{eV}$,    $ \mee = 0.0300\, \mbox{eV}$ with the  corresponding mass matrix (in eV):
\be
M_\n = \begin{pmatrix}
-0.0221-0.0204i &  -0.0148-0.0236i &    0.0231+0.0260i\\
-0.0148-0.0236i &   0.0221+0.0204i &   0.0041-0.0138i\\
 0.0231+0.0260i &  0.0041-0.0138i  &  0.0146+0.0072i\\
\end{pmatrix}.
\ee

We see, from Table(\ref{tab_pred_non_sing}), that $m_3$ can reach zero in inverted type, so we expect a viable corresponding singular pattern. Again,  from Table(\ref{tab_pred_non_sing}), $J$ at $1$-$2\s$-levels for normal ordering and $1\s$ for inverted ordering is negative so the corresponding $\d$ is in third or fourth quarters. For normal ordering, the values of the phase $\r$ are restricted to fall in the range $[64^\circ,126^\circ]$  at the  $1\s$-level, and in $[56^\circ,129^\circ]$  at the  $2\s$-level, and in  $[52^\circ,130^\circ]$  at the  $3\s$-level, but, in contrast, there is almost no restriction for $\s$.  For inverted ordering, there is a restriction for the phase $\r$ range: $[57^\circ,132^\circ]$ at $1$-$\s$-level, $[15^\circ,170^\circ]$  at $2$-$\s$-level  and  $[1.5^\circ,173^\circ]$
at $3$-$\s$-level. In contrast, there is a forbidden gap for $\s$ which is $[94^\circ,142^\circ]$ at $1$-$\s$-level,  $[86^\circ,106^\circ]$ at $2$-$\s$-level and $[83^\circ,100^\circ]$ at $3$-$\s$-level.

\begin{figure}[H]
\begin{minipage}{.5\linewidth}
           \centering
        \begin{tabular}{c}
             \includegraphics[height=25cm,width=8.5cm]{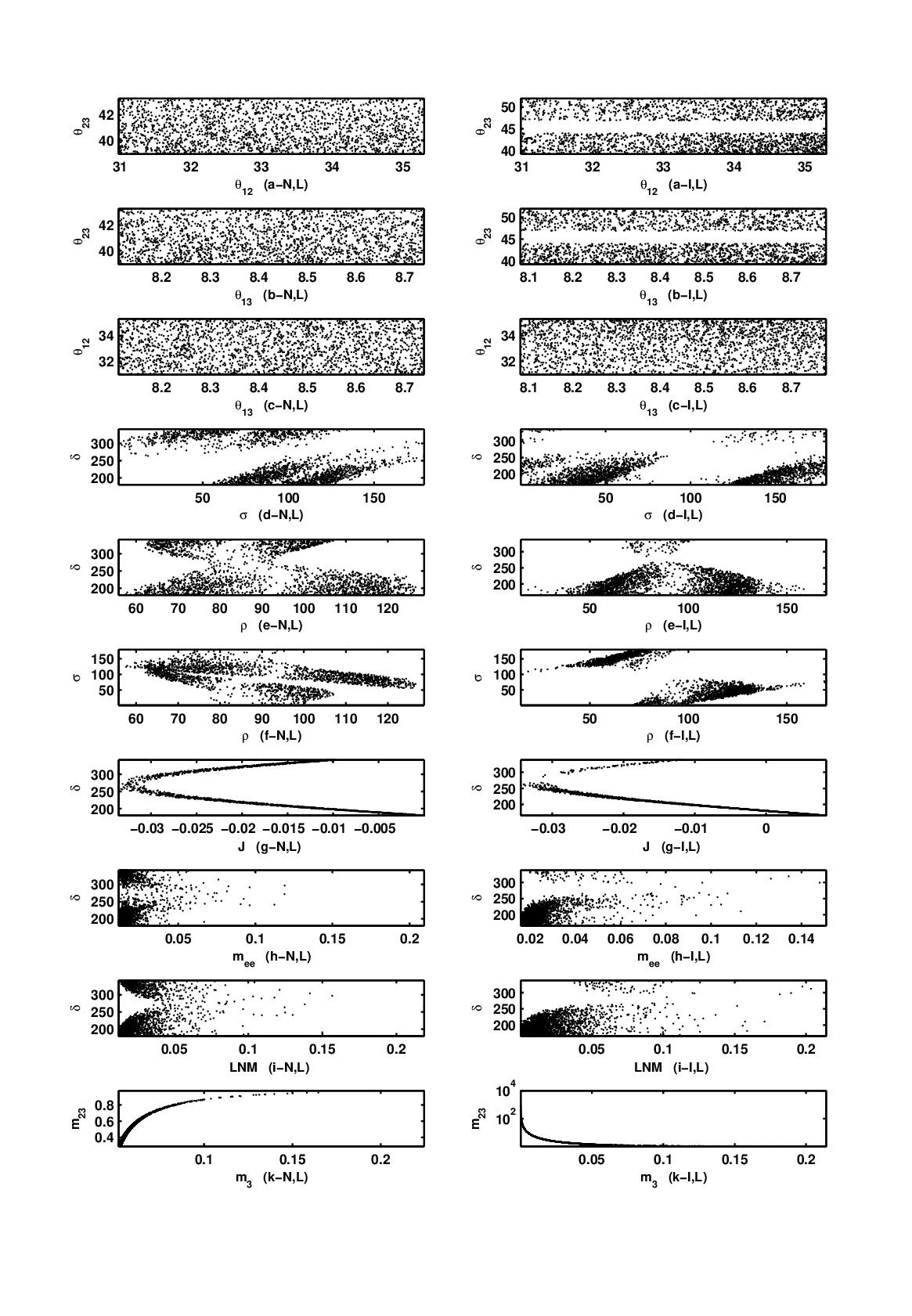}
        \end{tabular}
\end{minipage}
    \begin{minipage}{.5\linewidth}
      \centering
         \begin{tabular}{c}
            \includegraphics[height=25cm,width=8.5cm]{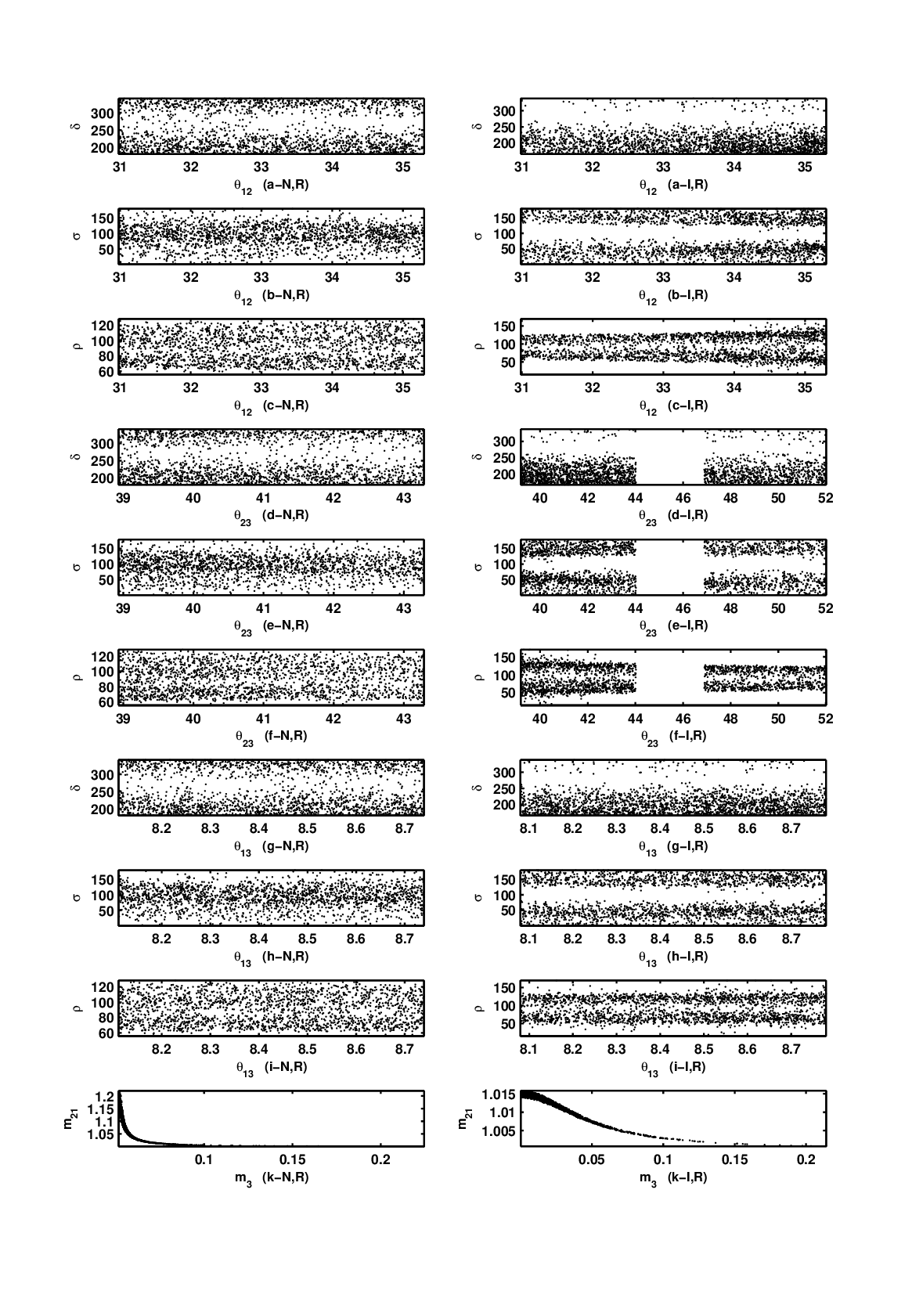}
\end{tabular}
    \end{minipage}
    \vspace{-2.6cm}
     \caption{\footnotesize Pattern $\mathbf{C_{33}} \equiv \mathbf M_{\n\,11} + \mathbf M_{\n\,22}=0$ for non-singular mass matrices: The left panel (the left two columns) shows three correlations amidst the mixing angles, three correlations amidst the phase angles and three correlations of $\d$ with $J, m_{ee}$, LNM (LNM=least neutrino mass) and finally the correlation ($m_3,m_{23}\equiv \frac{m_2}{m_3}$) for normal (N) and Inverted (I) hierarchy. The right panel (the right two columns) shows all the nine inter-correlations phase angles and mixing angles, and the correlation ($m_3,m_{21}\equiv \frac{m_2}{m_1}$)for normal (N) and Inverted (I) hierarchy. {  Angles (masses) are evaluated in degrees (eV).}}
 \label{fig1_Tr1122_2s}
     \end{figure}

For the plots,  Fig.(\ref{fig1_Tr1122_2s}), in normal ordering, we find   narrow forbidden bands for ($\r$) and a mild  mass hierarchy characterized by $(0.35 \le m_{23} \le 0.9)$. As to the plots,  Fig.(\ref{fig1_Tr1122_2s}), in inverted type, we also find forbidden bands for both $\r$ and $\s$, but the hierarchy can be severe with $m_{23}$ reaching up to $10^4$ indicating the possibility of vanishing $m_3$.

\section{Phenomenological analysis for singular textures}
Experimental data allow for one neutrino mass to vanish. The Eqs.(\ref{massratio}) are not valid when the neutrino mass matrix is singular, where instead we should use Eq. (\ref{m1=0Delta}, \ref{m3=0Delta}) to calculate the mass spectrum given the mixing and phase (Dirac and one Majorana) angles and the solar squared mass splitting. The analytic formulae we get are simpler than when the mass matrix is invertible, but still they are too cumbersome  to write them down, even if one restricts to first order in powers of $s_z$.

The mass spectrum in the normal ordering is given by
\be
m_1=0,\; m_2 = \sqrt{\d m^2},\; m_3 = \sqrt{\Delta m^2 + \d m^2/2},  \;\;\;  \Delta m^2 = \d m^2 \left(\left| \frac{A_2}{A_3}\right|^2-\frac{1}{2}\right).
\ee
Numerically, no singular texture of normal type could accommodate data.

In the inverted ordering the mass spectrum is given by
\be
 m_3=0,\; m_1 = \sqrt{\Delta m^2 -  \d m^2/2},\; m_2 = \sqrt{\Delta m^2 + \d m^2/2}, \;\;\;
\Delta m^2 = \frac{1}{2}  \d m^2 \left( \frac{\left|\frac{A_1}{A_2}\right|^2+1}{\left|\frac{A_1}{A_2}\right|^2-1}\right).
\ee
Four ``acceptable'' textures ($\mathbf {C_{12}, C_{13}, C_{22}, C_{33}}$) are found able to accommodate data.

We follow the same methodology in generating numerical results (random sampling) and the same nomenclature  in presenting results as in the case of non-singular mass matrices. All various predictions concerning the ranges spanned  by mixing angles, phase angles, neutrino masses, $\me$, $\mee$ and $J$ are summarized in Table~(\ref{tab_pred_sing}).  We note that the textures $\mathbf{C_{22}, C_{33}}$ do not pass the experimental constraints at $1\s$-level.  We present for each viable singular texture the neutrino mass matrix obtained at one representative point chosen from the accepted points out of those generated randomly in the corresponding parameter space at the $3$-$\s$-level. The choice of the  representative point is made in such a way to be as close as possible to the best fit values for mixing and Dirac phase angles.

Briefly, we see that $J<0$  at all  $\s$-levels for the texture $\mathbf C_{13}$, putting $\d$ in the third and fourth quarters.  The same applies for the texture $\mathbf C_{12}$  at $1$-$2\s$-levels, and for the texture $\mathbf C_{22}$ at $2\s$-level, specifying equally the $\d$-quarters for these acceptable textures. Positive values for $J$ can be achieved at  $3\s$-level for the textures $\mathbf C_{12}$ and $\mathbf C_{22}$ and also at $1$-$2\s$-levels for the texture $\mathbf C_{33}$.

Finally, we plot for each texture the possible correlations at the $2\s$-level showing eighteen correlations grouped into two panels. The left panel shows three correlations amidst the mixing angles, three correlations amidst the phase angles and two correlations of $\d$ with $(J, m_{ee})$  and finally the correlation ($m_{12}\equiv m_1/m_2,m_2$). The right panel includes all the nine inter-correlations between phase-angles and mixing angles.

In all four acceptable textures, the mass spectrum is almost degenerate ($m_1 \approx m_2$), and there is a strong linear correlation between ($\r, \s$) depicting two linear ribbons of positive slope. Also, there is a linear correlation between $(J,\d)$ in the four textures and this is due to the small allowed range for $\d$ which renders the sine curve $(J \propto \sin{\d})$ looking like a  linear one. In this respect, especially clear is the positive (negative) slope in the texture $\mathbf C_{22}$ ($\mathbf C_{33}$).

\subsection{Singular Pattern of $\mathbf C_{12}$: Vanishing of $M_{\n\,21} + M_{\n\, 33}$ and $m_3$}
We see, from Table(\ref{tab_pred_sing}), that $J$ is negative at $1-2\s$-levels and the corresponding $\d$ is in the third quarter.

For a representative point we take with $m_3=0$: $\theta_{12} = 33.8683^\circ$,    $\theta_{23} = 40.9412^\circ$,    $\theta_{13} = 8.7098^\circ$,    $\delta = 255.0672^\circ$,    $\rho = 26.7494^\circ$,    $\sigma = 174.1569^\circ$,    $ m_1 = 0.0490\, \mbox{eV}$,    $ m_2 = 0.0498\, \mbox{eV}$,     $ \me = 0.0487\, \mbox{eV}$,    $ \mee = 0.0417\, \mbox{eV}$ with the  corresponding mass matrix (in eV):
\be
M_\n = \begin{pmatrix}
0.0344+0.0235i &   0.0113+0.0086i &   -0.0168-0.0122i\\
 0.0113+0.0086i &   -0.0222-0.0167i  &  0.0170+0.0127i\\
-0.0168-0.0122i  &  0.0170+0.0127i &  -0.0113-0.0086i \\
\end{pmatrix}.
\ee

\begin{figure}[H]
             \includegraphics[height=25cm,width=17cm]{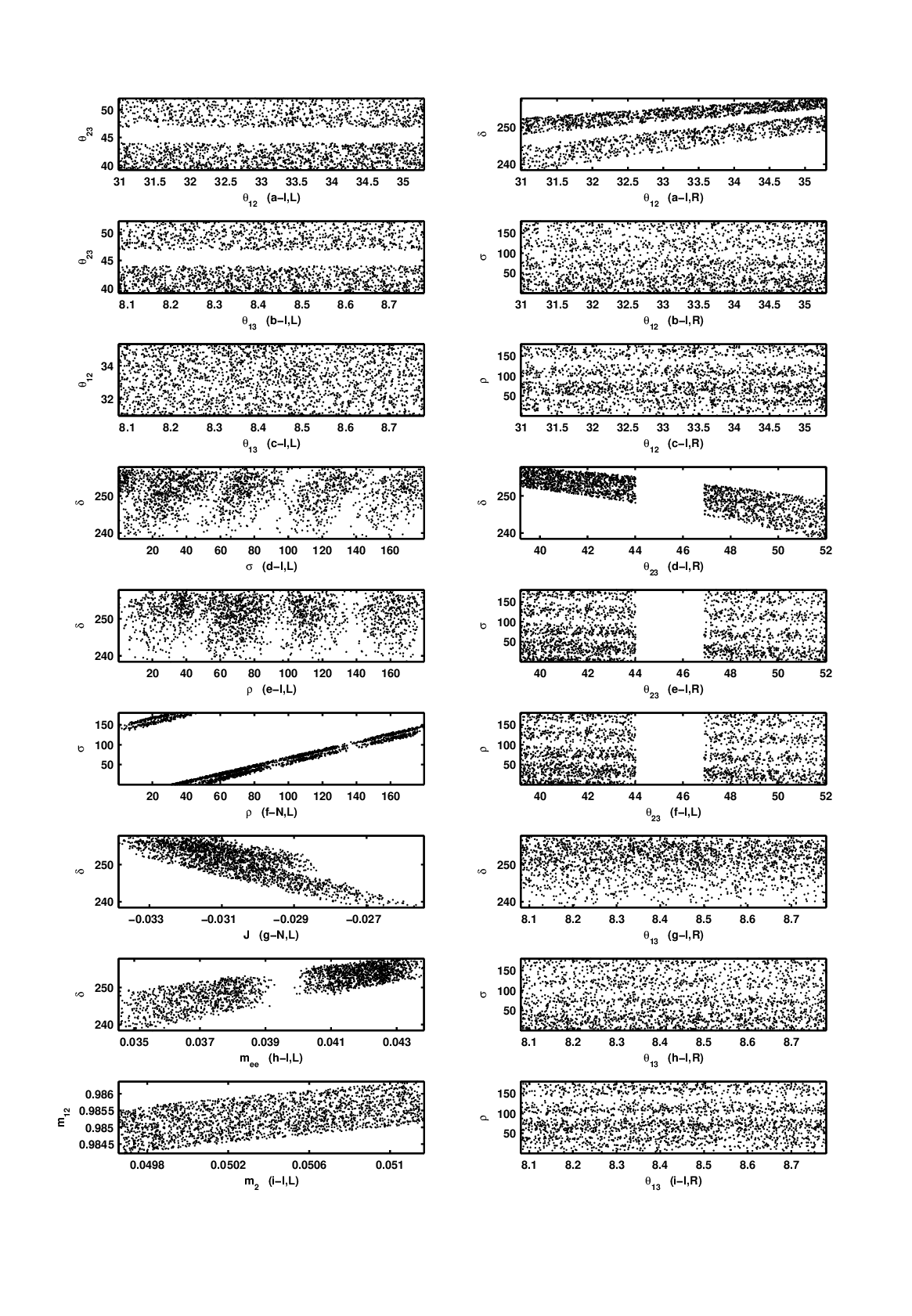}
\vspace{-2.6cm}
     \caption{\footnotesize Pattern $\mathbf C_{12}$ for singular mass matrices with inverted ordering: The left panel shows three correlations amidst the mixing angles, three correlations amidst the phase angles and two correlations of $\d$ with $J, m_{ee}$ and finally the correlation ($m_{12}\equiv \frac{m_1}{m_2},m_2$). The right panel shows all the nine correlations inter phase-angles and mixing angles. Angles (masses) are evaluated in degrees (eV).}
\label{figm3_Tr2133_2s}
 \end{figure}

For the plots,  Fig.(\ref{figm3_Tr2133_2s}), when $J$ increases $\delta$ tends to decrease in a linear manner. A strong positive linear correlation between $(\r, \s)$ exists with two ribbons. There is a forbidden gap for $m_{ee}$: $[0.0395,0.0400]$ eV. The mass spectrum is almost degenerate ($m_1\approx m_2$).


\subsection{Singular Pattern of $\mathbf C_{13}$: Vanishing of $M_{\n\,12} + M_{\n\, 23}$ and $m_3$}
We see, from Table(\ref{tab_pred_sing}), that $J$ is negative at all levels and the corresponding $\d$ is in the fourth quarter.

For a representative point we take with $m_3=0$: $\theta_{12} = 33.8148^\circ$,    $\theta_{23} = 40.7781^\circ$,    $\theta_{13} = 8.4919^\circ$,    $\delta = 284.0999^\circ$,    $\rho = 53.2226^\circ$,    $\sigma = 85.4376^\circ$,    $ m_1 = 0.0496\, \mbox{eV}$,    $ m_2 = 0.0504\, \mbox{eV}$,     $ \me = 0.0493\, \mbox{eV}$,    $ \mee = 0.0424\, \mbox{eV}$  with the  corresponding mass matrix (in eV):
\be
M_\n = \begin{pmatrix}
-0.0246+0.0346i  &   0.0127-0.0187i  & -0.0061+0.0093i\\
 0.0127-0.0187i  &   0.0118-0.0174i &   -0.0127+0.0187i\\
-0.0061+0.0093i  &  -0.0127+0.0187i &    0.0122-0.0180i\\
\end{pmatrix}.
\ee

\begin{figure}[H]
             \includegraphics[height=25cm,width=17cm]{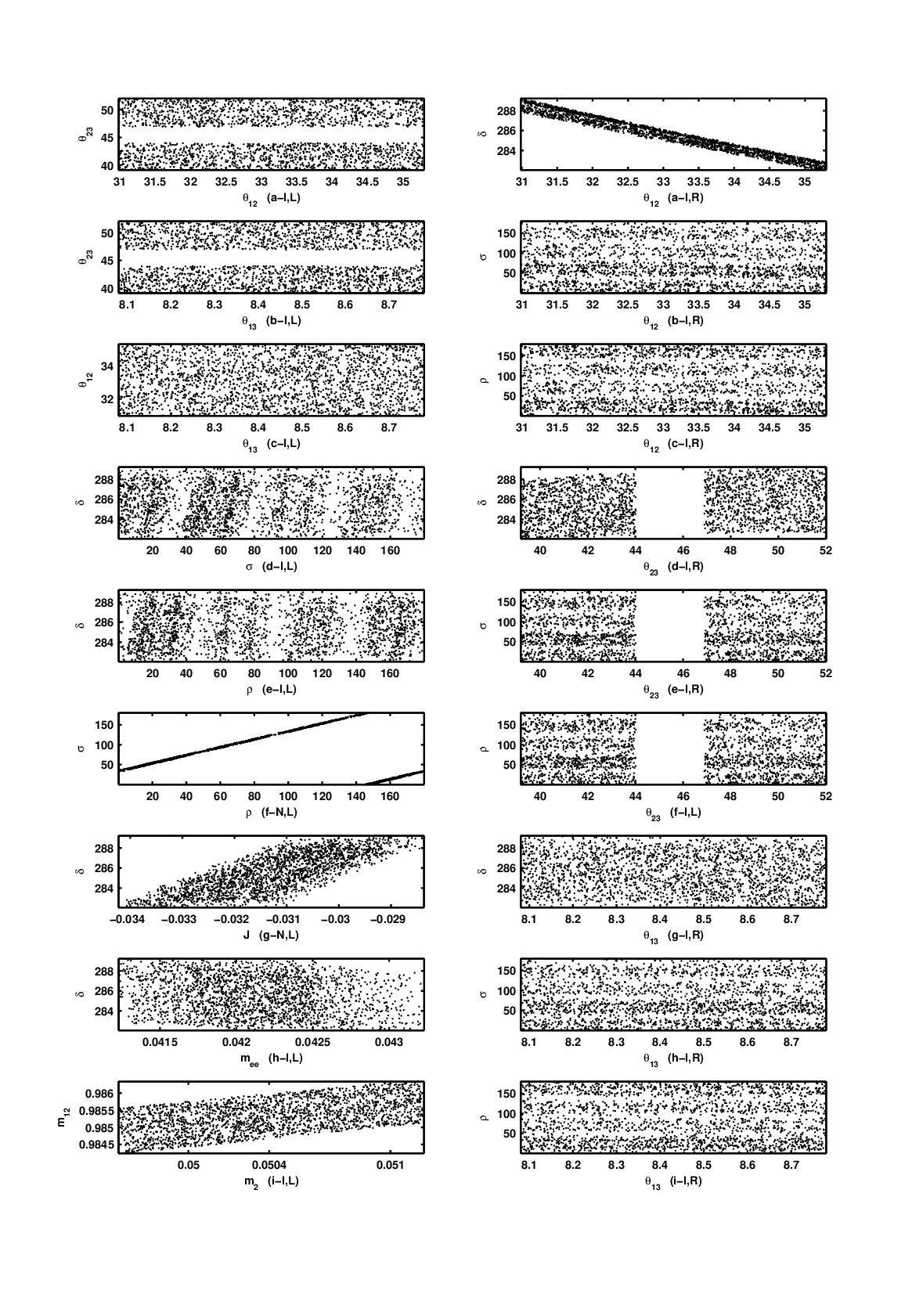}
\vspace{-2.6cm}
     \caption{\footnotesize Pattern $\mathbf C_{13}$ for singular mass matrices with inverted ordering: The left panel shows three correlations amidst the mixing angles, three correlations amidst the phase angles and two correlations of $\d$ with $J, m_{ee}$ and finally the correlation ($m_{12}\equiv \frac{m_1}{m_2},m_2$). The right panel shows all the nine correlations inter phase-angles and mixing angles. Angles (masses) are evaluated in degrees (eV).}
\label{figm3_Tr1223_2s}
     \end{figure}

For the plots in  Fig.(\ref{figm3_Tr1223_2s}), a linear correlation between $J$ and $\delta$ exists where $J$ tends to increase as $\d$ increases. The plots in  Fig.(\ref{figm3_Tr1223_2s}) also reveal a strong linear correlation between $(\r, \s)$ with two narrow ribbons exists. Also, there is a negative-slope linear dependence between ($\d,\t_{12}$). The neutrino masses are almost degenerate ($m_1 \approx m_2$).

\subsection{Singular Pattern of $\mathbf C_{22}$: Vanishing of $M_{\n\,11} + M_{\n\, 33}$ and $m_3$}

We see, from Table(\ref{tab_pred_sing}), that at $1\s$-level, the singular pattern is not viable. We also note that $J$ is negative at $2\s$-level and the corresponding $\d$ is in the fourth quarter.

\begin{figure}[H]
             \includegraphics[height=25cm,width=17cm]{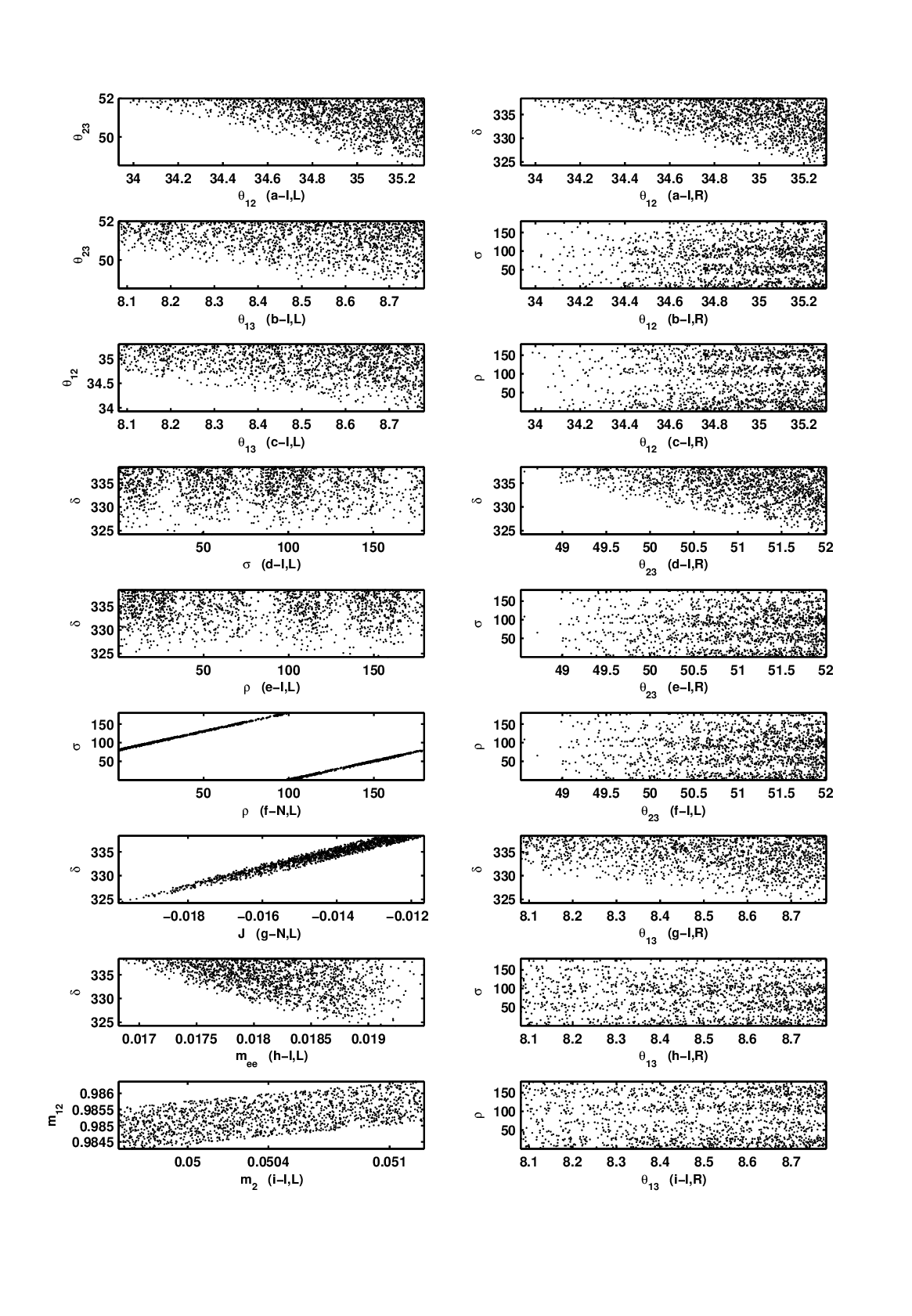}
\vspace{-2.6cm}
     \caption{\footnotesize Pattern $\mathbf C_{22}$ for singular mass matrices with inverted ordering: The left panel shows three correlations amidst the mixing angles, three correlations amidst the phase angles and two correlations of $\d$ with $J, m_{ee}$ and finally the correlation ($m_{12}\equiv \frac{m_1}{m_2},m_2$). The right panel shows all the nine correlations inter phase-angles and mixing angles. Angles (masses) are evaluated in degrees (eV).}
\label{figm3_Tr1133_2s}
     \end{figure}

For a representative point we take with $m_3=0$: $\theta_{12} = 34.5161^\circ$,    $\theta_{23} = 50.8655^\circ$,    $\theta_{13} = 8.5346^\circ$,    $\delta = 339.6445^\circ$,    $\rho = 124.3227^\circ$,    $\sigma = 26.7978^\circ$,    $ m_1 = 0.0492\, \mbox{eV}$,    $ m_2 = 0.0499\, \mbox{eV}$,     $ \me = 0.0489\, \mbox{eV}$,    $ \mee = 0.0180\, \mbox{eV}$ with the  corresponding mass matrix (in eV):
\be
M_\n = \begin{pmatrix}
-0.0026-0.0178i &   0.0046+0.0304i &  -0.0050-0.0331i\\
 0.0046+0.0304i  &  0.0000+0.0007i &  -0.0011-0.0081i \\
-0.0050-0.0331i  &  -0.0011-0.0081i  &  0.0026+0.0178i \\
\end{pmatrix}.
\ee
For the plots in Fig.(\ref{figm3_Tr1133_2s}), $J$ and  $\d$ are correlated quasi linearly and positively. A strong linear correlation with two ribbons between $(\r, \s)$ exists. The mass spectrum is almost degenerate ($m_1 \approx m_2$).


\subsection{Singular Pattern of $\mathbf C_{33}$: Vanishing of $M_{\n\,11} + M_{\n\, 22}$ and $m_3$}
As in the previous case $\mathbf C_{22}$, the singular pattern $\mathbf C_{33}$ is not viable  at $1\s$-level as evident from Table(\ref{tab_pred_sing}). In contrast to the previous case  $\mathbf C_{22}$,  $J$ can assume positive as well as negative values  at $2\s$-level and the corresponding $\d$ lies in the second and third  quarters.

For a representative point we take with $m_3=0$: $\theta_{12} = 35.9702^\circ$,    $\theta_{23} = 42.1759^\circ$,    $\theta_{13} = 8.4675^\circ$,    $\delta = 204.6858^\circ$,    $\rho = 127.4906^\circ$,    $\sigma = 45.5467^\circ$,    $ m_1 = 0.0487\, \mbox{eV}$,    $ m_2 = 0.0495\, \mbox{eV}$,     $ \me = 0.0485\, \mbox{eV}$,    $ \mee = 0.0159\, \mbox{eV}$ with the  corresponding mass matrix (in eV):
\be
M_\n = \begin{pmatrix}
-0.0084-0.0135i &  -0.0169-0.0275i &    0.0170+0.0276i\\
-0.0169-0.0275i  &  0.0084+0.0135i &   -0.0042-0.0067i\\
 0.0170+0.0276i &  -0.0042-0.0067i &    0.0004+0.0005i\\
\end{pmatrix}.
\ee

\begin{figure}[H]
             \includegraphics[height=25cm,width=17cm]{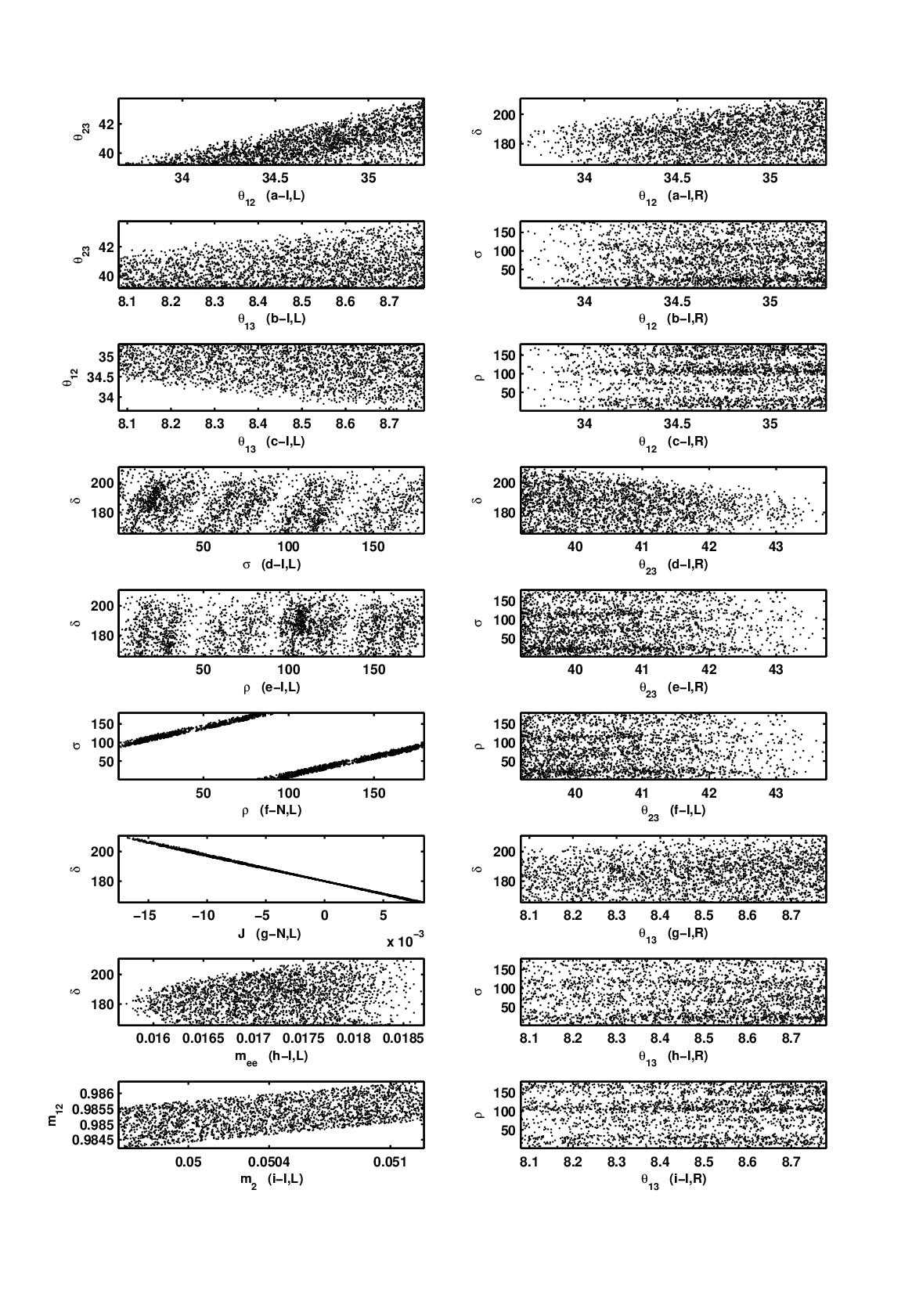}
\vspace{-2.6cm}
     \caption{\footnotesize Pattern $\mathbf C_{33}$ for singular mass matrices with inverted ordering: The left panel shows three correlations amidst the mixing angles, three correlations amidst the phase angles and two correlations of $\d$ with $J, m_{ee}$ and finally the correlation ($m_{12}\equiv \frac{m_1}{m_2},m_2$). The right panel shows all the nine correlations inter phase-angles and mixing angles. Angles (masses) are evaluated in degrees (eV).}
\label{figm3_Tr1122_2s}
 \end{figure}

For the plots in Fig.(\ref{figm3_Tr1122_2s}), we see that ($J, \d$) are strongly correlated linearly and negatively. A strong linear correlation between $(\r, \s)$ exists with two ribbons. The masses ($m_1, m_2$) are almost degenerate.

\begin{landscape}
\begin{table}[h]
 \begin{center}
\scalebox{0.7}{
{\tiny
 \begin{tabular}{c|c|c|c|c|c|c|c|c|c|c|c}
 \hline
 \hline
\multicolumn{12}{c}{\mbox{Model: Singular} $\mathbf{C_{12}}\equiv M_{\n\,21} + M_{\n\,33} = 0$\; \mbox{and}  $m_3 =0$} \\
\hline
\hline
  \mbox{quantity} & $\th_{12}^{\circ}$ & $\th_{23}^{\circ}$& $\th_{13}^{\circ}$ & $m_1$ $(10^{-2} \text{eV})$ & $m_2$ $(10^{-2} \text{eV})$ &  $\d^{\circ}$ & $\r^{\circ}$ & $\s^{\circ}$ & $\me$ $(10^{-2} \text{eV})$
 & $\mee$ $(10^{-2} \text{eV})$ & $J$ $(10^{-2})$\\
 \hline
 $1\, \sig$ &$32.02 -  34.09$ &$40.23 - 42.02 \cup 48.86 -   51.06$ &$8.27  -   8.59$ &$4.93  -  5.00$ &$5.01  -   5.08$ &$242.16 - 249.20 \cup 251.70 -  255.79$ &$0.09 -  178.95$ &$0.05 -    179.33$ &$4.90  - 4.97$ &$3.54  -     4.29$ &$-3.24  -   -2.76$ \\
 \hline
 $2\, \sig$ &$30.98 -  35.30$ &$39.19  - 44.01 \cup 46.90 -  52.01$ &$8.08 -   8.78$ &$4.89 -  5.04$ &$4.97 -   5.12$ &$238.49 - 257.75$ &$0.19 -  179.72$ &$0.03 -   179.99$ &$4.86 - 5.01$ &$3.45 - 4.38$ &$-3.39 -  -2.54$ \\
 \hline
 $3\, \sig$ &$30.00  -  36.51$ &$38.30 -  52.90$ &$7.92 -  8.94$ &$4.85 - 5.08$ &$4.92 -  5.16$ &$124.21 - 125.50 \cup 234.30 -  259.48$ &$0.22 -  179.78$ &$0.03 -  179.92$ &$4.81  -  5.06$ &$3.35 -  4.44$ &$-3.51 - -2.35 \cup  2.34 -  2.60$ \\
 \hline
 \hline
\multicolumn{12}{c}{\mbox{Model: Singular} $\mathbf{C_{13}}\equiv M_{\n\,12} + M_{\n\,23} = 0$\; \mbox{and}  $m_3 =0$} \\
\hline
  \mbox{quantity} & $\th_{12}^{\circ}$ & $\th_{23}^{\circ}$& $\th_{13}^{\circ}$ & $m_1$ $(10^{-2} \text{eV})$ & $m_2$ $(10^{-2} \text{eV})$  & $\d^{\circ}$ & $\r^{\circ}$ & $\s^{\circ}$ & $\me$ $(10^{-2} \text{eV})$
 & $\mee$ $(10^{-2} \text{eV})$ & $J$ $(10^{-2})$\\
 \hline
 $1\, \sig$ &$32.02  -  34.09$ &$40.23 - 42.02 \cup  48.87  -  51.06$ &$8.27  -   8.59$ &$4.93 -  5.00$ &$5.01  -  5.08$ &$283.68  - 287.46$ &$0.30 - 179.99$ &$0.06  -  179.97$ &$4.90  -  4.97$ &$4.16 -  4.28$ &$-3.27  -    -2.97$ \\
 \hline
 $2\, \sig$ & $30.98 -  35.30$ &$39.18 - 44.01 \cup 46.90 - 52.01$ &$8.08 -  8.78$ &$4.89  -  5.04$ &$4.97 -  5.12$ &$282.06 - 289.23$ &$0.09 -  179.97$ &$0.04 -  179.90$ &$4.86  - 5.02$ &$4.12 -  4.32$ &$-3.42  -   -2.84$ \\
 \hline
 $3\, \sig$ &$30.00  -  36.50$ &$38.31  -  52.90$ &$7.92 -  8.94$ &$4.85 -  5.08$ &$4.93 -  5.16$ &$280.43 - 291.00$ &$0.01 -  179.72$ &$0.00 - 179.90$ &$4.81  -  5.06$ &$4.08 -  4.37$ &$-3.55 -   -2.70$ \\
 \hline
 \hline
\multicolumn{12}{c}{\mbox{Model: Singular} $\mathbf{C_{22}}\equiv M_{\n\,11} + M_{\n\,33} = 0$\; \mbox{and} $m_3 =0$} \\
\hline
  \mbox{quantity} & $\th_{12}^{\circ}$ & $\th_{23}^{\circ}$& $\th_{13}^{\circ}$ & $m_1$ $(10^{-2} \text{eV})$ & $m_2$ $(10^{-2} \text{eV})$  & $\d^{\circ}$ & $\r^{\circ}$ & $\s^{\circ}$ & $\me$ $(10^{-2} \text{eV})$
 & $\mee$ $(10^{-2} \text{eV})$ & $J$ $(10^{-2})$\\
 \hline
 $1\, \sig$ &$\times$& $\times$ &$\times$ &$\times$ &$\times$ &
  $\times$ &$\times$ & $\times$ &$\times$ &$\times$ & $\times$  \\
 \hline
 $2\, \sig$ &$33.93  - 35.30$ &$48.52 -  52.01$ &$8.08 -   8.78$ &$4.89 -  5.04$ &$4.97 -  5.12$ &$324.31 - 338.39$ &$0.02 -  179.51$ &$0.04 - 179.97$ &$4.86  - 5.02$ &$1.68 - 1.95$ &$-1.99  -  -1.17$ \\
 \hline
 $3\, \sig$ &$32.46 - 36.51$ &$42.31  - 52.90$ &$7.92 -  8.94$ &$4.85 - 5.08$ &$4.92 -  5.16$ &$0.00 - 26.95 \cup 315.00 -  359.96$ &$0.26 - 179.92$ &$0.00 -   179.99$ &$4.82      - 5.06$ &$1.37 -  2.06$ &$-2.42 - 1.60$ \\
 \hline
 \hline
\multicolumn{12}{c}{\mbox{Model: Singular} $\mathbf{C_{33}}\equiv M_{\n\,11} + M_{\n\,22} = 0$\; \mbox{and}  $m_3 =0$} \\
\hline
 \mbox{quantity} & $\th_{12}^{\circ}$ & $\th_{23}^{\circ}$& $\th_{13}^{\circ}$ & $m_1$ $(10^{-2} \text{eV})$ & $m_2$ $(10^{-2} \text{eV})$ & $\d^{\circ}$ & $\r^{\circ}$ & $\s^{\circ}$ & $\me$ $(10^{-2} \text{eV})$
 & $\mee$ $(10^{-2} \text{eV})$ & $J$ $(10^{-2})$\\
 \hline
 $1\, \sig$ &$\times$& $\times$ &$\times$ &$\times$ &$\times$ &
  $\times$ &$\times$ & $\times$ &$\times$ &$\times$ & $\times$  \\
 \hline
 $2\, \sig$ &$33.66  - 35.30$ &$39.18 -  43.75$ &$8.08 -  8.78$ &$4.89 - 5.04$ &$4.97 - 5.12$ &$165.61 -  210.72$ &$0.26 - 179.36$ &$0.12 - 179.85$ &$4.86  - 5.02$ &$1.57 - 1.87$ &$-1.75 -  0.85$ \\
 \hline
 $3\, \sig$ &$33.01  -  36.51$ &$38.29 -   47.68$ &$7.92   -   8.94$ &$4.85 - 5.08$ &$4.92 - 5.16$ &$136.33 -  223.33$ &$0.31 -  179.87$ &$0.08 - 179.88$ &$4.82 - 5.06$ &$1.38 - 1.96$ &$-2.42 -  2.40$ \\
 \hline
 \end{tabular}
 }}
 \end{center}
  \caption{\footnotesize  The various predictions for the patterns of
  one vanishing subtrace textures and vanishing $m_3$  designated by $\mathbf{C_{12}, C_{13}, C_{22}}$ and
  $\mathbf C_{33}$.}
 \label{tab_pred_sing}
 \end{table}
\end{landscape}




\section{Theoretical Realization of the textures}
We present in this section, theoretical realizations of some of the one vanishing subtrace textures, where symmetry assignments at high scale impose this texture in the ``gauge'' basis. However, one way to find these assignments is to start from another symmetry imposing zero textures and relate these two symmetries by a rotation. As to the symmetry responsible for imposing the zero elements at high scale, we can just follow the analysis of \cite{0texture}. We shall find that four vanishing subtrace textures, out of six, are able to be amended by ``rotating'' zero-textures. In section 7, we explain the general strategy of relating the two symmetries, which would  be of great help in this method of indirect realization. In section 8, we discuss the notion of flavor basis due to its paramount relevance into our study. In section 9, making use of ``rotating'' zero-textures, we adopt a type I seesaw scenario with discrete symmetry ($Z_8 \times Z_2$) in order to generate nonsingular vanishing subtrace textures. We repeat the work for singular vanishing subtrace textures in section 10, but with discrete symmetry ($Z_{12} \times Z_2$). In section 11, we present an implementation of one vanishing subtrace texture using type II seesaw scenario supplemented with ($Z_5$) discrete symmetry, and following the same strategy of `rotation' from zero textures to vanishing subtrace.  In section 12, we present a direct way of realization for type I seesaw scenario implementation  with ($Z_6 \times Z_2$) discrete symmetry not related to zero textures.  In section 13, we pursue the direct method of realization but now for  type II seesaw scenario implementation  with ($Z_2^\prime \times Z_2$) discrete symmetry.  One can consider these  outlined sections as  an exercise in model building aiming to show that the studied texture of vanishing subtrace can be generated at the Lagrangian level by symmetry considerations
in which the symmetry is exact but broken spontaneously. {  The two ``indirect'' and ``direct'' methods are on equal footing, and one should not discriminate one against the other. It is just that the fields assignments in the ``indirect'' method turn out to be more complex, so we sought a ``mathematical'' method in order to find them. } As a final remark, the presented method of `rotation' is applicable to any specific pattern which can be generated from the zero-texture pattern via a unitary transformation.

 \section{`Rotating' Strategy: from zero-texture to vanishing subtrace texture} \label{strategysubsection}
We need to find a unitary matrix $S$ which when acted on the symmetric neutrino matrix:
 \be
 M_\n =
\begin{pmatrix}
  A & B & C \\
  B & D & E \\
 C & E & F\\
\end{pmatrix}.
 \label{MABC}
  \ee
 gives the combination which define the subtrace patterns  $\left[(\mathbf C_{11}): D+F\right]$, $\left[(\mathbf C_{12}): B+F\right]$, $\left[ (\mathbf C_{13}): B+E\right]$,
$\left[ (\mathbf C_{22}): A+F\right]$, $\left[ (\mathbf C_{23}): A+E\right]$,  and $\left[(\mathbf C_{33}): A+D \right]$ in one of the elements of the transformed matrix
$(\tilde{M} = S^T\; M_\n \;S)$, where $M_\n$ is the effective Majorana neutrino mass matrix. More specifically, for the texture $\mathbf C_{33}$, if we take the unitary matrix
\be
S_{33}=
\frac{1}{\sqrt{2}}
\begin{pmatrix}
 i & -1 & 0 \\
 i & 1 & 0\\
0 & 0 & \sqrt{2}\\
\end{pmatrix}
\label{33}
\ee
then we find that
\be
S_{33} \; M_\n\; S^T_{33} = -\frac{1}{2}
\begin{pmatrix}
A+2\,i\,B-D&A+D& -\sqrt{2}\,(i\,C-E)\\
A+D&A-2\,i\,B-D& -\sqrt{2}\,(i\,C+E) \\
 -\sqrt{2}(i\,C-E)& -\sqrt{2}\,(i\,C+E)&-2\,F\\
\end{pmatrix}
 \label{16}
 \ee
 and so the combination $(A+D)$ appears in the element $(12)$ of the transformed matrix
\be
M_{\n 0}=S_{33}\;M_{\n}\;S_{33}^T
\label{massTransform}
\ee
Thus, if by some symmetry $S_{Y0}$ applied on the transformed matrix $M_{\n0}$ one can impose a zero element:
 \be
S_{Y0}^T\; M_{\n 0} \; S_{Y0} = M_{\n 0} \; \Rightarrow\;
M_{{\n 0},{12}}=0,
\label{sy0}
\ee
then we see that we have
 \be
S_{Y}^T\; M_\n \; S_Y = M_\n \;\Rightarrow\; M_{\n 11}+M_{\n 22}=0
\label{fi}
\ee
where the new symmetry implementing the vanishing subtrace of the texture $C_{33}$ is
 \be
 S_{Y} =  S_{33}^{T} \;S_{Y0}\;S_{33}^{T\dagger}.
\label{conjugation}
\ee

 Let's define $u^{ij}$ as the matrix resulting by swapping the $i^{th}$ and the $j^{th}$ columns of the identity matrix $(I)$. Then we have the properties:
\be
u^{ij} = u^{ijT} = u^{ij\dagger}\; ,\;  u^{ij}\; u^{ij\dagger} = I
\ee

Then, for any matrix $M$, we see that
$(M\;u^{ij})$ swaps the $i^{th}$ and $j^{th}$ columns of $(M)$, whereas $(u^{ij}\;\;M)$ swaps the $i^{th}$ and $j^{th}$ rows of ($M$).
Note that $u^{ij}\; M\;u^{ij} $ has the effect of swapping first the ($i^{th}$ and $j^{th}$) columns, followed by the ($i^{th}$ and $j^{th}$) rows,
or the other way round. We note now that the six vanishing one subtrace textures can be divided into three classes
\begin{itemize}
\item {\bf Class of textures $\{\mathbf C_{11}, \mathbf C_{22},\mathbf C_{33}\}$:}
In the sense that if I find a unitary transformation $\tilde{S}$ giving me one of them, then directly I get the unitary transformation giving me the other two textures. This comes because
\be
\left.
\begin{array}{lll}
\left(u^{13}\;M\;u^{13}\right)_{22}=M_{22},\;  \left(u^{13}\;M\;u^{13}\right)_{11}=M_{33} &\Rightarrow &  u^{13}\;\tilde{S}_{33}\;u^{13}=\tilde{S}_{11}, \\\\
\left(u^{12}\;M\;u^{12}\right)_{33}=M_{33},\;  \left(u^{12}\;M\;u^{12}\right)_{11}=M_{22} & \Rightarrow &  u^{12}\;\tilde{S}_{22}\;u^{12}=\tilde{S}_{11},
\end{array}
\right\}
\label{class1}
\ee
where $\tilde{S}_{ij}$ is a unitary matrix which, provided its action on $M_\n$ keeps the latter invariant, imposes the texture defined by the subtrace $(C_{i,j})$ (see Eq. \ref{fi} where $S_Y$ plays the role of $\tilde{S}_{33}$).

\item {\bf Class of textures $\{\mathbf C_{13}\}$:}
Actually, we can take:
\be
 S_{13}=
\frac{1}{\sqrt{2}}
\left (\begin{array}{ccc} 1 & 0 & 1 \\
0 & \sqrt{2} & 0\\
-1 & 0 & 1
\end{array}
 \right ),
\label{13}
\ee
because the $(B+E)$ combination appears in the $(1,2)$ element of
\be
S_{13} \; M_\n\; S_{13}^T = \frac{1}{2}
\begin{pmatrix}
A+2\,C+F&\sqrt{2}\, \left( B+E
 \right) &F-A\\
 \sqrt{2}\, \left( B+E
 \right) &2D&\sqrt{2}\, \left( E - B \right) \\
F-A&\sqrt{2}\; \left( E- B \right) &A-2\,C+F
\end{pmatrix}.
\ee

\item {\bf Class of textures $\{\mathbf C_{12}, \mathbf C_{32}\}$:}
\be
\left(u^{13}\;M\;u^{13}\right)_{33}=M_{11} ,\;  \left(u^{13}\;M\;u^{13}\right)_{21}=M_{23}\;\Rightarrow\; u^{13}\;\tilde{S}_{12}\;u^{13}=\tilde{S}_{32}
\ee
However, one can algebraically show  that the transformation $S \; M_\n\; S^T$ can not bring in the sole  combination  $ A+E$, corresponding to the texture $\mathbf {C_{32}}$,  at any entry.\\
\end{itemize}

Our strategy for the realization of the vanishing subtrace texture is that one imposes a starting symmetry, with corresponding transformations on the Higgs and the  lepton fields
at the Lagrangian level (``gauge'' basis), known to impose some zero elements for the neutrino mass matrix $  M_{\nu 0}$.  We then transform this symmetry by applying some ``rotation'' so that to get a new second symmetry, with new transformations on the fields (also at the gauge basis), which would imply the vanishing subtrace texture for $  M_{\nu}$. {  We stress here that $M_{\nu}$ and $M_{\nu 0}$ are not mass matrices for the same system at different bases related by rotation. Rather, they are mass matrices of two systems, satisfying two different symmetries, where the matrices are defined in the same Lagrangian gauge -or ``symmetry''- bases. The two symmetries are related by `rotation'.} By following the previous discussion, we may find the rotation which, when applied on neutrino mass matrix, allows to go from zero texture to vanishing subtrace, so now this rotation would help to define, by (Eq. \ref{conjugation}), how to move from the first symmetry field transformations to the second symmetry ones by the following ``adjoint action" rule:
\be
T_f = S^T \; T_f^{\,0 }\; S^{T\dagger},
\label{star}
\ee
where $T_f(T_f^{\,0})$ defines the transformation on the field $f$ satisfying the new (old) symmetry, and $S$ is the unitary transformation relating the two symmetries. {  We remind here that this ``rotation'' method is just to find some ``complex'' field assignments by relating them to other more ``trivial'' ones, and had we been able to ``guess'' the complicated assignments then we would have dispensed with the whole idea of ``rotation''.}

 However, one should be sure that the new symmetry transformations assure that we are at the flavor basis, or approximately so. The point is that we should get a generic charged lepton mass matrix by the first symmetry, so that to get also a generic one by the second symmetry. Then, by adopting some `natural' assumptions on the fields vacuum expectation values (vevs), without the need of unnatural
 constraints on the Yukawa couplings, one can diagonalize the `generic' charged lepton mass matrix by an infinitesimal rotation, and so one can, with a good approximation, assume that the new symmetry puts us in the flavor basis. We shall give some examples for this strategy within both type I and II seesaw scenarios, where the symbol $0$ will be consigned for quantities corresponding to the first `unrotated' symmetry\footnote{  Concretely, a discrete symmetry of the form $Z^0_n \times Z^0_m$ is imposed on one system leading to zero textures in the neutrino mass matrix $M_{\n 0}$ in the ``symmetry-gauge'' basis, whereas another system has a symmetry $Z_n \times Z_m$ leading to vansihing subtrace in the corresponding $M_\n$ defined again in the ``symmetry-gauge'' basis. The unitary transformation $S$ helps to relate the field assignments of both systems under the two symmetries.}. The method based on `rotated' symmetry can be considered as an indirect method for realizing  the texture of vanishing subtrace.

\section{Flavor basis}
\label{fb}
The notion of basis is intricate and needs to be clarified since a variety of bases could arise in our discussion such as gauge, flavor and mass basis.  In order to delve into the notion of different types of basis, we take for simplicity the following Lagrangian piece responsible for the mass terms in the leptonic sector expressed in the gauge basis as:
\be
\mathcal{L}_M  \supset Y^g_{1\,{ij}}\; \overline{D}_{Li}\; H\, e_{Rj} + Y^g_{2\,{ij}}\; \overline{D}_{Li}\; \tilde{H}\; \nu_{Rj} + Y^g_{3\,{ij}}\; \nu_{Ri}^T \, \mathcal{C}^{-1}\; \chi\; \nu_{Rj},
\ee
where $D_{Li}$ is the left-handed lepton doublet $(\nu_{Li}, e_{Li})^T$, $e_{Ri}$ is the right-handed charged lepton, $\nu_{Ri}$ is the right-handed neutrino, $\chi$ is a scalar singlet, $H$ is the Higgs doublet and $\tilde{H} \equiv i\,\s_2\, H^*$. The relevant Yukawa coupling matrices are denoted by  ($Y^g_1, Y^g_2, Y^g_3$) which are defined in this ``gauge" basis, whence the superscript ($g$). The indices $i,j$ are the family ones while $\mathcal{C}$ is the charge conjugation.

When the Higgs doublet $H$ and the singlet $\chi$ take a vev, then we get the mass term which can be cast into the form,
\be
\bar{e}_{Li}\; M_{\ell\,ij}\; e_{Rj} + \bar{e}_{Ri}\; M_{\ell\,ij}\; e_{Lj} +
 \left (\begin{array}{cc}
\nu_{Li}^T & \nu_{Rj}^T
\end{array} \right ) \,\mathcal{C}^{-1}\,
\left (\begin{array}{cc} 0 & M_{D\,ij}  \\ M_{D\,ji} & M_{R\,ij}
\end{array}
\right )
\left (\begin{array}{c} \nu_{Li} \\ \nu_{Rj}  \end{array} \right ),
\ee
which, via the seesaw mechanism, gives approximately, after decoupling the right-handed  neutrinos,
\be
\mathcal{L}_M \supset   \bar{e}_{L_i}\; M_{\ell\,ij}\; e_{Rj} + \nu_{li}^T\,\mathcal{C}^{-1}\; M_{\nu\,ij} \; \nu_{lj},
\ee
 with $M_\nu = M_D \; M_R^{-1}\;  M_D^T$, and $\nu_{li}$ are approximately left-handed ($\approx \nu_{Li}$).

By diagonalizing, we get the ``mass'' basis which is denoted by the superscript $m$:
\be
\left.
\begin{array}{lll}
\mathcal{L}_M &\supset& \bar{e}^m_L\;  U_L^\dagger\; M_{\ell}\; U_R\; e^m_{R} + \nu^m_{l}\; V^T\; M_{\nu}\; V\; \nu^m_{l}, \\\\
&\supset & \bar{e}^m_L\; M_\ell^{\mbox{diag}}\; e^m_R + \nu^m_l\;  M^{\mbox{diag}}_{\nu}\; \nu^m_{l},\\\\
 & & e^m_L = U_L^\dagger\; e_L,\;\; e^m_R = U_R^\dagger\; e_R,\;\; \nu^m_{l} = V^\dagger\;  \nu_{l},
\end{array}
\right\}
\ee
where  $U_L^\dagger\; M_{\ell}\; U_R$ and $V^T \; M_{\nu}\; V$ are diagonal.

In the gauge basis, the interaction (say, $\bar{e}_L\, W^- \n_l$) between the charged lepton sector and the neutrino sector, when expressed in terms of the mass bases ($\bar{e}^m_L\, U_L^\dagger\, V\, W^-\, \n^m_l$) would involve the experimentally measurable $V_{\mbox{\tiny PMNS}}= U_L^\dagger V$ expressing the mismatch between the rotations of the left-handed charged leptons and of the left-handed  neutrinos. The ``flavor basis'', by definition, occurs when by convention we assume, without loss of generality, the left-handed charged leptons to be pure states, i.e. $U_L = 1$ and $e^m_L = e_L$. This can always be taken, since one can use the freedom in defining the fields in a way to attribute the whole rotation, appearing when expressing the interaction term in terms of mass states, entirely to the left-handed neutrinos. The situation is exactly the same for the quark sectors when one can take by convention the up sector as pure states and the flavor mixing is described in terms of the rotation CKM matrix operating on the down sector only \cite{ckm}.

{  In the realization models we shall construct in next sections, the field assignments are given in the ``symmetry-gauge'' basis at the level of the Lagrangian, and thus we get a charged lepton field mass matrix which is not necessarily diagonal. We shall examine at which conditions one can have diagonal, or almost diagonal to a very good approximation, mass matrix for the charged leptons, in a way to say that the symmetry leading to the sought for texture in the neutrino sector puts us  also, nearly, in the flavor basis for the charged leptons.}

{  The question arises as to whether or not one should update the phenomenological analysis upon carrying out the infinitesimally small, under these conditions, rotation $R_\eps$ of the charged leptons from the ``symmetry-gauge'' into the ``flavor (mass)'' basis. Actually, the phenomenology study was carried out in the flavor basis, which means it is valid up to small corrections of the order of the small rotation $R_\eps$\footnote{  One should be aware not to mix the two ``rotations". The ``mathematical rotation" $S$ relating two systems with two different discrete symmetries, which is ``large (finite)" normally, and the ``physical rotation" $R_\eps$ which is ``small (infinitesimal)" and is applied on the charged leptons to go from the symmetry-gauge basis into the flavor-mass basis.}. With this in mind, this small correction should be added to the already anticipated one stemming from the renormalization group loop effects upon running from the high scale, when the symmetry was imposed, to the low scale of the experimental data. }

\section{Indirect realization of type I seesaw with $Z_8 \times Z_2$ symmetry for nonsingular textures}
We implement here a discrete symmetry within type I seesaw scenario  in order to generate one vanishing subtrace texture following the `Rotating' strategy.

\subsection{Indirect realization of $\mathbf C_{33}$ (Type I nonsingular): Vanishing of $M_{\n\,11} + M_{\n\, 22}$}
We saw that the matrix $S_{33}$ conjures, when acted on $M_\n$, the combination $M_{\nu\,22}+M_{\nu\,33}$ in the element ($1,2$) of the transformed $\tilde{M}_\n$. Thus we follow \cite{0texture} and impose $Z_8\times Z_2$ symmetry to have a zero in the $(1,2)$ entry of the `unrotated'  $M_{\n0}$ and check that the `rotated' mass matrix  $M_\n = S^T\; M_{\n0} \;S $ has a texture $\mathbf C_{33}$ with $  S= S_{33}$
\be
S= \frac{1}{\sqrt{2}}\left(
\begin {array}{ccc}
 i&-1& 0\\
i& 1& 0\\
0&0&\sqrt{2}
\end {array}
\right)
\Rightarrow
S^T\; \left(
\begin {array}{ccc}
A &0& C\\
0& D& E\\
C&E&F
\end {array}
\right) \; S = \left(
\begin {array}{ccc}
-\frac{1}{2}(A+D) &\frac{i}{2}(D-A)& \frac{i}{\sqrt{2}}(E+C)\\
\frac{i}{2}(D-A) & \frac{1}{2}(A+D))& \frac{1}{\sqrt{2}}(E-C)\\
\frac{i}{\sqrt{2}}(E+C) &\frac{1}{\sqrt{2}}(E-C) &F
\end {array}
\right),
\label{SC33}
\ee

First, we show how one can impose the zero texture. We introduce five Standard Model (SM) Higgs doublets $\Phi_a(a=1,\ldots,5)$, three real scalar singlets $\chi_{i}(i=1,2,3)$, and denote the left-handed lepton Doublet of the first (second, third) family by $D_{L1}$ ($D_{L2},D_{L3}$). The right-handed charged lepton and neutrino singlets are denoted by  ($\ell_R, \n_{R}$). We assume the following transformations in Table~(\ref{tablec33typeI}) under $Z_8^{ 0} \times Z_2^{ 0}$ for the fields:

\begin{table}[hbtp]
\begin{center}
\begin{tabular}{ccccccccccccccc}
\toprule
\multicolumn{15}{c}{$ \mbox{Symmetry under}\; Z_8^0\; \mbox{factor}$}\\
\midrule
    $\Phi_1$  & $\Phi_2$ & $\Phi_3$ & $\Phi_4$ & $\Phi_5$ & $D_{L1} $ &$D_{L2} $ & $D_{L3} $ &$\n_{R1}$ &$\n_{R2}$& $\n_{R3}$ & $\chi_{1}$ &
   $\chi_{2}$ & $\chi_{3}$ & $\ell_R$    \\
\midrule
 $1$ & $\omega^4$ & $\omega$ & $1$ & $\omega^7$ & $1$ & $\omega^4$ & $\omega$ & $1$ & $\omega^4$ & $\omega$ &  $1$ & $\omega^3$ & $\omega^6$ & $1$
\\
\toprule
\multicolumn{15}{c}{$ \mbox{Symmetry under}\; Z_2^0\; \mbox{factor}$}\\
\midrule
$1$& $1$& $1$& $\omega^4$& $\omega^4$& $1$ &$1$& $1$& $\omega^4$& $\omega^4$& $\omega^4$& $1$ & $1$ & $1$ & $1$  \\
\bottomrule
\end{tabular}
\end{center}
 \caption{\small  The $Z_8^0\times Z_2^0$ symmetry realization of the
 one zero texture at $(1,2)$-entry corresponding upon `rotation' to vanishing subtrace $\mathbf C_{33}$. The index $D_{L1}$ indicates
the left-handed lepton doublet first
 family and so on. The $\chi_{k}$ denotes a scalar singlet which produces
 an entry in the right-handed Majorana mass matrix when acquiring a VEV
 at the see-saw scale.
 $\omega$ denotes $e^{{i\,\pi/4}}$. }
\label{tablec33typeI}
 \end{table}
By forming bilinear terms of $\overline{D}_{Li}\, \ell_{Rj}$ and $\overline{D}_{Li}\, \n_{Rj}$, relevant for Dirac mass matrices of
neutrino and charged leptons, and of $\nu_{Ri} \, \nu_{Rj}$, relevant for the Majorana neutrino mass matrix
$M_R$ in the Lagrangian ($Y_{ij}^a$ are the Yaukawa coupling constants, the indices ($i,j$) are flavor ones, the indices ($a$, $b$) run respectively over the Higgs doublet and Scalar singlet fields, $\mathcal{C}$ is the charge conjugation matrix and $\tilde{\Phi} = i\, \s_2\,  \Phi^{*}$):
\be
\mathcal{L}_M \supset \sum_{i,j=1}^3\sum_{a=1}^5\sum_{b=1}^3\,Y_{0\chi ij}^{b}\; \chi_b\;  \n_{Ri}^T\; \mathcal{C}^{-1}\; \n_{Rj} + Y^{a}_{0D ij}\; \overline{D}_{Li}\; \tilde{\Phi}_a\;  \n_{Rj}
+ Y_{0\ell ij}^{\,a}\; \overline{D}_{Li}\; \Phi_a \; \ell_{Rj} \,
\label{genlagm}
\ee
and examining how they transform under $Z_8^{ 0} \times Z_2^{ 0}$, we see that the invariance under the symmetry implies the following forms
\be
M_{D0}=\left(
\begin {array}{ccc}
\times &0& 0\\
0 & \times & 0\\
\times & 0 &\times
\end {array}
\right),\; M_{R0}=\left(
\begin {array}{ccc}
\times &0& 0\\
0& \times& \times\\
0& \times &\times
\end {array}
\right) \Rightarrow M_{\n0} = M_{D0}\, M_{R0}^{-1}\, M_{D0}^T =
\left(
\begin {array}{ccc}
\times &0& \times\\
0& \times& \times\\
\times&\times&\times
\end {array}
\right).
\ee
Note that, in contrast to \cite{0texture} where we introduced only three Higgs doublets, we introduce here five Higgs doublets otherwise we would have got as in \cite{0texture} a diagonal charged lepton mass matrix before proceeding to the `rotation' defined by $S$ of Eq.(\ref{SC33}). Had we done this then we shall get field transformations corresponding to the `rotated' symmetry by adjoint acting on the `unrotated' transformations by the rotation $S$, which will produce a non-diagonal matrix for the charged leptons, which means that  upon `rotating' and getting the vanishing subtrace texture we would have left  the flavor basis. Actually, we added the extra Higgs fields exactly in order to get a `generic' charged lepton mass matrix in the `unrotated' basis while keeping the form of the Dirac neutrino mass matrix. The fields $\Phi_{4,5}$ are responsible for the desired form of $M_{D0}$, whereas the fields ($\Phi_{1,2,3}$) produce generic $M_{\ell 0}$.

In order to find the new `rotated' symmetry, we need to find then how all the fields would transform. Thus, we should explicitly write down the form of the mass matrices in terms of the Yukawa couplings when the Higgs and singlet scalar fields get vevs. Actually the invariance of the Majorana term under $Z_8^{ 0}\times Z_2^{ 0}$ implies the following constraint

\be
(Y^{b}_{0\chi}) = T^{0Z}_{\chi ab}\; (T^{0Z}_{\n R})^T \;(Y^{a}_{0\chi})\; (T^{0Z}_{\n_R})
\label{constr1}
\ee
where ($a,b=1,2,3$), $(Y^{b}_{0\chi})$ is a matrix in flavor space with element $Y^{b}_{0\chi ij}$ at its $(i,j)^{th}$ entry, and $T^{0Z}_{f}$($f=\chi, \n_R$) is a matrix (diagonal by construction) defining the transformation of the field $f$ under the considered symmetry factor $Z$ ($Z=Z_8^{ 0}$ or $Z_2^{ 0}$). This constraint (Eq. \ref{constr1}) can be solved for both symmetry factors and leads to the following form, when $\chi_a$ gets a vev $v_{0\chi_a}$:
\be
M_{R0} = \left(
\begin {array}{ccc}
Y^1_{0\chi 11}\; v_{0\chi_1} &0& 0\\
0& Y^1_{0\chi 22}\; v_{0\chi_1}& Y^2_{0\chi 23} \;v_{0\chi_2}\\
0&Y^2_{0\chi 23} \;v_{0\chi_2}&Y^3_{0\chi 33}\; v_{0\chi_3}
\end {array}
\right).
\label{M_R0}
\ee
The invariance of the Dirac neutrino mass term
under $Z_8^{ 0}\times Z_2^{ 0}$ implies the following constraint,
\be
 (Y^b_{0D}) = {(T^{0Z}_{\Phi})^\dagger}_{ab}\; (T^{0Z}_{D_L})^\dagger\; (Y^a_{0D}) \; (T^{0Z}_{\n_R}
) \label{constr2}
\ee
where  ($a,b=1,\ldots,5$), $(Y^a_{0D})$ is a matrix in flavor space with element $Y^a_{0D ij}$ at its $(i,j)^{\mbox{th}}$ entry, and $(T^{0Z}_{f})(f= \Phi, D_L, \n_R)$ is a diagonal -by construction- matrix defining the transformation of the field $f$ under the considered symmetry factor $Z$ ($Z$=$Z_8^{ 0}$ or $Z_2^{ 0}$) which leads when solved for both $Z_8^{ 0}$ and $Z_2^{ 0}$ to the following form, when $\Phi_a$ gets a vev $v_{0\Phi_a}$:
\be
M_{D0} = \left(
\begin {array}{ccc}
Y^4_{0D 11}\; v_{0\Phi_4} &0& 0\\
0& Y^4_{0D 22} \;v_{0\Phi_4}& 0\\
Y^5_{0D 31}\; v_{0\Phi_5}&0&Y^4_{0D 33}\; v_{0\Phi_4}
\end {array}
\right).
\label{M_D0}
\ee
And we get $M_{\n0} = M_{D0}\; M_{R0}^{-1}\; M_{D0}^T$ of the desired form with vanishing element at the ($2,1$)$^{\mbox{th}}$ entry. As to the charged lepton mass matrix, the invariance of the corresponding mass term give
\be
(Y^b_{0\ell}) = T^{0Z}_{\Phi ab}\; (T^{0Z}_{D_L})^\dagger \; (Y^a_{0\ell})\;(T^{0Z}_{\ell_R})
\label{constr3}
\ee
where  ($a,b=1,\ldots,5$), ($Y^a_{0\ell}$) is a matrix in flavor space with element $Y^a_{0\ell ij}$ at its $(i,j)^{\mbox{th}}$ entry, and $T^{0Z}_{f}(f=\Phi, D_L, \ell_R)$ is a matrix defining the transformation of the field $f$ under the considered symmetry factor $Z$ ($Z$=$Z_8^{ 0}$ or $Z_2^{ 0}$) which leads to a generic form for the charged lepton matrix:
\be
M_{\ell 0} = \left(
\begin {array}{ccc}
Y^1_{0\ell 11}\; v_{0\Phi_1} &Y^1_{0\ell 12}\; v_{0\Phi_1}&Y^1_{0\ell 13} \;v_{0\Phi_1} \\
Y^2_{0\ell 21}\; v_{0\Phi_2}& Y^2_{0\ell 22}\; v_{0\Phi_2}& Y^2_{0\ell 23}\; v_{0\Phi_2}\\
Y^3_{0\ell 31}\; v_{0\Phi_3}&Y^3_{0\ell 32}\; v_{0\Phi_3}&Y^3_{0\ell 33}\; v_{0\Phi_3}
\end {array}
\right).
\label{M_ell0}
\ee

In order to find the field transformations corresponding to the new `rotated' symmetry defined by $S$ (Eq. \ref{SC33}), we apply the same rule as in  Eq. (\ref{conjugation}) or Eq.(\ref{star}), with caution,  for all the fields $f$:
\be
 T^Z_f = S^\dagger\; T^{0Z}_f\;  S
\label{rule}
\ee
and extending in the case of the 5-dimensional $\Phi$ the matrix $S$ to be $S_{ex}=\mbox{ {diag}} \left( S, 1_{2\times2}\right)$. {  We state in Table~(\ref{newtableC33}) the resulting assignments for the fields under $Z_8 \times Z_2$.

\begin{table}[hbtp]
\begin{center}
{\small
\begin{tabular}{cccc}
\toprule
\multicolumn{4}{c}{$ \mbox{Symmetry under}\; Z_8\; \mbox{factor}$}\\
\midrule
    $\boldsymbol{\Phi} =\left(\Phi_1, \Phi_2, \Phi_3, \Phi_4, \Phi_5 \right)^T$ & $\boldsymbol{D_L} = \left(D_{L1}, D_{L2}, D_{L3}\right)^T $ & $\boldsymbol{\n_R} = \left( \n_{R1}, \n_{R2}, \n_{R3}\right)^T$  & $\boldsymbol{\chi} = \left(\chi_{1}, \chi_{2}, \chi_{3}\right)^T$     \\
\midrule
$\left( \begin{array}{ccccc} 0 & i & 0 & 0 & 0 \\
-i & 0 & 0 & 0 & 0 \\ 0 & 0 & \omega & 0 & 0 \\ 0 & 0 & 0 & 1 & 0 \\ 0 & 0 & 0 & 0 & \omega^7 \end{array}   \right)$ &
\multicolumn{2}{c}{
$\left( \begin{array}{ccc} 0 & i & 0\\
-i & 0 & 0 \\ 0 & 0 & \omega  \end{array}   \right)$ }
&
$\left( \begin{array}{ccc} \frac{(1+\omega^3)}{2} & \frac{i(1-\omega^3)}{2} & 0\\
\frac{i(-1+\omega^3)}{2} & \frac{(1+\omega^3)}{2} & 0 \\ 0 & 0 & \omega^6  \end{array}   \right)$
 \\
\toprule
\multicolumn{4}{c}{$ \mbox{Symmetry under}\; Z_2\; \mbox{factor}$}\\
\midrule
 $\mbox{ diag}(1,1,1,-1,-1)$ & $\mbox{ diag}(1,1,1)$ & $\mbox{ diag}(-1,-1,-1)$ & $\mbox{\small diag}(1,1,1)$  \\
\bottomrule
\end{tabular}
}
\end{center}
 \caption{  \small   The $Z_8\times Z_2$ symmetry realization of the vanishing subtrace $\mathbf C_{33}$. The  $D_{L1}$ indicates
the left-handed lepton doublet first
 family and so on. The $\chi_{k}$ denotes a scalar singlet which produces
 an entry in the right-handed Majorana mass matrix when acquiring a VEV
 at the see-saw scale. The right-handed charged leptons $\ell_R$ are assumed singlets under the discrete symmetry.
 $\omega$ denotes $e^{{i\,\pi/4}}$. }
\label{newtableC33}
 \end{table}

}

Note that we do not get generally diagonal matrices $T^Z_f$ because of the rotation $S$.  Thus one can write down similar constraints to those of Eqs.(\ref{constr1}, \ref{constr2}, \ref{constr3}) corresponding to the rotated symmetry, albeit with Yukawa couplings and vevs without the subscript $0$, and by solving them we get:
\be
M_{R} = \left(
\begin {array}{ccc}
-Y^1_{\chi 22}\; (v_{\chi_1}+i\,v_{\chi_2})  &-i\,Y^2_{\chi 12}\; (v_{\chi_1}+i\,v_{\chi_2})&  Y^2_{\chi 23} \;(-v_{\chi_1}+i\,v_{\chi_2})\\
\rule{0.7cm}{0.1mm} & Y^1_{\chi 22}\; (v_{\chi_1}+i\,v_{\chi_2})& -i\,Y^2_{\chi 23} \;(-v_{\chi_1}+i\,v_{\chi_2})\\
\rule{0.7cm}{0.1mm}&\rule{0.7cm}{0.1mm}&Y^3_{\chi 33}\; v_{\chi_3}
\end {array}
\right).
\label{M_R33}
\ee
where $\rule[1mm]{0.7cm}{0.1mm}$ denotes an element deduced by symmetry property of the matrix ($M = M^T$) and this convention will be used from now on.
\be
M_{D} = \left(
\begin {array}{ccc}
Y^4_{D 22}\; v_{\Phi_4} &-Y^4_{D 21} \;v_{\Phi_4}& 0\\
Y^4_{D 21} \; v_{\Phi_4}& Y^4_{D 22}\; v_{\Phi_4}& 0\\
-i\,Y^5_{D 32}\; v_{\Phi_5}&  Y^5_{D 32} \;v_{\Phi_5}&Y^4_{D 33}\; v_{\Phi_4}
\end {array}
\right).
\label{M_D33}
\ee
One can check that the resulting $M_\n$ satisfies the texture $\mathbf C_{33}$.
Note also that all the Yukawa couplings and the vevs in Eqs (\ref{M_R33},\ref{M_D33}) are different from those in Eqs (\ref{M_R0},\ref{M_D0}) since each set of Yukawa couplings and vevs correspond to the Lagrangian under a specific symmetry. However, they are related through the transformation:
\be
M= S^T\;M_0\;S
\label{masstransformS}
\ee
 which should be valid for ($M_\n, M_R, M_D$), and one can check that the form of $M_{(\n,\; D,\;R)}$ is the same as that of $S^T\; M_{(\n0,\; D0,\; R0)}\;S$.

{   We need to show now that the symmetry-gauge basis for the charged leptons, in which the symmetry was given, can under natural assumptions be taken, to a very good approximation, to be the flavor basis.}
Actually, we get a generic mass matrix $M_\ell$:
\be
M_{\ell } =  \left(
\begin {array}{ccc}
Y^2_{\ell 21}\;  v_{\Phi_1} - Y^1_{\ell 21}\; v_{\Phi_2} &Y^2_{\ell 22} \;v_{\Phi_1} - Y^1_{\ell 22} \;v_{\Phi_2} & Y^2_{\ell 23}\; v_{\Phi_1} - Y^1_{\ell 23} \;v_{\Phi_2} \\
Y^1_{\ell 21} \; v_{\Phi_1} + Y^2_{\ell 21}\; v_{\Phi_2} &Y^1_{\ell 22} \; v_{\Phi_1} + Y^2_{\ell 22}\; v_{\Phi_2} & Y^1_{\ell 23}\; v_{\Phi_1} + Y^2_{\ell 23}\; v_{\Phi_2}\\
Y^3_{\ell 31} \; v_{\Phi_3} & Y^3_{\ell 32}\; v_{\Phi_3} &Y^3_{\ell 33}\; v_{\Phi_3}
\end {array}
\right),
\label{M_ellgeneral}
\ee
so if assume the related vevs are comparable $v_{\Phi_1}\approx v_{\Phi_2} \approx v_{\Phi_3} \approx v$ then we get
\be
\label{M_ell}
M_{\ell } \approx v \left(
\begin {array}{ccc}
Y^2_{\ell 21} - Y^1_{\ell 21} &Y^2_{\ell 22}  - Y^1_{\ell 22} & Y^2_{\ell 23} - Y^1_{\ell 23} \\
Y^1_{\ell 21}  + Y^2_{\ell 21} &Y^1_{\ell 22}  + Y^2_{\ell 22} & Y^1_{\ell 23} + Y^2_{\ell 23}\\
Y^3_{\ell 31}  & Y^3_{\ell 32} &Y^3_{\ell 33}
\end {array}
\right) = v \left( \begin {array}{c}
{\bf a}^T\\{\bf b}^T\\{\bf c}^T
\end {array}
\right),
\ee
where $\bf a$,  $\bf b$ and  $\bf c$ stand for column vectors extracted from the corresponding rows, formed of Yukawa couplings,  in the  matrix $M_{\ell }$, and this abbreviation will be used from now  on. The dot product refers to the usual Hermitian inner product defined as ${\bf a.b} = \sum_{i=1}^{3} a_i\, b_i^*$.
Thus
\be
M_{\ell } M_{\ell}^\dagger \approx v^2 \left(\begin {array}{ccc}
{\bf a.a} &{\bf a.b}&{\bf a.c} \\
{\bf b.a} &{\bf b.b}&{\bf b.c}\\
{\bf c.a} &{\bf c.b}&{\bf c.c}
\end {array} \right),
\label{M_ell2}
\ee
so taking only the following natural assumption on the norms of the vectors
\bea \parallel {\bf a} \parallel /\parallel {\bf c} \parallel = m_e/m_\tau \sim 3 \times 10^{-4} &,&  \parallel {\bf b} \parallel /\parallel {\bf c} \parallel = m_\mu/m_\tau \sim 6 \times 10^{-2}\eea
one can diagonalize $M_{\ell } M_{\ell}^\dagger$ by an infinitesimal rotation as was done in \cite{0texture}, which proves that we are to a good approximation in the flavor basis.

{
Some remarks are in order here. First, one would naturally assume Yukawa couplings of the same order, and the assumption $||\bf a| \ll ||\bf b|| \ll ||\bf c ||$ can not be met unless there is a fine tuning in the Yukawas. We find nothing wrong with the needed finetuning, especially that an analogous finetuning, to enforce the charged lepton mass hierarchies, is needed in many similar models, and even in the SM \cite{grimus}. Second, as said earlier and in line with \cite{ma2004}, the subtrace texture is zero by construction in the symmetry-gauge basis of the neutrino fields, whereas the gauge-symmetry basis of the charged leptons is deviated infinitesimally from the flavor basis, and this deviation is of the order of the ``acute" charged lepton masses hierarchies, which means we are to a very good approximation in the flavor basis.

}

\subsection{Indirect realization of $\mathbf C_{11}$(Type I nonsingular): Vanishing of $M_{\n\,22} + M_{\n\, 33}$}
Following the same procedure as for the case $\mathbf C_{33}$, we just state briefly the results.
The `rotation' matrix which moves a zero texture at $(2,3)$ to the texture $\mathbf C_{11}$ is given by:
\be
S= \frac{1}{\sqrt{2}}\left(
\begin {array}{ccc}
 \sqrt{2}&0& 0\\
0& i& -1\\
0&i&1
\end {array}
\right)\Rightarrow S^T\; \left(
\begin {array}{ccc}
A &B& C\\
B& D& 0\\
C&0&F
\end {array}
\right) \; S = \left(
\begin {array}{ccc}
A &\frac{i}{\sqrt{2}}(B+C)& -\frac{1}{\sqrt{2}}(B-C)\\
\rule{0.7cm}{0.1mm} & -\frac{1}{2}(D+F))& -\frac{i}{2}(D-F)\\
\rule{0.7cm}{0.1mm}&\rule{0.7cm}{0.1mm} &\frac{1}{2}(D+F)
\end {array}
\right),
\label{SC11}
\ee
and we check that the sum of elements at $(2,2)$ and $(3,3)$ vanishes.
At the Lagrangian level, the symmetry transformations for the fields which imposes a zero texture neutrino mass matrix with generic charged lepton mass matrix are given in Table~(\ref{tablec11typeI})
\begin{table}[hbtp]
\begin{center}
\begin{tabular}{ccccccccccccccc}
\toprule
\multicolumn{15}{c}{$ \mbox{Symmetry under}\; Z_8^0\; \mbox{factor}$}\\
\midrule
    $\Phi_1$  & $\Phi_2$ & $\Phi_3$ & $\Phi_4$ & $\Phi_5$ & $D_{L1} $ &$D_{L2} $ & $D_{L3} $ &$\n_{R1}$ &$\n_{R2}$& $\n_{R3}$ & $\chi_{1}$ &
   $\chi_{2}$ & $\chi_{3}$ & $\ell_R$    \\
\midrule
 $1$ & $\omega^4$ & $\omega$ & $1$ & $\omega$ & $1$ & $\omega^4$ & $\omega$ & $1$ & $\omega^4$ & $\omega$ &  $1$ & $\omega^4$ & $\omega^6$ & $1$
\\
\toprule
\multicolumn{15}{c}{$ \mbox{Symmetry under}\; Z_2^0\; \mbox{factor}$}\\
\midrule
$1$& $1$& $1$& $\omega^4$& $\omega^4$& $1$ &$1$& $1$& $\omega^4$& $\omega^4$& $\omega^4$& $1$ & $1$ & $1$ & $1$  \\
\bottomrule
\end{tabular}
\end{center}
 \caption{ \small  The $Z_8^0\times Z_2^0$ symmetry realization of the
 one zero texture at $(2,3)$-entry corresponding upon rotation to vanishing subtrace $\mathbf C_{11}$. The $D_{L1}$ indicates the left-handed lepton doublet first family and so on. The $\chi_{k}$ denotes a scalar singlet which produces
 an entry in the right-handed Majorana mass matrix when acquiring a VEV
 at the see-saw scale. The right-handed charged leptons $\ell_R$ are assumed singlets under the discrete symmetry. $\omega$ denotes $e^{{i\,\pi/4}}$. }
\label{tablec11typeI}
 \end{table}

By forming bilinear terms of the fields we see that the above transformations force a neutrino mass matrix with zero texture at $(2,3)$ entry. Again we define the new transformations for the fields corresponding to the new symmetry imposing the vanishing subtrace by the rule in Eq. (\ref{rule}), but with $S$ as given in Eq.(\ref{SC11}).  {  We state in Table~(\ref{newtableC11}) the resulting assignments for the fields under $Z_8 \times Z_2$.

\begin{table}[hbtp]
\begin{center}
{\small
\begin{tabular}{cccc}
\toprule
\multicolumn{4}{c}{$ \mbox{Symmetry under}\; Z_8\; \mbox{factor}$}\\
\midrule
    $\boldsymbol{\Phi} =\left(\Phi_1, \Phi_2, \Phi_3, \Phi_4, \Phi_5 \right)^T$ & $\boldsymbol{D_L} = \left(D_{L1}, D_{L2}, D_{L3}\right)^T $ & $\boldsymbol{\n_R} = \left( \n_{R1}, \n_{R2}, \n_{R3}\right)^T$  & $\boldsymbol{\chi} = \left(\chi_{1}, \chi_{2}, \chi_{3}\right)^T$      \\
\midrule
$\left( \begin{array}{ccccc} 1 & 0 & 0 & 0 & 0 \\
 0& \frac{-1+\omega}{2} & \frac{-i(1+\omega)}{2} & 0 & 0 \\ 0 & \frac{i(1+\omega)}{2} & \frac{-1+\omega}{2} & 0 & 0 \\ 0 & 0 & 0 & 1 & 0 \\ 0 & 0 & 0 & 0 & \omega \end{array}   \right)$ &
\multicolumn{2}{c} {$\left( \begin{array}{ccc} 1 & 0 & 0\\
0 & \frac{-1+\omega}{2} & \frac{-i(1+\omega)}{2} \\ 0 & \frac{i(1+\omega)}{2} &  \frac{-1+\omega}{2} \end{array}   \right)$}
&
$\left( \begin{array}{ccc} 1 & 0 & 0\\
0 & \frac{-1+\omega^6}{2} & \frac{-i(1+\omega^6)}{2} \\ 0 & \frac{i(1+\omega^6)}{2} &  \frac{-1+\omega^6}{2} \end{array}   \right)$
 \\
\toprule
\multicolumn{4}{c}{$ \mbox{Symmetry under}\; Z_2\; \mbox{factor}$}\\
\midrule
 $\mbox{ diag}(1,1,1,-1,-1)$ & $\mbox{ diag}(1,1,1)$ & $\mbox{ diag}(-1,-1,-1)$ & $\mbox{\small diag}(1,1,1)$ \\
\bottomrule
\end{tabular}
}
\end{center}
 \caption{  \small  The $Z_8\times Z_2$ symmetry realization of the vanishing subtrace $\mathbf C_{11}$. The $D_{L1}$ indicates
the left-handed lepton doublet first
 family and so on. The $\chi_{k}$ denotes a scalar singlet which produces
 an entry in the right-handed Majorana mass matrix when acquiring a VEV
 at the see-saw scale. The right-handed charged leptons are assumed to be singlets under the discrete symmetry.
 $\omega$ denotes $e^{{i\,\pi/4}}$. }
\label{newtableC11}
 \end{table}

}

The `rotated' symmetry imposes some constraints on the Yukawa couplings and the vevs, which when solved give the following results for $M_R$ and $M_D$.
\be
M_{R} = \left(
\begin {array}{ccc}
Y^1_{\chi 11}\; v_{\chi_1}  &Y^2_{\chi 12} \;(v_{\chi_2}+i\,v_{\chi_3})& i\, Y^2_{\chi 12}\; (v_{\chi_2}+i\,v_{\chi_3})\\
\rule{0.7cm}{0.1mm}& i \,(-Y^1_{\chi 23}\; v_{\chi_1}+i \,Y^2_{\chi 33}\; v_{\chi_2} + Y^2_{\chi 33}\; v_{\chi_3} )& Y^1_{\chi 23}\; v_{\chi_1}+i\,Y^2_{\chi 33} \;v_{\chi_2} + Y^2_{\chi 33}\; v_{\chi_3} \\
\rule{0.7cm}{0.1mm}&\rule{0.7cm}{0.1mm}&-i\, (-Y^1_{\chi 23}\; v_{\chi_1}+i\, Y^2_{\chi 33}\; v_{\chi_2} + Y^2_{\chi 33}\; v_{\chi_3} )
\end {array}
\right).
\label{M_R11}
\ee
and,
\be
M_{D} = \left(
\begin {array}{ccc}
Y^4_{D 11}\; v_{\Phi_4} &i\, Y^5_{D 13}\; v_{\Phi_5}&  Y^5_{D 13}\; v_{\Phi_5} \\
0& Y^4_{D 22}\; v_{\Phi_4}&  Y^4_{D 23}\; v_{\Phi_4}\\
0&-Y^4_{D 23}\; v_{\Phi_4}&Y^4_{D 22}\; v_{\Phi_4}
\end {array}
\right).
\label{M_D11}
\ee
 One can check that the resulting $M_\n$ satisfies the texture $\mathbf C_{11}$.

 As to $M_\ell$ we get,
\be
M_{\ell } = \left(
\begin {array}{ccc}
Y^1_{\ell 11}\; v_{\Phi_1}  &Y^1_{\ell 12} \; v_{\Phi_1}&Y^1_{\ell 13}\; v_{\Phi_1} \\
Y^3_{\ell 13}\; v_{\Phi_2} + Y^3_{\ell 21}\; v_{\Phi_3} &Y^2_{\ell 22}\; v_{\Phi_2} - Y^2_{\ell 32}\; v_{\Phi_3}&Y^3_{\ell 33} \; v_{\Phi_2}+ Y^3_{\ell 23}\; v_{\Phi_3}\\
-Y^3_{\ell 21}\; v_{\Phi_2}+ Y^3_{\ell 31}\; v_{\Phi_3} &Y^2_{\ell 32}\; v_{\Phi_2} + Y^2_{\ell 22} \; v_{\Phi_3}&-Y^3_{\ell 23}\; v_{\Phi_2}+
Y^3_{\ell 33}\; v_{\Phi_3}
\end {array}
\right),
\label{M_ell_11}
\ee
 then we see that if assume all the related vevs are comparable $v_{\Phi_1}\approx v_{\Phi_2} \approx v_{\Phi_3} \approx v$ then we get
\be
M_{\ell } \approx v \left(
\begin {array}{ccc}
Y^1_{\ell 11}  &Y^1_{\ell 12} &Y^1_{\ell 13} \\
Y^3_{\ell 13} + Y^3_{\ell 21} &Y^2_{\ell 22} - Y^2_{\ell 32}&Y^3_{\ell 33} + Y^3_{\ell 23}\\
-Y^3_{\ell 21}+ Y^3_{\ell 31} &Y^2_{\ell 32} + Y^2_{\ell 22} &-Y^3_{\ell 23}+Y^3_{\ell 33}
\end {array}
\right) = v \left( \begin {array}{c}
{\bf a}^T\\{\bf b}^T\\{\bf c}^T
\end {array}
\right)
\label{M_ell_11f}
\ee
which can be diagonalized by an infinitesimal rotation under some natural assumptions on the amplitudes of the Yukawa vectors as done for the case of $\mathbf C_{33}$.

\subsection{Indirect realization of $\mathbf C_{22}$(Type I nonsingular): Vanishing of $M_{\n\,11} + M_{\n\,33}$}
The `rotation' matrix which moves a zero texture at $(1,3)$ to the texture $\mathbf C_{22}$ is given by:
\be
S= \frac{1}{\sqrt{2}}\left(
\begin {array}{ccc}
i&0& -1\\
0& \sqrt{2}& 0\\
i&0&1
\end {array}
\right)\Rightarrow S^T\; \left(
\begin {array}{ccc}
A &B& 0\\
B& D& E\\
0&E&F
\end {array}
\right)\; S = \left(
\begin {array}{ccc}
-\frac{1}{2} (A+F)&\frac{i}{\sqrt{2}}(B+E)& \frac{i}{\sqrt{2}}(F-A)\\
\rule{0.7cm}{0.1mm}& D& \frac{1}{\sqrt{2}}(E-B)\\
\rule{0.7cm}{0.1mm}&\rule{0.7cm}{0.1mm}&\frac{1}{2}(A+F)
\end {array}
\right),
\label{SC22}
\ee
and we check that the sum of elements at $(1,1)$ and $(3,3)$ vanishes.

At the Lagrangian level, the symmetry transformations for the fields which imposes a zero texture $M_{\n0}$ with generic $M_{\ell0}$ are given
in Table~(\ref{tablec2typeI})
\begin{table}[hbtp]
\begin{center}
\begin{tabular}{ccccccccccccccc}
\toprule
\multicolumn{15}{c}{$ \mbox{Symmetry under}\; Z_8^0\; \mbox{factor}$}\\
\midrule
    $\Phi_1$  & $\Phi_2$ & $\Phi_3$ & $\Phi_4$ & $\Phi_5$ & $D_{L1} $ &$D_{L2} $ & $D_{L3} $ &$\n_{R1}$ &$\n_{R2}$& $\n_{R3}$ & $\chi_{1}$ &
   $\chi_{2}$ & $\chi_{3}$ & $\ell_R$    \\
\midrule
 $1$ & $\omega^4$ & $\omega$ & $1$ & $\omega^5$ & $\omega$ & $\omega^4$ & $1$ & $\omega$ & $\omega^4$ & $1$ &  $1$ & $\omega^6$ & $\omega^6$ & $1$
\\
\toprule
\multicolumn{15}{c}{$ \mbox{Symmetry under}\; Z_2^0\; \mbox{factor}$}\\
\midrule
$1$& $1$& $1$& $\omega^4$& $\omega^4$& $1$ &$1$& $1$& $\omega^4$& $\omega^4$& $\omega^4$& $1$ & $1$ & $1$ & $1$  \\
\bottomrule
\end{tabular}
\end{center}
 \caption{\small  The $Z_8^0\times Z_2^0$ symmetry realization of the
 one zero texture at $(1,3)$-entry corresponding upon rotation to vanishing subtrace $\mathbf C_{22}$. The  $D_{L1}$ indicates the left-handed lepton doublet first family and so on. The $\chi_{k}$ denotes a scalar singlet which produces
 an entry in the right-handed Majorana mass matrix when acquiring a VEV at the see-saw scale. $\omega$ denotes $e^{{i\,\pi/4}}$. }
\label{tablec2typeI}
 \end{table}

By forming bilinear terms of the fields we see that the above transformations force the $(1,3)$ entry in $M_{\n 0}$ to vanish. Again we define the new transformations for the fields corresponding to the new symmetry imposing the vanishing subtrace by the  rule in  Eq.(\ref{rule}), but with $S$ given by Eq. (\ref{SC22}).

{  We state in Table~(\ref{newtableC22}) the resulting assignments for the fields under $Z_8 \times Z_2$.
\begin{table}[hbtp]
\begin{center}
{\small
\begin{tabular}{cccc}
\toprule
\multicolumn{4}{c}{$ \mbox{Symmetry under}\; Z_8\; \mbox{factor}$}\\
\midrule
    $\boldsymbol{\Phi} =\left(\Phi_1, \Phi_2, \Phi_3, \Phi_4, \Phi_5 \right)^T$ & $\boldsymbol{D_L} = \left(D_{L1}, D_{L2}, D_{L3}\right)^T $ & $\boldsymbol{\n_R} = \left( \n_{R1}, \n_{R2}, \n_{R3}\right)^T$  & $\boldsymbol{\chi} = \left(\chi_{1}, \chi_{2}, \chi_{3}\right)^T$      \\
\midrule
$\left( \begin{array}{ccccc} \frac{1+\omega}{2} & 0 & \frac{i(1-\omega)}{2} & 0 & 0 \\
  0 & -1 & 0 & 0 & 0 \\ \frac{-i(1-\omega)}{2} & 0 & \frac{1+\omega}{2} & 0 & 0 \\ 0 & 0 & 0 & 1 & 0 \\ 0 & 0 & 0 & 0 & \omega^5 \end{array}   \right)$ &
\multicolumn{2}{c} {$\left( \begin{array}{ccc} \frac{1+\omega}{2} & 0 & \frac{-i(1-\omega)}{2}\\
0 & -1 & 0 \\ \frac{i(1-\omega)}{2} & 0 &  \frac{1+\omega}{2} \end{array}   \right)$}
&
 $\left( \begin{array}{ccc} 0& 0 & i\\
0 & \omega^6 & 0 \\ -i & 0 &  0 \end{array}   \right)$
 \\
\toprule
\multicolumn{4}{c}{$ \mbox{Symmetry under}\; Z_2\; \mbox{factor}$}\\
\midrule
 $\mbox{ diag}(1,1,1,-1,-1)$ & $\mbox{ diag}(1,1,1)$ & $\mbox{ diag}(-1,-1,-1)$ & $\mbox{\small diag}(1,1,1)$ \\
\bottomrule
\end{tabular}
}
\end{center}
 \caption{  \small  The $Z_8\times Z_2$ symmetry realization of the vanishing subtrace $\mathbf C_{22}$. The  $D_{L1}$ indicates
the left-handed lepton doublet first
 family and so on. The $\chi_{k}$ denotes a scalar singlet which produces
 an entry in the right-handed Majorana mass matrix when acquiring a VEV
 at the see-saw scale. The right-handed charged leptons are assumed to be singlets under the discrete symmetry.
 $\omega$ denotes $e^{{i\,\pi/4}}$. }
\label{newtableC22}
 \end{table}

}

The `rotated' symmetry imposes some constraints on the Yukawa couplings and the vevs, which when solved give the following results,
\be
M_{R} =\left(
\begin {array}{ccc}
i\, [Y^3_{\chi 33}\; (v_{\chi_1}+i\, v_{\chi_3}) - Y^2_{\chi 13}\; v_{\chi_2} ] &Y^3_{\chi 23} \;(-v_{\chi_1}+i\,v_{\chi_3})&  Y^3_{\chi 33}\; (v_{\chi_1}+i\,v_{\chi_3}) + Y^2_{\chi 13}\; v_{\chi_2}\\
\rule{0.7cm}{0.1mm}& -i Y^3_{\chi 22}\; (v_{\chi_1}+i\,v_{\chi_3})& -i\, Y^3_{\chi 23}\; (-v_{\chi_1}+i\,v_{\chi_3}) \\
\rule{0.7cm}{0.1mm}&\rule{0.7cm}{0.1mm}&-i\, [Y^3_{\chi 33}\; (v_{\chi_1}+i\, v_{\chi_3}) - Y^2_{\chi 13} \;v_{\chi_2} ]
\end {array}
\right).
\label{M_R_22}
\ee
and
\be
M_{D} = \left(
\begin {array}{ccc}
Y^4_{D 11}\; v_{\Phi_4} &0& - Y^4_{D 31}\; v_{\Phi_4} \\
-i Y^5_{D 23} \;v_{\Phi_5}& Y^4_{D 22}\; v_{\Phi_4}&  Y^5_{D 23} \;v_{\Phi_5}\\
 Y^4_{D 31}\; v_{\Phi_4}&0&Y^4_{D 11}\; v_{\Phi_4}
\end {array}
\right).
\label{M_D_22}
\ee
One can check that the resulting $M_\n$ satisfies the texture $\mathbf C_{22}$.
 As to $M_\ell$ one gets,
\be
M_{\ell } = \left(
\begin {array}{ccc}
Y^1_{\ell 11}\;v_{\Phi_1}+ Y^3_{\ell 11}\;v_{\Phi_3} &Y^1_{\ell 12}\;v_{\Phi_1}+ Y^3_{\ell 12}\;v_{\Phi_3} &Y^1_{\ell 13}\;v_{\Phi_1} + Y^3_{\ell 13}\;v_{\Phi_3} \\
Y^2_{\ell 21}\;v_{\Phi_2} &Y^2_{\ell 22}\;v_{\Phi_2} &Y^2_{\ell 23}\;v_{\Phi_2}\\
Y^3_{\ell 11}\;v_{\Phi_1}- Y^1_{\ell 11}\;v_{\Phi_3} &Y^3_{\ell 12} \;v_{\Phi_1}- Y^1_{\ell 12}\;v_{\Phi_3} &Y^3_{\ell 13}\;v_{\Phi_1}-Y^1_{\ell 13}\;v_{\Phi_3}
\end {array}
\right) ,
\label{M_ell_22}
\ee
 then we see that if assume all the related vevs are comparable $v_{\Phi_1}\approx v_{\Phi_2} \approx v_{\Phi_3} \approx v$ then we get
\be
M_{\ell } \approx v \left(
\begin {array}{ccc}
Y^1_{\ell 11}+ Y^3_{\ell 11} &Y^1_{\ell 12}+ Y^3_{\ell 12} &Y^1_{\ell 13} + Y^3_{\ell 13} \\
Y^2_{\ell 21} &Y^2_{\ell 22} &Y^2_{\ell 23}\\
Y^3_{\ell 11}- Y^1_{\ell 11} &Y^3_{\ell 12} - Y^1_{\ell 12} &Y^3_{\ell 13}-Y^1_{\ell 13}
\end {array}
\right) = v \left( \begin {array}{c}
{\bf a}^T\\{\bf b}^T\\{\bf c}^T
\end {array}
\right),
\label{M_ell_22f}
\ee
which can be diagonalized by an infinitesimal rotation under some natural assumptions on the amplitudes of the Yukawa vectors as done for the previous two cases.

\subsection{Indirect realization of $\mathbf C_{31}$(Type I nonsingular): Vanishing of $M_{\n\,12} + M_{\n\,23}$}
The `rotation' matrix which moves a zero texture at $(2,3)$ to the texture $\mathbf C_{23}$ is given by:
\be
S= \frac{1}{\sqrt{2}}\left(
\begin {array}{ccc}
1&0& -1\\
0& \sqrt{2}& 0\\
1&0&1
\end {array}
\right)\Rightarrow S^T\; \left(
\begin {array}{ccc}
A &B& C\\
B& D& 0\\
C&0&F
\end {array}
\right) \; S = \left(
\begin {array}{ccc}
\frac{1}{2} (A+F) + C&\frac{1}{\sqrt{2}}B& \frac{1}{\sqrt{2}}(F-A)\\
\rule{0.7cm}{0.1mm}& D& -\frac{1}{\sqrt{2}}B\\
\rule{0.7cm}{0.1mm}&\rule{0.7cm}{0.1mm}&\frac{1}{2}(A+F)-C
\end {array}
\right),
\label{SC31}
\ee
and we check that the sum of elements at $(1,2)$ and $(2,3)$ vanishes.

At the Lagrangian level, the symmetry transformations for the fields which imposes a zero texture $M_{\n 0}$ at $(2,3)$ entry with generic $M_{\ell 0}$ are given in Table~(\ref{tablec31typeI})
\begin{table}[hbtp]
\begin{center}
\begin{tabular}{ccccccccccccccc}
\toprule
\multicolumn{15}{c}{$ \mbox{Symmetry under}\; Z_8^0\; \mbox{factor}$}\\
\midrule
    $\Phi_1$  & $\Phi_2$ & $\Phi_3$ & $\Phi_4$ & $\Phi_5$ & $D_{L1} $ &$D_{L2} $ & $D_{L3} $ &$\n_{R1}$ &$\n_{R2}$& $\n_{R3}$ & $\chi_{1}$ &
   $\chi_{2}$ & $\chi_{3}$ & $\ell_R$    \\
\midrule
 $1$ & $\omega^4$ & $\omega$ & $1$ & $\omega$ & $1$ & $\omega^4$ & $\omega$ & $1$ & $\omega^4$ & $\omega$ &  $1$ & $\omega^4$ & $\omega^6$ & $1$
\\
\toprule
\multicolumn{15}{c}{$ \mbox{Symmetry under}\; Z_2^0\; \mbox{factor}$}\\
\midrule
$1$& $1$& $1$& $\omega^4$& $\omega^4$& $1$ &$1$& $1$& $\omega^4$& $\omega^4$& $\omega^4$& $1$ & $1$ & $1$ & $1$  \\
\bottomrule
\end{tabular}
\end{center}
 \caption{\small  The $Z_8^0\times Z_2^0$ symmetry realization of the
 one zero texture at $(2,3)$-entry corresponding upon rotation to vanishing subtrace $\mathbf C_{31}$. The $D_{L1}$ indicates the left-handed lepton doublet first
 family and so on. The $\chi_{k}$ denotes a scalar singlet which produces
 an entry in the right-handed Majorana mass matrix when acquiring a VEV at the see-saw scale.
 $\omega$ denotes $e^{{i\,\pi/4}}$. }
\label{tablec31typeI}
 \end{table}

 Again we define the new transformations for the fields corresponding to the new symmetry imposing the vanishing subtrace by the adjoint action rule (Eq. \ref{rule}), but with $S$ given by Eq. (\ref{SC31}).
 {  We state in Table~(\ref{newtableC31}) the resulting assignments for the fields under $Z_8 \times Z_2$.

\begin{table}[hbtp]
\begin{center}
{\small
\begin{tabular}{cccc}
\toprule
\multicolumn{4}{c}{$ \mbox{Symmetry under}\; Z_8\; \mbox{factor}$}\\
\midrule
    $\boldsymbol{\Phi} =\left(\Phi_1, \Phi_2, \Phi_3, \Phi_4, \Phi_5 \right)^T$ & $\boldsymbol{D_L} = \left(D_{L1}, D_{L2}, D_{L3}\right)^T $ & $\boldsymbol{\n_R} = \left( \n_{R1}, \n_{R2}, \n_{R3}\right)^T$  & $\boldsymbol{\chi} = \left(\chi_{1}, \chi_{2}, \chi_{3}\right)^T$      \\
\midrule
$\left( \begin{array}{ccccc} \frac{1+\omega}{2} & 0 & \frac{-1+\omega}{2} & 0 & 0 \\
  0 & -1 & 0 & 0 & 0 \\ \frac{-1+\omega}{2} & 0 & \frac{1+\omega}{2} & 0 & 0 \\ 0 & 0 & 0 & 1 & 0 \\ 0 & 0 & 0 & 0 & \omega \end{array}   \right)$ &
\multicolumn{2}{c} {$\left( \begin{array}{ccc} \frac{1+\omega}{2} & 0 & \frac{-1+\omega}{2}\\
0 & -1 & 0 \\ \frac{-1+\omega}{2} & 0 &  \frac{1+\omega}{2} \end{array}   \right)$}
&
 $\left( \begin{array}{ccc} \frac{1+\omega^6}{2} & 0 & \frac{-1+\omega^6}{2}\\
0 & -1 & 0 \\ \frac{-1+\omega^6}{2} & 0 &  \frac{1+\omega^6}{2} \end{array}   \right)$
 \\
\toprule
\multicolumn{4}{c}{$ \mbox{Symmetry under}\; Z_2\; \mbox{factor}$}\\
\midrule
 $\mbox{ diag}(1,1,1,-1,-1)$ & $\mbox{ diag}(1,1,1)$ & $\mbox{ diag}(-1,-1,-1)$ & $\mbox{\small diag}(1,1,1)$ \\
\bottomrule
\end{tabular}
}
\end{center}
 \caption{  \small  The $Z_8\times Z_2$ symmetry realization of the vanishing subtrace $\mathbf C_{31}$. The  $D_{L1}$ indicates
the left-handed lepton doublet first
 family and so on. The $\chi_{k}$ denotes a scalar singlet which produces
 an entry in the right-handed Majorana mass matrix when acquiring a VEV
 at the see-saw scale. The right-handed charged leptons are assumed to be singlets under the discrete symmetry.
 $\omega$ denotes $e^{{i\,\pi/4}}$. }
\label{newtableC31}
 \end{table}

}

{  We repeat that if we could guess the ``non-diagonal'' transformations under $Z_8 \times Z_2$ of Tables (\ref{newtableC33}, \ref{newtableC11}, \ref{newtableC22} and \ref{newtableC31}), then we would not have needed to resort to the `rotation' method relating them to simpler ones under $Z_8^0 \times Z_2^0$. However, as clear from the tables, the Higgs and scalar fields transformations in particular are difficult to guess directly.}

 The `rotated' symmetry imposes some constraints on the Yukawa couplings and the vevs, which when solved give the following results.
\be
M_{R} = \left(
\begin {array}{ccc}
Y^1_{\chi 33}\; v_{\chi_1}+Y^1_{\chi 13}\; v_{\chi_3}  &-Y^2_{\chi 23}\; v_{\chi_2}&  Y^1_{\chi 13}\; v_{\chi_1}+Y^1_{\chi 33}\; v_{\chi_3}\\
\rule{0.7cm}{0.1mm}& - Y^3_{\chi 22}\;(v_{\chi_1}-v_{\chi_3})&  Y^2_{\chi 23}\; v_{\chi_2} \\
\rule{0.7cm}{0.1mm}&\rule{0.7cm}{0.1mm}&Y^1_{\chi 33}\; v_{\chi_1} + Y^1_{\chi 13}\; v_{\chi_3}
\end {array}
\right),
\label{M_R_31}
\ee
and
\be
M_{D} = \left(
\begin {array}{ccc}
Y^4_{D 11}\; v_{\Phi_4}+ Y^5_{D 11} \;v_{\Phi_5} &0& Y^4_{D 31}\; v_{\Phi_4}+ Y^5_{D 11}\; v_{\Phi_5} \\
0& Y^4_{D 22}\; v_{\Phi_4}&  0\\
- Y^5_{D 11}\; v_{\Phi_5}+ Y^4_{D 31}\; v_{\Phi_4}&0&Y^4_{D 11}\; v_{\Phi_4} - Y^5_{D 11}\; v_{\Phi_5}
\end {array}
\right).
\label{M_D_31}
\ee
One can check that the resulting $M_\n$ satisfies the texture $C_{31}$.
 As to $M_\ell$, we get
\be
M_{\ell } =  \left(
\begin {array}{ccc}
Y^1_{\ell 11} \; v_{\Phi_1}+ Y^1_{\ell 31} \;v_{\Phi_3} &Y^1_{\ell 12}\; v_{\Phi_1} + Y^1_{\ell 32}\; v_{\Phi_3} & Y^1_{\ell 13}\; v_{\Phi_1} + Y^1_{\ell 33} \;v_{\Phi_3} \\
Y^2_{\ell 21} \; v_{\Phi_2}  &Y^2_{\ell 22}\; v_{\Phi_2} & Y^2_{\ell 23}\; v_{\Phi_2}\\
Y^1_{\ell 31} \; v_{\Phi_1} + Y^1_{\ell 11}\; v_{\Phi_3} &Y^1_{\ell 32}\; v_{\Phi_1} + Y^1_{\ell 12}\; v_{\Phi_3} & Y^1_{\ell 33}\; v_{\Phi_1} + Y^1_{\ell 13} \;v_{\Phi_3}
\end {array}
\right),
\label{M_ell_31}
\ee
then we see that if assume   $v \approx v_{\Phi_1}\approx v_{\Phi_2} \gg v_{\Phi_3} $ then we get
\be
\label{M_ell}
M_{\ell } \approx v \left(
\begin {array}{ccc}
Y^1_{\ell 11} &Y^1_{\ell 12}&Y^1_{\ell 13}  \\
Y^2_{\ell 21} &Y^2_{\ell 22} &Y^2_{\ell 23}\\
Y^1_{\ell 31} &Y^1_{\ell 32}  &Y^1_{\ell 33}
\end {array}
\right) = v \left( \begin {array}{c}
{\bf a}^T\\{\bf b}^T\\{\bf c}^T
\end {array}
\right),
\ee
which can be diagonalized by an infinitesimal rotation under some natural assumptions on the amplitudes of the vectors as done in the previous cases.

\section{Indirect realization of type I seesaw with $Z_{12} \times Z_2$ symmetry for singular textures}
We shall adopt the same strategy of moving from the symmetry imposing a zero texture where $M_D$ is singular to the symmetry imposing a vanishing subtrace with again $M_D$ singular, which gives via seesaw type I a singular neutrino mass matrix. Again, we follow \cite{0texture} to find the symmetry transformations leading to zero elements at singular $M_\n$, but will add in new fields so that to get a generic charged lepton mass matrix and not a diagonal one as was the case in \cite{0texture}, in such a way that the new `rotated' symmetry, as defined in Eq.(\ref{rule}), leads to vanishing subtraces at singular $M_\n$ and to another generic $M_\ell$. The latter under some reasonable assumptions can be diagonalized via infinitesimal rotations, which put us to a good approximation in the flavor basis.

\subsection{Indirect realization of $\mathbf C_{33}$ (Type I singular): Vanishing of $M_{\n\,11} + M_{\n\, 22}$}
As in the nonsingular cases, we move from  zero texture at $(1,2)$ to the texture $C_{33}$  by $S$ of Eq. (\ref{SC33}).

At the Lagrangian level, the symmetry transformations for the fields which imposes a zero texture neutrino mass matrix with generic charged lepton mass matrix and singular Dirac neutrino mass matrix are given in Table~(\ref{33singular}).
\begin{table}[hbtp]
\begin{center}
\begin{tabular}{ccccccccccccccccc}
\toprule
\multicolumn{17}{c}{$ \mbox{Symmetry under}\; Z_{12}^0\; \mbox{factor}$}\\
\midrule
    $\Phi_1$  & $\Phi_2$ & $\Phi_3$ & $\Phi_4$ & $\Phi_5$ & $\Phi_6$ & $\Phi_7$ & $D_{L1} $ &$D_{L2} $ & $D_{L3} $ &$\n_{R1}$ &$\n_{R2}$& $\n_{R3}$ & $\chi_{1}$ &
   $\chi_{2}$ & $\chi_{3}$ & $\ell_R$    \\
\midrule
 $\theta^{11}$ & $\theta^{9}$ & $\theta^{4}$ & $\theta^{2}$ & $\theta^{8}$ & $\theta^{9}$ & $\theta$ & $\theta^{11}$ & $\theta^{9}$ & $\theta^{4}$ & $\theta$ &  $\theta^{2}$ & $\theta^{5}$ & $\theta^{10}$ & $\theta^{8}$ & $\theta^{2}$ & $1$
\\
\toprule
\multicolumn{17}{c}{$ \mbox{Symmetry under}\; Z_2^0\; \mbox{factor}$}\\
\midrule
$1$& $1$& $1$ & $\theta^{6}$& $\theta^{6}$ & $\theta^{6}$ & $\theta^{6}$& $1$& $1$& $1$& $\theta^{6}$ & $\theta^{6}$ & $\theta^{6}$ & $1$ & $1$ & $1$ & $1$  \\
\bottomrule
\end{tabular}
\end{center}
 \caption{\small  The $Z_{12}^0\times Z_2^0$ symmetry realization of the
 one zero singular texture at $(1,2)$-entry corresponding upon rotation to singular vanishing subtrace $C_{33}$. The  $D_{L1}$ indicates the left-handed  lepton doublet first
 family and so on. The $\chi_{k}$ denotes a scalar singlet which produces
 an entry in the right-handed Majorana mass matrix when acquiring a VEV
 at the see-saw scale.
 $\theta$ denotes $e^{{i\,\pi/6}}$. }
\label{33singular}
 \end{table}

By forming bilinear terms of the fields we see that the above transformations force a neutrino mass matrix with zero texture at $(1,2)$ entry.
Actually, we get:
\be
M_{R0} = \left(
\begin {array}{ccc}
Y^1_{0\chi 11}\; v_{0\chi_1} &0& 0\\
0& Y^2_{0\chi 22}\; v_{0\chi_2}& 0\\
0&0&Y^3_{0\chi 33}\; v_{0\chi_3}
\end {array}
\right) ,\;\;\;
M_{D0} = \left(
\begin {array}{ccc}
Y^4_{0D 11}\; v_{0\Phi_4} &0& 0\\
0& 0& Y^5_{0D 23}\; v_{0\Phi_5}\\
Y^6_{0D 31}\; v_{0\Phi_6}&0&Y^7_{0D 33}\; v_{0\Phi_7}
\end {array}
\right).
\label{M_RD0singular33}
\ee
We see that $M_{D0}$ is singular, and $M_{\n0} = M_{D0}\; M_{R0}^{-1}\;M_{D0}^T$ is singular with the desired form of vanishing element at the ($1,2$)$^{\mbox{th}}$ entry. We can check that $M_{\ell 0}$ is of generic form as the one presented in Eq.(\ref{M_ell0}).

Again, in order to find the field transformations corresponding to the new `rotated' symmetry defined by $S$ (Eq. \ref{SC33}), we apply the  rule in Eq. (\ref{rule}) for all the fields $f$
and extending in the case of the 7-dimensional $\Phi$ the matrix $S$ to be $S_{ex}=\mbox{ {diag}} \left( S, 1_{4\times 4}\right)$, in such a way that we do not get generally diagonal matrices $T^Z_f$ because of the rotation $S$.  As in nonsingular cases, one can write down constraints involving the Yukawa couplings and vevs (now without the subscript $0$), and by solving them we get:
\be
\begin{array}{lll}
M_{R}& = &\left(
\begin {array}{ccc}
-Y^1_{\chi 22}\; v_{\chi_1}- Y^2_{\chi 22}\; v_{\chi_2}  &-Y^2_{\chi 22}\; v_{\chi_1}+ Y^1_{\chi 22}\; v_{\chi_2}& 0\\
 -Y^2_{\chi 22}\; v_{\chi_1}+ Y^1_{\chi 22}\;v_{\chi_2}& Y^1_{\chi 22}\; v_{\chi_1}+ Y^2_{\chi 22}\; v_{\chi_2} &0\\
0&0&Y^3_{\chi 33}\; v_{\chi_3}
\end {array}
\right) ,\\
M_{D} &=&\left(
\begin {array}{ccc}
-i\,Y^4_{D 12}\; v_{\Phi_4} &Y^4_{D 12}\; v_{\Phi_4}& -i\, Y^5_{D 23}\; v_{\Phi_5} \\
-Y^4_{D 12}\; v_{\Phi_4}& -i\, Y^4_{D 12}\; v_{\Phi_4}& Y^5_{D 23}\; v_{\Phi_5} \\
-i\, Y^6_{D 32}\; v_{\Phi_6}&Y^6_{D 32}\; v_{\Phi_6}&Y^7_{D 33}\; v_{\Phi_7}
\end {array}
\right).
\end{array}
\label{M_RDsingular33}
\ee
One can check that the resulting $M_\n$ is singular and satisfies the texture $\mathbf C_{33}$.

As to $M_\ell$, we get a generic mass matrix:
\be
M_{\ell } =  \left(
\begin {array}{ccc}
Y^2_{\ell 21}\;  v_{\Phi_1}+ Y^2_{\ell 11}\; v_{\Phi_2} &Y^2_{\ell 22}\; v_{\Phi_1} + Y^2_{\ell 12}\; v_{\Phi_2} & Y^2_{\ell 23}\; v_{\Phi_1} + Y^2_{\ell 13} \;v_{\Phi_2} \\
-Y^2_{\ell 11}\;  v_{\Phi_1} + Y^2_{\ell 21}\; v_{\Phi_2} &- Y^2_{\ell 12}\; v_{\Phi_1} + Y^2_{\ell 22}\; v_{\Phi_2} & - Y^2_{\ell 13}\; v_{\Phi_1} + Y^2_{\ell 33}\; v_{\Phi_2}\\
Y^3_{\ell 31} \; v_{\Phi_3} & Y^3_{\ell 32}\; v_{\Phi_3} &Y^3_{\ell 33}\; v_{\Phi_3}
\end {array}
\right).
\label{M_ellgeneralsingular33}
\ee
If we assume the related vevs are comparable $v_{\Phi_1}\approx v_{\Phi_2} \approx v_{\Phi_3} \approx v$ then we get
\bea
\label{M_ell-singular33}
M_{\ell } \approx v \left(
\begin {array}{ccc}
Y^2_{\ell 21} + Y^1_{\ell 11} &Y^2_{\ell 22} + Y^2_{\ell 12}&Y^2_{\ell 23} + Y^2_{\ell 13} \\
-Y^2_{\ell 11} + Y^2_{\ell 21} &-Y^2_{\ell 12} + Y^2_{\ell 22}&-Y^2_{\ell 13} + Y^2_{\ell 23}\\
Y^3_{\ell 31} &Y^3_{\ell 32} &Y^3_{\ell 33}
\end {array}
\right) &=& v \left( \begin {array}{c}
{\bf a}^T\\{\bf b}^T\\{\bf c}^T
\end {array}
\right),
\eea
whereas if we assume $v\approx v_{\Phi_1}\approx v_{\Phi_3} \gg v_{\Phi_2}$ we get
\bea
\label{M_ell-singular-second33}
M_{\ell } \approx v \left(
\begin {array}{ccc}
Y^2_{\ell 21}  &Y^2_{\ell 22} &Y^2_{\ell 23} \\
-Y^2_{\ell 11} &-Y^2_{\ell 12}&-Y^2_{\ell 13} \\
Y^3_{\ell 31} &Y^3_{\ell 32} &Y^3_{\ell 33}
\end {array}
\right) &=& v \left( \begin {array}{c}
{\bf a}^T\\{\bf b}^T\\{\bf c}^T
\end {array}
\right)
\eea
In both cases, one can naturally diagonalize $M_\ell$ by an infinitesimal rotation, which means that we are to a good approximation in the flavor basis.

\subsection{Indirect realization of $\mathbf C_{11}$ (Type I singular): Vanishing of $M_{\n\,22} + M_{\n\, 33}$}
We move from  zero texture at $(2,3)$ to the texture $\mathbf C_{11}$  by $S$ of Eq.(\ref{SC11}).

The symmetry transformations for the fields which imposes a zero texture $M_{\n 0}$ with generic $M_{\ell 0}$ and singular $M_{D0}$ are given
 in Table~(\ref{11singular}).
\begin{table}[hbtp]
\begin{center}
\begin{tabular}{ccccccccccccccccc}
\toprule
\multicolumn{17}{c}{$ \mbox{Symmetry under}\; Z_{12}^0\; \mbox{factor}$}\\
\midrule
    $\Phi_1$  & $\Phi_2$ & $\Phi_3$ & $\Phi_4$ & $\Phi_5$ & $\Phi_6$ & $\Phi_7$ & $D_{L1} $ &$D_{L2} $ & $D_{L3} $ &$\n_{R1}$ &$\n_{R2}$& $\n_{R3}$ & $\chi_{1}$ &
   $\chi_{2}$ & $\chi_{3}$ & $\ell_R$    \\
\midrule
 $\theta^{11}$ & $\theta^{9}$ & $\theta^{4}$ & $\theta^{2}$ & $\theta^{6}$ & $\theta^{4}$ & $\theta$ & $\theta^{11}$ & $\theta^{9}$ & $\theta^{4}$ & $\theta$ &  $\theta^{2}$ & $\theta^{5}$ & $\theta^{10}$ & $\theta^{8}$ & $\theta^{2}$ & $1$\\
\toprule
\multicolumn{17}{c}{$ \mbox{Symmetry under}\; Z_2^0\; \mbox{factor}$}\\
\midrule
$1$& $1$& $1$ & $\theta^{6}$& $\theta^{6}$ & $\theta^{6}$ & $\theta^{6}$& $1$& $1$& $1$& $\theta^{6}$ & $\theta^{6}$ & $\theta^{6}$ & $1$ & $1$ & $1$ & $1$  \\
\bottomrule
\end{tabular}
\end{center}
 \caption{\small  The $Z_{12}^0\times Z_2^0$ symmetry realization of the
 one zero singular texture at $(2,3)$-entry corresponding upon rotation to singular vanishing subtrace $\mathbf C_{11}$. The  $D_{L1}$ indicates the left-handed lepton doublet first
 family and so on. The $\chi_{k}$ denotes a scalar singlet which produces
 an entry in the right-handed Majorana mass matrix when acquiring a VEV
 at the see-saw scale.
 $\theta$ denotes $e^{{i\,\pi/6}}$. }
\label{11singular}
 \end{table}
By forming bilinear terms of the fields we see that the above transformations force a neutrino mass matrix with zero texture at $(2,3)$ entry. Again we define the new transformations for the fields corresponding to the new symmetry imposing the vanishing subtrace by the  rule in Eq.(\ref{rule}) with $S$ given by Eq. (\ref{SC11}) or its extension $S_{ex}$ to the $7$-dim space of $\Phi$'s. The `rotated' symmetry imposes some constraints on the Yukawa couplings and the vevs, which when solved give the following results,
\be
\begin{array}{lll}
M_{R} &=& \left(
\begin {array}{ccc}
Y^1_{\chi 11} \;v_{\chi_1} &0& 0\\
 0& -Y^3_{\chi 23}\; v_{\chi_1}+ Y^2_{\chi 23}\; v_{\chi_3} & Y^2_{\chi 23}\; v_{\chi_2}+ Y^3_{\chi 23}\; v_{\chi_3}\\
0& Y^2_{\chi 23}\; v_{\chi_1}+ Y^3_{\chi 23}\; v_{\chi_3} &Y^3_{\chi 21}\; v_{\chi_2} - Y^2_{\chi 23}\; v_{\chi_3}
\end {array}
\right) , \\\\
M_{D} &=& \left(
\begin {array}{ccc}
Y^4_{D 11}\; v_{\Phi_4} &i\, Y^5_{D 13}\; v_{\Phi_5}& Y^5_{D 13}\; v_{\Phi_5} \\
i\, Y^6_{D 31}\; v_{\Phi_6}&  Y^7_{D 33}\; v_{\Phi_7}& -i\, Y^7_{D 33}\; v_{\Phi_7} \\
Y^6_{D 31}\; v_{\Phi_6}&  i\, Y^7_{D 33}\;v_{\Phi_7}& Y^7_{D 33} \;v_{\Phi_7}
\end {array}
\right).
\end{array}
\label{M_RDsingular11}
\ee
One can check that $\mbox{det}(M_D)=0$, and that the resulting $M_\n$ is singular and satisfies the texture $\mathbf C_{11}$.

As to $M_\ell$, we get a generic mass matrix:
\be
M_{\ell } =  \left(
\begin {array}{ccc}
 Y^1_{\ell 11}\;  v_{\Phi_1} & Y^1_{\ell 12} \;v_{\Phi_1} &Y^1_{\ell 13} \;v_{\Phi_1} \\
Y^3_{\ell 31} \; v_{\Phi_2} + Y^3_{\ell 21} \;v_{\Phi_3} & Y^3_{\ell 32} \;v_{\Phi_2} + Y^3_{\ell 22}\; v_{\Phi_3} &  Y^3_{\ell 33}\; v_{\Phi_2} + Y^3_{\ell 23}\; v_{\Phi_3}\\
  - Y^3_{\ell 21} \; v_{\Phi_2}+ Y^3_{\ell 31}\; v_{\Phi_3} &-Y^3_{\ell 22}\;v_{\Phi_2} + Y^3_{\ell 32}\; v_{\Phi_3} & -Y^3_{\ell 23}\; v_{\Phi_2} + Y^3_{\ell 33}\; v_{\Phi_3}
\end {array}
\right).
\label{M_ellgeneralsingular11}
\ee
Assuming the related vevs are comparable $v_{\Phi_1}\approx v_{\Phi_2} \approx v_{\Phi_3} \approx v$ then we get
\be
\label{M_ell-singular11}
M_{\ell } \approx v \left(
\begin {array}{ccc}
Y^1_{\ell 11}  & Y^1_{\ell 12} &Y^1_{\ell 13}  \\
Y^3_{\ell 31}  + Y^3_{\ell 21}  & Y^3_{\ell 32} + Y^3_{\ell 22} &  Y^3_{\ell 33} + Y^3_{\ell 23}\\
  - Y^3_{\ell 21}  + Y^3_{\ell 31}  &-Y^3_{\ell 22}  + Y^3_{\ell 32} & -Y^3_{\ell 23}  + Y^3_{\ell 33}
\end {array}
\right) = v \left( \begin {array}{c}
{\bf a}^T\\{\bf b}^T\\{\bf c}^T
\end {array}
\right),
\ee
whereas if we assume $v\approx v_{\Phi_1}\approx v_{\Phi_3} \gg v_{\Phi_2}$ we get
\be
\label{M_ell-singular-second11}
M_{\ell } \approx v \left(
\begin {array}{ccc}
Y^1_{\ell 11}  &Y^1_{\ell 12} &Y^1_{\ell 13} \\
Y^3_{\ell 21} &Y^3_{\ell 22}&Y^3_{\ell 23} \\
Y^3_{\ell 31} &Y^3_{\ell 32} &Y^3_{\ell 33}
\end {array}
\right) = v \left( \begin {array}{c}
{\bf a}^T\\{\bf b}^T\\{\bf c}^T
\end {array}
\right)
\ee
Again, in both cases, one can naturally diagonalize $M_\ell$ by an infinitesimal rotation, which means that we are to a good approximation in the flavor basis.

\subsection{Indirect realization of $\mathbf C_{22}$ (Type I singular): Vanishing of $M_{\n\,11} + M_{\n\, 33}$}
We move from  zero texture at $(1,3)$ to the texture $\mathbf C_{22}$  by $S$ of Eq.(\ref{SC22}).
The symmetry transformations for the fields which imposes a zero texture $M_{\n 0}$ at entry ($1,3$), with generic $M_{\ell 0}$ and singular $M_{D0}$ are given in Table~(\ref{22singular}).
\begin{table}[hbtp]
\begin{center}
\begin{tabular}{ccccccccccccccccc}
\toprule
\multicolumn{17}{c}{$ \mbox{Symmetry under}\; Z_{12}^0\; \mbox{factor}$}\\
\midrule
    $\Phi_1$  & $\Phi_2$ & $\Phi_3$ & $\Phi_4$ & $\Phi_5$ & $\Phi_6$ & $\Phi_7$ & $D_{L1} $ &$D_{L2} $ & $D_{L3} $ &$\n_{R1}$ &$\n_{R2}$& $\n_{R3}$ & $\chi_{1}$ &
   $\chi_{2}$ & $\chi_{3}$ & $\ell_R$    \\
\midrule
 $\theta^{11}$ & $\theta^{9}$ & $\theta^{4}$ & $\theta^{2}$ & $\theta^{4}$ & $\theta^{8}$ & $\theta$ & $\theta^{11}$ & $\theta^{9}$ & $\theta^{4}$ & $\theta$ &  $\theta^{2}$ & $\theta^{5}$ & $\theta^{10}$ & $\theta^{8}$ & $\theta^{2}$ & $1$
\\
\midrule
\multicolumn{17}{c}{$ \mbox{Symmetry under}\; Z_2^0\; \mbox{factor}$}\\
\toprule
$1$& $1$& $1$ & $\theta^{6}$& $\theta^{6}$ & $\theta^{6}$ & $\theta^{6}$& $1$& $1$& $1$& $\theta^{6}$ & $\theta^{6}$ & $\theta^{6}$ & $1$ & $1$ & $1$ & $1$  \\
\bottomrule
\end{tabular}
\end{center}
 \caption{\small  The $Z_{12}^0\times Z_2^0$ symmetry realization of the
 one zero singular texture at $(1,3)$-entry corresponding upon rotation to singular vanishing subtrace $\mathbf C_{22}$. The  $D_{L1}$ indicates the left-handed lepton doublet first
 family and so on. The $\chi_{k}$ denotes a scalar singlet which produces
 an entry in the right-handed  Majorana mass matrix when acquiring a VEV
 at the see-saw scale.
 $\theta$ denotes $e^{{i\,\pi/6}}$. }
\label{22singular}
 \end{table}

  Once more, we define the new transformations for the fields corresponding to the new symmetry imposing the vanishing subtrace by applying the rule in
Eq.(\ref{rule}) with $S$ given by Eq.(\ref{SC22}) or its extension $S_{ex}$ to the $7$-dim space of $\Phi$'s. The `rotated' symmetry imposes some constraints on the Yukawa couplings and the vevs, which when solved give the following results,
\be
\label{M_RDsingular22}
\begin{array}{lll}
M_{R} &=&  \left(
\begin {array}{ccc}
-Y^3_{\chi 13}\; v_{\chi_1} + Y^3_{\chi 11}\; v_{\chi_3} &0& Y^3_{\chi 11}\;v_{\chi_1}+ Y^3_{\chi 13}\;v_{\chi_3}\\
 0& Y^2_{\chi 22}\;v_{\chi_2} & 0\\
Y^3_{\chi 11}\;v_{\chi_1}+Y^3_{\chi 13}\; v_{\chi_3} & 0 &Y^3_{\chi 13}\; v_{\chi_1} - Y^3_{\chi 11}\;v_{\chi_3}
\end {array}
\right) , \\\\
M_{D} &=&  \left(
\begin {array}{ccc}
Y^7_{D 33}\; v_{\Phi_7} + Y^4_{D 33}\; v_{\Phi_4} &0& -i\,Y^7_{D 33}\; v_{\Phi_7}+i\,Y^4_{D 33}\; v_{\Phi_4} \\
-i\, Y^5_{D 23}\;v_{\Phi_5} + i \, Y^6_{D 23}\; v_{\Phi_6} &  0& Y^6_{D 23}\; v_{\Phi_6} + Y^5_{D 23}\; v_{\Phi_5} \\
i\,Y^7_{D 33}\; v_{\Phi_7} - i \, Y^4_{D 33}\; v_{\Phi_4} &  0& Y^7_{D 33}\; v_{\Phi_7} + Y^4_{D 33}\; v_{\Phi_4}
\end {array}
\right).
\end{array}
\ee
One can check that $\mbox{det}(M_D)=0$, and that the resulting $M_\n$ is singular and satisfies the texture $\mathbf C_{22}$.

As to $M_\ell$, we get a generic mass matrix:
\be
\label{M_ellgeneralsingular22}
M_{\ell } =\left(
\begin {array}{ccc}
 Y^3_{\ell 31}\;  v_{\Phi_1}+  Y^3_{\ell 11} \; v_{\Phi_3} & Y^3_{\ell 32} \;v_{\Phi_1} + Y^3_{\ell 12} \; v_{\Phi_3} &Y^3_{\ell 33} \;v_{\Phi_1} + Y^3_{\ell 13}\;  v_{\Phi_3} \\
Y^2_{\ell 21} \; v_{\Phi_2}  & Y^2_{\ell 22} \;v_{\Phi_2}  &  Y^2_{\ell 23}\;v_{\Phi_2} \\
  - Y^3_{\ell 11} \; v_{\Phi_1}+ Y^3_{\ell 31} \;v_{\Phi_3} &-Y^3_{\ell 12}\; v_{\Phi_1} + Y^3_{\ell 32}\; v_{\Phi_3} & -Y^3_{\ell 13}\; v_{\Phi_1} + Y^3_{\ell 33}\; v_{\Phi_3}
\end {array}
\right)
\ee
When $v_{\Phi_1}\approx v_{\Phi_2} \approx v_{\Phi_3} \approx v$ then we get
\be
\label{M_ell-singular22}
M_{\ell } \approx v \left(
\begin {array}{ccc}
Y^3_{\ell 31}  +  Y^3_{\ell 11}  & Y^3_{\ell 32} + Y^3_{\ell 12}   &Y^3_{\ell 33} + Y^3_{\ell 13}   \\
Y^2_{\ell 21}    & Y^2_{\ell 22}   &  Y^2_{\ell 23}  \\
  - Y^3_{\ell 11}  + Y^3_{\ell 31}  &-Y^3_{\ell 12}  + Y^3_{\ell 32}  & -Y^3_{\ell 13}  + Y^3_{\ell 33}
\end {array}
\right) = v \left( \begin {array}{c}
{\bf a}^T\\{\bf b}^T\\{\bf c}^T
\end {array}
\right),
\ee
whereas when $v\approx v_{\Phi_3}\approx v_{\Phi_2} \gg v_{\Phi_1}$ we get
\be
\label{M_ell-singular-second22}
M_{\ell } \approx v \left(
\begin {array}{ccc}
Y^3_{\ell 11}  &Y^3_{\ell 12} &Y^3_{\ell 13} \\
Y^2_{\ell 21} &Y^2_{\ell 22}&Y^2_{\ell 23} \\
Y^3_{\ell 31} &Y^3_{\ell 32} &Y^3_{\ell 33}
\end {array}
\right) = v \left( \begin {array}{c}
{\bf a}^T\\{\bf b}^T\\{\bf c}^T
\end {array}
\right).
\ee
In both cases, one can naturally diagonalize $M_\ell$ by an infinitesimal rotation, which means that we are approximately in the flavor basis.

\subsection{Indirect realization of $\mathbf C_{31}$ (Type I singular): Vanishing of $M_{\n\,12} + M_{\n\, 23}$}
We move from  zero texture at $(2,3)$ to the texture $\mathbf C_{31}$  by $S$ of Eq.(\ref{SC31}).
The symmetry transformations for the fields which imposes a zero texture at entry $(2,3)$ of $M_{\n 0}$ with generic $M_{\ell}$ and singular $M_{D0}$ are given in Table~(\ref{31singular}).
\begin{table}[hbtp]
\begin{center}
\begin{tabular}{ccccccccccccccccc}
\toprule
\multicolumn{17}{c}{$ \mbox{Symmetry under}\; Z_{12}^0\; \mbox{factor}$}\\
\midrule
    $\Phi_1$  & $\Phi_2$ & $\Phi_3$ & $\Phi_4$ & $\Phi_5$ & $\Phi_6$ & $\Phi_7$ & $D_{L1} $ &$D_{L2} $ & $D_{L3} $ &$\n_{R1}$ &$\n_{R2}$& $\n_{R3}$ & $\chi_{1}$ &
   $\chi_{2}$ & $\chi_{3}$ & $\ell_R$    \\
\midrule
 $\theta^{11}$ & $\theta^{9}$ & $\theta^{4}$ & $\theta^{2}$ & $\theta^{6}$ & $\theta^{4}$ & $\theta$ & $\theta^{11}$ & $\theta^{9}$ & $\theta^{4}$ & $\theta$ &  $\theta^{2}$ & $\theta^{5}$ & $\theta^{10}$ & $\theta^{8}$ & $\theta^{2}$ & $1$
\\
\bottomrule
\multicolumn{17}{c}{$ \mbox{Symmetry under}\; Z_2^0\; \mbox{factor}$}\\
\midrule
$1$& $1$& $1$ & $\theta^{6}$& $\theta^{6}$ & $\theta^{6}$ & $\theta^{6}$& $1$& $1$& $1$& $\theta^{6}$ & $\theta^{6}$ & $\theta^{6}$ & $1$ & $1$ & $1$ & $1$  \\
\bottomrule
\end{tabular}
\end{center}
 \caption{\small  The $Z_{12}^0\times Z_2^0$ symmetry realization of the
 one zero singular texture at $(2,3)$-entry corresponding upon rotation to singular vanishing subtrace $\mathbf C_{31}$. The  $D_{L1}$ indicates the left-handed lepton doublet first
 family and so on. The $\chi_{k}$ denotes a scalar singlet which produces
 an entry in the right-handed Majorana mass matrix when acquiring a VEV
 at the see-saw scale.
 $\theta$ denotes $e^{{i\,\pi/6}}$. }
\label{31singular}
 \end{table}

  In order to define the new transformations for the fields corresponding to the new symmetry imposing the vanishing subtrace, we apply the  rule of Eq.(\ref{rule}) with $S$ given by Eq.(\ref{SC31}) or its extension $S_{ex}$ to the $7$-dim space of $\Phi$'s. Solving the constraints on the Yukawa couplings and the vevs resulting from the `rotated' symmetry, we get:
\be
\label{M_RDsingular31}
\begin{array}{lll}
M_{R} &=&  \left(
\begin {array}{ccc}
Y^1_{\chi 11}\; v_{\chi_1} + Y^1_{\chi 13}\; v_{\chi_3} &0& Y^1_{\chi 13}\; v_{\chi_1}+ Y^1_{\chi 11} \; v_{\chi_3}\\
 0& Y^2_{\chi 22}\; v_{\chi_2} & 0\\
Y^1_{\chi 13}\; v_{\chi_1}+Y^1_{\chi 11}\; v_{\chi_3} & 0 &Y^1_{\chi 11}\; v_{\chi_1} + Y^1_{\chi 13}\; v_{\chi_3}
\end {array}
\right) , \\\\
M_{D} &=&  \left(
\begin {array}{ccc}
Y^4_{D 33}\; v_{\Phi_4} -   Y^5_{D 31}\; v_{\Phi_5} +Y^7_{D 33} \; v_{\Phi_7} &0& -Y^4_{D 33}\; v_{\Phi_4} -  Y^5_{D 31}\; v_{\Phi_5} + Y^7_{D 33} \; v_{\Phi_7} \\
- Y^6_{D 23}\; v_{\Phi_6} &  0& Y^6_{D 23}\;  v_{\Phi_6} \\
- Y^4_{D 33} \; v_{\Phi_4} +  Y^5_{D 31}\; v_{\Phi_5} + Y^7_{D 33}\; v_{\Phi_7} &  0& Y^4_{D 33}\;v_{\Phi_4} + Y^5_{D 31}\; v_{\Phi_5} +  Y^7_{D 33} \; v_{\Phi_7}
\end {array}
\right).
\end{array}
\ee
One can check that $\mbox{det}(M_D)=0$, and that the resulting $M_\n$ is singular and satisfies the texture $\mathbf C_{31}$.

As to $M_\ell$, we get a generic mass matrix:
\be
\label{M_ellgeneralsingular31}
M_{\ell } =  \left(
\begin {array}{ccc}
 Y^3_{\ell 31} \; v_{\Phi_1}+  Y^3_{\ell 11}\;  v_{\Phi_3} & Y^3_{\ell 32}\; v_{\Phi_1} + Y^3_{\ell 12}\;  v_{\Phi_3} &Y^3_{\ell 33}\; v_{\Phi_1} + Y^3_{\ell 13}\;  v_{\Phi_3} \\
Y^2_{\ell 21}\;  v_{\Phi_2}  & Y^2_{\ell 22}\; v_{\Phi_2}  &  Y^2_{\ell 23}\; v_{\Phi_2} \\
  Y^3_{\ell 11} \; v_{\Phi_1}+ Y^3_{\ell 31}\; v_{\Phi_3} &Y^3_{\ell 12}\; v_{\Phi_1} + Y^3_{\ell 32}\; v_{\Phi_3} & Y^3_{\ell 13}\; v_{\Phi_1} + Y^3_{\ell 33}\; v_{\Phi_3}
\end {array}
\right).
\ee
When $v \approx v_{\Phi_1}\approx v_{\Phi_2} \gg v_{\Phi_3}$ then we get
\be
\label{M_ell-singular31}
M_{\ell } \approx v \left(
\begin {array}{ccc}
Y^3_{\ell 31}    & Y^3_{\ell 32} &Y^3_{\ell 33}   \\
Y^2_{\ell 21}    & Y^2_{\ell 22}   &  Y^2_{\ell 23}  \\
   Y^3_{\ell 11}   &Y^3_{\ell 12}   & Y^3_{\ell 13}
\end {array}
\right) = v \left( \begin {array}{c}
{\bf a}^T\\{\bf b}^T\\{\bf c}^T
\end {array}
\right),
\ee
whereas if we assume $v\approx v_{\Phi_3}\approx v_{\Phi_2} \gg v_{\Phi_1}$ we get
\be
\label{M_ell-singular-second31}
M_{\ell } \approx v \left(
\begin {array}{ccc}
Y^3_{\ell 11}  &Y^3_{\ell 12} &Y^3_{\ell 13} \\
Y^2_{\ell 21} &Y^2_{\ell 22}&Y^2_{\ell 23} \\
Y^3_{\ell 31} &Y^3_{\ell 32} &Y^3_{\ell 33}
\end {array}
\right) = v \left( \begin {array}{c}
{\bf a}^T\\{\bf b}^T\\{\bf c}^T
\end {array}
\right).
\ee
In both cases, one can naturally diagonalize $M_\ell$ by an infinitesimal rotation, which means that we are in the flavor basis approximately.

\section{Indirect realization of type II seesaw with $Z_5$ symmetry}
To fix the ideas, we treat here in some details the case of $\mathbf C_{33}$ vanishing subtrace which can be related to zero texture, noting that the procedure can be generalized to all other textures ($\mathbf C_{11},\mathbf C_{22}$ and $\mathbf C_{31}$) that also can be related to zero textures. We follow the same `Rotating' strategy outlined in section (\ref{strategysubsection}).

As we saw in Eq.(\ref{SC33}), the matrix $S$ allows to move from one zero texture at ($1,2$)$^{\mbox{\tiny{th}}}$ entry to vanishing subtrace texture $\mathbf C_{33}$. Again, we use a subscript (or superscript) $0$ to denote the gauge basis satisfying the `unrotated' symmetry $Z^0_5$, whereas we drop this subscript (superscript) for the `rotated' $Z_5$.

\subsection{Matter Content}

Following the conventions of \cite{0texture}, we extend the SM extended by
introducing several $SU(2)_L$  scalar triplets $H_a$, $(a=1,2,\cdots N)$,
\be
H_a\equiv \left[H_a^{++}, H_a^+, H_a^0\right].
\ee
The gauge invariant Yukawa interaction relevant for neutrino mass takes the form,
\be \label{type2Lagrangian}
\mathcal{L}_{H,L} = \sum_{i,j=1}^3\sum_{a=1}^N\,Y_ {ij}^{\n a}\,\left[ H_a^0 \n_{Li}^T\, \mathcal{C}^{-1}\, \n_{Lj} + H_a^+ \left(\n_{Li}^T\, \mathcal{C}^{-1}\, \ell_{Lj}
+ \ell_{Lj}^T \,\mathcal{C}^{-1}\, \n_{Li}\right) + H_a^{++} \ell_{Li}^T \,\mathcal{C}^{-1}\, \ell_{Lj}\right],
\ee
where $Y_{ij}^a$ are the corresponding Yukawa coupling constants, the indices $i,j$ are flavor ones, and $\mathcal{C}$ is the charge conjugation matrix.

The field $H_a^0$ could acquire a small vev, $\langle H_a^0\rangle = v^H_a$ that gives rise to a Majorana neutrino mass
matrix of the following form,
\be
M_{\n\;ij} =  \sum_{a=1}^N\,Y_{ij}^{\n a}\,v^H_a.
\ee
The smallness of the vev $v^H_a$ is attributed to the largeness of the triplet scalar mass scale\cite{seesaw2}.

As to the charged lepton mass, we introduce, in contrast to \cite{0texture}, various Higgs doublets $\Phi_a$, $a=1, \dots, K$
 \be \label{type2LagrangianCharged}
\mathcal{L}_{\ell} = \sum_{i,j=1}^3\sum_{a=1}^K\,Y_ {ij}^{\ell a}\,   \overline{D}_{Li}\; \Phi_a \; \ell_{Rj}   ,
\ee
Note that we did not consider only one SM Higgs, otherwise we would have got, as in \cite{0texture}, a diagonal charged lepton mass matrix $M_{\ell 0}$ when the neutrino mass matrix $M_{\n 0}$ had a zero texture. We would like to get a generic $M_{\ell 0}$ corresponding to zero texture $M_{\n 0}$, so that when we `rotate' and get a vanishing subtrace texture for the neutrino mass matrix $M_\n$ we get also another generic charged lepton mass matrix $M_\ell$. This latter can under suitable assumptions be diagonalized by infinitesimal rotations. Had we restricted our SM Higgs to only one Higgs doublet, then the diagonal $M_{\ell 0}$ corresponding to zero texture $M_{\n 0}$ will give, upon `rotation' by $S$, a nondiagonal charged mass matrix $M_\ell$ that is diagonalizable by a finite rotation $S$, which means that the vanishing subtrace texture does not correspond to the flavor basis.

\subsection{$Z^0_5$ symmetry for zero texture $M_{\n 0}$ characterized by $M_{\n0\,12}=0$}
In order to impose a zero texture $M_{\n 0}$ by $Z^0_5$ symmetry with a generic $M_{\ell 0}$, we introduce four scalar triplets $H_a$ and three Higgs doublets $\Phi_b$, with the following assignments under $Z^0_5$ defined in
Table \ref{33type2}.
\begin{table}[hbtp]
\begin{center}
\begin{tabular}{ccccccccccc}
\toprule
\multicolumn{11}{c}{$ \mbox{Symmetry under}\; Z_5^0 $}\\
\midrule
    $H_1$  & $H_2$ & $H_3$ & $H_4$ & $1_F$ & $2_F$ & $3_F$ & $\ell_R$ &$ \Phi_1 $ & $\Phi_2 $ &$\Phi_3$   \\
\midrule
 $1$ & $\Omega^{3}$ & $\Omega^{2}$ & $\Omega$ & $1$ & $\Omega$ & $\Omega^{2}$ & $1$ & $1$ & $\Omega$ & $\Omega^2$
\\
\bottomrule
\end{tabular}
\end{center}
 \caption{\small  The $Z^0_5$ symmetry seesaw type II realization of the
 one zero texture at $(1,2)$-entry corresponding upon rotation to vanishing subtrace $\mathbf C_{33}$. $H_a$ are triplet scalars, whereas $1_F$ refers to the fermions, apart from the right-handed charged leptons $\ell_R$, in the first generation and so on. The $\Phi_{b}$ denote SM Higgs doublets.
 $\Omega$ denotes $e^{{i\,2\pi/5}}$. }
\label{33type2}
 \end{table}

By forming bilinear terms of  $H_a \n^T_{L} \n_{L}$ we can find out the invariant Lagrangian terms under $Z^0_5$, which gives
\be
M_{\n 0}=\left(
\begin {array}{ccc}
\times &0& \times\\
0 & \times & \times\\
\times & \times &\times
\end {array}
\right)
\ee

 \subsection{$Z_5$ symmetry for $\mathbf C_{33}$ texture ($M_{\n\,11} + M_{\n\, 22}=0$) and Yukawa couplings constraints}

In order to find the new `rotated' symmetry $Z_5$, we need first to find how all the fields would transform. Here, we carefully use the rule of  (Eq. \ref{conjugation} or \ref{star}) for all the fields $f$, in that if $f$ transforms under $Z^0_5$ according to the diagonal, by construction, matrix $T^{0Z}_f$, then it transforms under $Z_5$ according to $T^{Z}_f=S_{ex}^{f\dagger} \; T^{0Z}_f \; S_{ex}^f$  (cf.  Eq. \ref{rule}) with $S_{ex}^f=\mbox{{diag}} \left( S, 1_{r\times r}\right)$ possibly an extension of $S$ to match the finite-dimensional space of the field $f$ of dimensions ($3+r$). The invariance of the Lagrangian terms under the symmetry will impose constraints on the Yukawa couplings that one can in principle solve to give the form of the mass matrices when the Higgs/scalar fields get a vev.

Actually, one can check that under both $Z^0_5$ and $Z_5$, defined by the transformations $T^{0Z}_f$ and $T^Z_f$ respectively, we have the following constraints:
\be
 (Y^{\n b}_0) =  T^{0Z}_{H ab}\; (T^{0Z}_{\n_L})^T \; (Y^{\n a}_0)\; (T^{0Z}_{\n_L}) ,
 \label{type2TransformUnrotated}
\ee
and
\be
(Y^{\n b}) = T^{Z}_{H ab}\; (T^{Z}_{\n_L})^T \;(Y^{\n a})\;(T^{Z}_{\n_L}),
\label{type2TransformRotated}
 \ee
where ($a,b=1,\ldots,4$), $(Y^{\n b}_{0})$ is a matrix in flavor space with element $Y^{\n b}_{0\, ij}$ at its $(i,j)^{th}$ entry. The two constraints of Eqs. (\ref{type2TransformUnrotated}, \ref{type2TransformRotated}) are related in that if we know the solution to one constraint we know it for the other. More specifically, one can check that if $(Y^{\n b}_0)$ was a solution of Eq. \ref{type2TransformUnrotated}, then
\be
(Y^{\n e}) = (S^T)\;(Y^{\n b}_0)\; (S)\; (S_{ex}^{H})_{b\,e}.
 \label{YrelationY0}
\ee
is a solution of Eq.(\ref{type2TransformRotated}).

 \subsection{$M_{\n0}$ and $M_{\n}$ resulting respectively from $Z_5^0$ and $Z_5$ invariance }
 By solving  Eqs. (\ref{type2TransformUnrotated}, \ref{type2TransformRotated}) we get when $H^0_a$'s get the vevs $v^H_{0\,a}$ under $Z_5^0$:
\be
\begin{array}{lll}
M_{\n0} &=& \left(
\begin {array}{ccc}
Y^{\n1}_{0\, 11}\; v^H_{0\,1} &0& Y^{\n2}_{0\, 13}\; v^H_{0\,2}\\
\rule{0.7cm}{0.1mm}& Y^{\n2}_{0\, 22}\; v^H_{0\,2}& Y^{\n3}_{0\, 23}\; v^H_{0\,3}\\
\rule{0.7cm}{0.1mm}&\rule{0.7cm}{0.1mm}&Y^{\n4}_{0\, 33}\; v^H_{0\, 4}
\end {array}
\right), \\\\
M_{\n} &=& \left(
\begin {array}{ccc}
-Y^{\n2}_{23}\; v^H_{1} -Y^{\n2}_{22} \;v^H_{2} &-Y^{\n2}_{22}\; v^H_{1} +Y^{\n2}_{12}\; v^H_{2}& Y^{\n2}_{23}\; v^H_{1} -i \,Y^{\n2}_{23}\; v^H_{2} + i\, Y^{\n3}_{23}\; v^H_{3}\\
\rule{0.7cm}{0.1mm}& Y^{\n2}_{12}\; v^H_{1} +Y^{\n2}_{22}\; v^H_{2}& i\,Y^{\n2}_{23}\; v^H_{1} + Y^{\n2}_{23}\;v^H_{2} +  Y^{\n3}_{23} \;v^H_{3}\\
\rule{0.7cm}{0.1mm}&\rule{0.7cm}{0.1mm}&Y^{\n4}_{33}\; v^H_{4}
\end {array}
\right).
\end{array}
\ee
 We see that the texture $\mathbf C_{33}$ is met in $M_\n$ while $(M_{\n0\,12}=0)$ for $M_{\n0}$.
One can deduce the relations between the Higgs vevs in the ``unrotated" system ($v^H_{0a}$) and the Higgs vevs in the ``rotated" system ($v^H_a$) by writing Eq.(\ref{masstransformS}) and considering Eq.(\ref{YrelationY0}).

\subsection{$M_{\ell 0}$ and $M_{\ell}$   resulting respectively from $Z_5^0$ and $Z_5$ invariance}
The introduction of three SM Higgs $\Phi$'s was needed essentially to produce a generic charged lepton matrix. Actually the bilinear of the relevant  term $\overline{D}_{Li}\,  \ell_{Rj}$
transforms under $Z^0_5$ as,
 \be
\begin{array}{lll}
\overline{D}_{Li}\, \ell_{Rj} &\cong &
\left(
\begin {array}{ccc}
1&1 & 1\\
\Omega^4 & \Omega^4  & \Omega^4 \\
\Omega^3 &\Omega^3 &\Omega^3
\end {array}
\right).
\end{array}
\label{biseesaw2}
\ee
We see now that the transformations of $\Phi$'s in Table~(\ref{33type2}) were chosen exactly to make all the entries in $M_{\ell 0}$ eligible. Again expressing the invariance under $Z^{0}_5$ give constraints on the Yukawa couplings:
\be
(Y^{\ell b}_0) = T^{0Z}_{\Phi ab}\; (T^{0Z}_{D_L})^\dagger \;(Y^{\ell a}_0)\; (T^{0Z}_{\ell_R}),
\label{type2TransformUnrotatedCharged}
\ee
and
\be
 (Y^{\ell b}) = T^{Z}_{\Phi ab}\; (T^{Z}_{D_L})^\dagger \; (Y^{\ell a})\; (T^{Z}_{\ell_R}),
 \label{type2TransformRotatedCharged}
\ee
where ($a,b=1,\ldots,3$), $(Y^{\ell b}_{0})$ is a matrix in flavor space with element $Y^{\ell b}_{0\, ij}$ at its $(i,j)^{th}$ entry. The two constraints of Eqs. (\ref{type2TransformUnrotatedCharged}, \ref{type2TransformRotatedCharged}) are related in that if $(Y^{\ell b}_0)$ was
a solution of Eq.(\ref{type2TransformUnrotatedCharged}), then
\be
(Y^{\ell g}) = (S^\dagger)\;(Y^{\ell b}_0)\; (S)\; (S_{ex}^{\Phi})_{bg}
\label{YrelationY0Charged}
\ee
is a solution of Eq.(\ref{type2TransformRotatedCharged}) where $S^\Phi_{ex}=S$ since we have three $\Phi$'s.

Solving the Yukawa constraints in Eqs.(\ref{type2TransformUnrotatedCharged},\ref{type2TransformRotatedCharged})  we see that when the $\Phi_a$'s
 get vevs $v^\Phi_{0\,a}$ we get the following $M_{\ell 0}$ and $M_{\ell }$,
 \be
M_{\ell 0} = \left(
\begin {array}{ccc}
Y^{\ell 1}_{0\, 11}\; v^\Phi_{0\,1} &Y^{\ell 1}_{0\, 12}\; v^\Phi_{0\,1}& Y^{\ell 1}_{0\, 13} v^\Phi_{0\,1}\\
Y^{\ell 2}_{0\, 21}\; v^\Phi_{0\,2} &Y^{\ell 2}_{0\, 22}\; v^\Phi_{0\,2}& Y^{\ell 2}_{0\, 23}\; v^\Phi_{0\,2}\\
Y^{\ell 3}_{0\,31}\; v^\Phi_{0\,3} &Y^{\ell 3}_{0\; 32} v^\Phi_{0\,3}& Y^{\ell 3}_{0\, 33} v^\Phi_{0\,3}
\end {array}
\right),
\label{M_ell0new}
\ee
and
\be
M_{\ell} = \left(
\begin {array}{ccc}
Y^{\ell 2}_{21}\; v^\Phi_{1} +Y^{\ell 2}_{11}\; v^\Phi_{2} & Y^{\ell 2}_{22}\; v^\Phi_{1} +Y^{\ell 2}_{12}\; v^\Phi_{2}  & Y^{\ell 2}_{23}\;v^\Phi_{1} +Y^{\ell 2}_{13}\; v^\Phi_{2} \\
-Y^{\ell 2}_{11} \; v^\Phi_{1} +Y^{\ell 2}_{21}\; v^\Phi_{2} & - Y^{\ell 2}_{12}\; v^\Phi_{1} +Y^{\ell 2}_{22}\; v^\Phi_{2}  & - Y^{\ell 2}_{13}\; v^\Phi_{1} +Y^{\ell 2}_{23}\; v^\Phi_{2} \\
Y^{\ell 3}_{31}\; v^\Phi_{3}& Y^{\ell 3}_{31}\;v^\Phi_{3} &Y^{\ell 3}_{33}\;v^\Phi_{3}
\end {array}
\right).
\label{M_ell}
\ee
Again, one can deduce the relations between the ``unrotated'' vevs ($v^\Phi_{oa}$) and the ``rotated'' vevs ($v^\Phi_a$) by writing (cf. Eq. \ref{masstransformS}) $M_\ell = S^\dagger . M_{\ell o}. S$ and considering Eq. (\ref{YrelationY0Charged}).

Thus if we assume the related vevs are comparable $v_{\Phi_1}\approx v_{\Phi_2} \approx v_{\Phi_3} \approx v$ then we get
\bea
\label{M_elltype2}
M_{\ell } \approx v \left(
\begin {array}{ccc}
Y^{\ell 2}_{21}  +Y^{\ell 2}_{11}  & Y^{\ell 2}_{22}  +Y^{\ell 2}_{12}   & Y^{\ell 2}_{23}  +Y^{\ell 2}_{13}  \\
-Y^{\ell 2}_{11}  +Y^{\ell 2}_{21}  & - Y^{\ell 2}_{12}  +Y^{\ell 2}_{22}   & - Y^{\ell 2}_{13}  +Y^{\ell 2}_{23}  \\
Y^{\ell 3}_{31} & Y^{\ell 3}_{31}  &Y^{\ell 3}_{33}
\end {array}
\right) &=& v \left( \begin {array}{c}
{\bf a}^T\\{\bf b}^T\\{\bf c}^T
\end {array}
\right)
\eea
Consequently,
\bea
\label{M_ell2}
M_{\ell } M_{\ell}^\dagger &\approx& v^2 \left(\begin {array}{ccc}
{\bf a.a} &{\bf a.b}&{\bf a.c} \\
{\bf b.a} &{\bf b.b}&{\bf b.c}\\
{\bf c.a} &{\bf c.b}&{\bf c.c}
\end {array} \right)
\eea
so taking only the following natural assumption on the norms of the vectors
\bea \parallel {\bf a} \parallel /\parallel {\bf c} \parallel = m_e/m_\tau \sim 3 \times 10^{-4} &,&  \parallel {\bf b} \parallel /\parallel {\bf c} \parallel = m_\mu/m_\tau \sim 6 \times 10^{-2}\eea
one can diagonalize $M_{\ell } M_{\ell}^\dagger$ by an infinitesimal rotation as was done in \cite{0texture}, which proves that we are to a good approximation in the flavor basis.
\section{Direct realization of type I seesaw with $Z_6 \times Z_2$-symmetry}
We present now another method which leads directly to the vanishing subtrace texture without relating it to zero textures by `rotation'. It is applicable again only for the four textures ($\mathbf C_{33}, \mathbf C_{11}, \mathbf C_{22}$ and $\mathbf C_{31}$).
\subsection{Type I seesaw Direct realization of $\mathbf C_{33}$: Vanishing  of $M_{\n\,11} + M_{\n\, 22}$}
Within type I seesaw scenario,  the Lagrangian  responsible for mass is similar to the one given in Eq.(\ref{genlagm}) which, after conveniently simplifying the notations by dropping the Yukawa $0$-subscript and the summation signs, is rewritten here
\be
\begin{array}{lll}
\mathcal{L}_M &\supset & Y_{\chi ij}^{b}\; \chi_b\;  \n_{Ri}^T\; \mathcal{C}^{-1}\; \n_{Rj} + Y^{a}_{D ij}\; \overline{D}_{Li}\; \tilde{\Phi}_a\;  \n_{Rj}
+ Y_{\ell ij}^{\,a}\; \overline{D}_{Li}\; \Phi_a \; \ell_{Rj}.
\end{array}
\label{genlagmf}
\ee
 We have for the pattern $\mathbf C_{33}$  the relation  $M_{\n\,11} + M_{\n\, 22}=0$, which can give a hint motivating the search for solutions involving  a permutation symmetry ($1\leftrightarrow 2$). Actually, we can  think of the vanishing subtrace constraint as arising from symmetry considerations leading to textures implementing these ``permutation" restrictions at the level of $M_R$ and $M_D$, which by seesaw scenario resurface at the level of  $M_\n$ which inherits the ``permutation" structure. One can try simple forms for both $M_R$ and $M_D$ with enough number of parameters in order to produce generic $M_\nu$ having the sole constraint $M_{\n\,11} + M_{\n\, 22}=0$ . To be concrete, one can assume the following forms for $M_R$ and $M_D$  as shown below together with the derived $M_\nu$ (through seesaw mechanism),
\be
\begin{array}{lll}
 M_R =
\left(
\begin {array}{ccc}
x&y & 0\\
y& -x  & 0 \\
0 & 0 & z
\end {array}
\right), &  M_D =
\left(
\begin {array}{ccc}
A&-B & i\, C\\
B & A  & - C \\
-i\,D& D & E
\end {array}
\right),&
 M_\nu = M_D\, M_R^{-1}\, M_D^T = \left(
\begin {array}{ccc}
\Delta &\times  &\times\\
\rule{0.7cm}{0.1mm} & -\Delta  & \times\\
\rule{0.7cm}{0.1mm}& \rule{0.7cm}{0.1mm} & \times
\end {array}
\right),
\end{array}
\label{MnMDMR}
\ee
where $A$, $B$, $C$, $D$, $E$, $x$, $y$ and  $z$ are generic independent parameters,  the $\times$ and $\Delta$ signs denote generic independent nonvanishing entries.  We stress here that these forms proposed for $M_R$ and $M_D$ are not necessarily the simplest choice, but they are just mere possibilities that can be derived from symmetry considerations.

The fields and their assigned symmetry transformations under $Z_6 \times Z_2$ are presented in  Table (\ref{33type1norotation}).
\begin{table}[hbtp]
\begin{center}
\scalebox{1}{
\begin{tabular}{cccccc}
\toprule
\multicolumn{6}{c}{Matter Content and symmetry transformation  (Pattern $\mathbf C_{33}$) }\\
\midrule
\multicolumn{6}{c}{$ \mbox{Symmetry under}\; Z_6 $}\\
\midrule
$ \nu_{R1}\rightarrow \nu_{R1} $ & $ \nu_{R2}\rightarrow \nu_{R2} $ & $ \nu_{R3}\rightarrow \omega\,\nu_{R3}$ & $
\chi_1 \rightarrow  \omega^4\,\chi_1 $ & $  \chi_2 \rightarrow  \chi_2 $ & $ \chi_3 \rightarrow  \chi_3 $ \\
\midrule
 $D_{L1}\rightarrow D_{L1} $ & $  D_{L2}\rightarrow D_{L2} $ & $  D_{L3}\rightarrow \omega D_{L3} $&
 $ \Phi_1 \rightarrow \Phi_1 $ & $ \Phi_2 \rightarrow \omega\,\Phi_2 $ & $   \Phi_3 \rightarrow \omega^5\,\Phi_3$ \\
\midrule
 $ \Phi_4 \rightarrow \omega^3\,\Phi_4 $ & $  \Phi_5 \rightarrow \omega^3\,\Phi_5 $ &
 $  \ell_{R1} \rightarrow \omega^3\, \ell_{R1} $ & $  \ell_{R2} \rightarrow \omega^3\, \ell_{R2} $ & $  \ell_{R3} \rightarrow \omega^4\, \ell_{R3}$ &
\\
\toprule
\multicolumn{6}{c}{$ \mbox{Symmetry under}\; Z_2 $}\\
\midrule
$\nu_{R1}\rightarrow i\,\nu_{R2} $ & $ \nu_{R2}\rightarrow -i\,\nu_{R1} $ & $ \nu_{R3}\rightarrow \nu_{R3}$ & $
\chi_1 \rightarrow  \chi_1 $ & $  \chi_2 \rightarrow  \chi_2 $ & $ \chi_3 \rightarrow  \chi_3 $ \\
\midrule
$  D_{L1}\rightarrow i\, D_{L2} $ & $  D_{L2}\rightarrow -i\, D_{L1} $ & $  D_{L3}\rightarrow  D_{L3} $ &
 $ \Phi_1 \rightarrow \Phi_1 $ & $ \Phi_2 \rightarrow - \Phi_2 $ & $   \Phi_3 \rightarrow \Phi_3 $ \\
\midrule
 $  \Phi_4 \rightarrow i\,\Phi_5 $ & $  \Phi_5 \rightarrow -i\,\Phi_4 $ & $  \ell_{R1} \rightarrow  \ell_{R1} $ &
$  \ell_{R2} \rightarrow  \ell_{R2} $ & $  \ell_{R3} \rightarrow  \ell_{R3}$ &
\\
\bottomrule
\end{tabular}}
\end{center}
 \caption{\small  The $Z_6 \times Z_2$ symmetry seesaw type I realization of the
  vanishing subtrace $\mathbf C_{33}$. $\Phi$ are five SM Higgs doublets, $D_L$ refers to the flavor three left handed lepton doublets, while the three right-handed charged lepton singlets are denoted by  $\ell_R$. $\omega$ denotes $e^{{i\,\pi/ 3}}$}
\label{33type1norotation}
 \end{table}

Forming the required  bilinears dictated by $Z_6$ symmetry, we obtain
\be
\begin{array}{lll}
 \n^T_{Ri}\;\n_{Rj} \stackrel{Z_6}{\cong}
\left(
\begin {array}{ccc}
1&1 & \omega\\
1 & 1  & \omega \\
\omega & \omega & \omega^2
\end {array}
\right), &    \overline{D}_{Li}\;\n_{Rj} \stackrel{Z_6}{\cong}
\left(
\begin {array}{ccc}
1&1 & \omega\\
1 & 1  & \omega \\
\omega^5 & \omega^5 & 1
\end {array}
\right), &
  \overline{D}_{Li}\;\ell_{Rj} \stackrel{Z_6}{\cong}
\left(
\begin {array}{ccc}
-1&-1 & \omega^4\\
-1 & -1  & \omega^4 \\
\omega^2 & \omega^2 & -1
\end {array}
\right).
\end{array}
\label{bilinz6type1c33}
\ee
When the resulting  bilinears combine with the appropriate scalar fields, we get under $Z_6$, keeping only the combinations that produce singlets, the following:
\be
\begin{array}{lll}
 \chi_1\,\n^T_{Ri}\;\n_{Rj} \stackrel{Z_6}{\cong}
\left(
\begin {array}{ccc}
\omega^4&\omega^4 & \omega^5\\
\omega^4&\omega^4 & \omega^5 \\
\omega^5&\omega^5 & 1
\end {array}
\right), &    \chi_2\,\n^T_{Ri}\;\n_{Rj} \stackrel{Z_6}{\cong}
\left(
\begin {array}{ccc}
1&1 & \omega\\
1 & 1  & \omega \\
\omega & \omega & \omega^2
\end {array}
\right), &
\chi_3\,\n^T_{Ri}\;\n_{Rj} \stackrel{Z_6}{\cong}
\left(
\begin {array}{ccc}
1&1 & \omega\\
1 & 1  & \omega \\
\omega & \omega & \omega^2
\end {array}
\right),\\\\
 \tilde{\Phi}_1\,  \overline{D}_{Li}\;\n_{Rj} \stackrel{Z_6}{\cong}
\left(
\begin {array}{ccc}
1&1 & \omega\\
1 & 1  & \omega \\
\omega^5 & \omega^5 & 1
\end {array}
\right), &   \tilde{ \Phi}_2\,  \overline{D}_{Li}\;\n_{Rj} \stackrel{Z_6}{\cong}
\left(
\begin {array}{ccc}
\omega^5&\omega^5 & 1\\
\omega^5&\omega^5 & 1\\
\omega^4 & \omega^4 & \omega^5
\end {array}
\right), &
\tilde{\Phi}_3\,  \overline{D}_{Li}\;\n_{Rj} \stackrel{Z_6}{\cong}
\left(
\begin {array}{ccc}
\omega&\omega & \omega^2\\
\omega&\omega & \omega^2\\
1 & 1 & \omega^5
\end {array}
\right),\\\\
 \Phi_4\,  \overline{D}_{Li}\;\ell_{Rj} \stackrel{Z_6}{\cong}
\left(
\begin {array}{ccc}
1&1 & \omega\\
1 & 1  & \omega \\
\omega^5 & \omega^5 & 1
\end {array}
\right), &   \Phi_5\,  \overline{D}_{Li}\;\ell_{Rj} \stackrel{Z_6}{\cong}
\left(
\begin {array}{ccc}
1&1 & \omega\\
1 & 1  & \omega \\
\omega^5 & \omega^5 & 1
\end {array}
\right). &
\end{array}
\label{z6norot_trilin}
\ee
Thus the resulting Lagrangian dictated  by $Z_6$ symmetry takes the form,
\be
\small
\begin{array}{lll}
\mathcal{L}_M^{Z_6} &\propto&   Y^{1}_{\chi\, 33}\, \chi_1\,  \n_{R3}^T\,\mathcal{C}^{-1}\, \n_{R3} \\\\
&&+   Y^{2}_{\chi\, 11}\, \chi_2\,  \n_{R1}^T\,\mathcal{C}^{-1}\, \n_{R1} + Y^{2}_{\chi\, 12}\, \chi_2\,  \n_{R1}^T\,\mathcal{C}^{-1}\, \n_{R2} +  Y^{2}_{\chi\, 12}\, \chi_2\,  \n_{R2}^T\,\mathcal{C}^{-1}\, \n_{R1} +  Y^{2}_{\chi\, 22}\, \chi_2\,  \n_{R2}^T\,\mathcal{C}^{-1}\, \n_{R2}\\\\
&& + Y^{3}_{\chi\, 11}\, \chi_3\,  \n_{R1}^T\,\mathcal{C}^{-1}\, \n_{R1} + Y^{3}_{\chi\, 12}\, \chi_3\,  \n_{R1}^T\,\mathcal{C}^{-1}\, \n_{R2} +  Y^{3}_{\chi\, 12}\, \chi_3\,  \n_{R2}^T\,\mathcal{C}^{-1}\, \n_{R1} +  Y^{3}_{\chi\, 22}\, \chi_3\,  \n_{R2}^T\,\mathcal{C}^{-1}\, \n_{R2}\\\\
&& +  Y^{1}_{D\, 11}\, \overline{D}_{L1}\, \tilde{\Phi}_1\,  \n_{R1} +  Y^{1}_{D\, 12}\, \overline{D}_{L1}\, \tilde{\Phi}_1\,  \n_{R2} +
 Y^{1}_{D\, 21}\, \overline{D}_{L2}\, \tilde{\Phi}_1\,  \n_{R1} +  Y^{1}_{D\, 22}\, \overline{D}_{L2}\, \tilde{\Phi}_1\,  \n_{R2}  +  Y^{1}_{D\, 33}\, \overline{D}_{L3}\, \tilde{\Phi}_1\,  \n_{R3}\\\\
&& +  Y^{2}_{D\, 13}\, \overline{D}_{L1}\, \tilde{\Phi}_2\,  \n_{R3} +  Y^{2}_{D\, 23}\, \overline{D}_{L2}\, \tilde{\Phi}_2\,  \n_{R3} +  Y^{3}_{D\, 31}\, \overline{D}_{L3}\, \tilde{\Phi}_3\,  \n_{R1} +  Y^{3}_{D\, 32}\, \overline{D}_{L3}\, \tilde{\Phi}_3\,  \n_{R2} \\\\
&&  +  Y^{4}_{\ell\, 11}\, \overline{D}_{L1}\, \Phi_4\,  \ell_{R1} +  Y^{4}_{\ell\, 12}\, \overline{D}_{L1}\, \Phi_4\,  \ell_{R2} +
 Y^{4}_{\ell\, 21}\, \overline{D}_{L2}\, \Phi_4\,  \ell_{R1} +  Y^{4}_{\ell\, 22}\, \overline{D}_{L2}\, \Phi_4\,  \ell_{R2}
 +  Y^{4}_{\ell\, 33}\, \overline{D}_{L3}\, \Phi_4\,  \ell_{R3}\\\\
&&  +  Y^{5}_{\ell\, 11}\, \overline{D}_{L1}\, \Phi_5\,  \ell_{R1} +  Y^{5}_{\ell\, 12}\, \overline{D}_{L1}\, \Phi_5\,  \ell_{R2} +  Y^{5}_{\ell\, 21}\, \overline{D}_{L2}\, \Phi_5\,  \ell_{R1}+  Y^{5}_{\ell\, 22}\, \overline{D}_{L2}\, \Phi_5\,  \ell_{R2}
+  Y^{5}_{\ell\, 33}\, \overline{D}_{L3}\, \Phi_5\,  \ell_{R3},
\end{array}
\label{lagc33z6type1}
\ee
which transforms under $Z_2$ as
\be
\small
\begin{array}{lll}
\mathcal{L}_M^{Z_6} &\stackrel{Z_2}{\rightarrow}&   Y^{1}_{\chi\, 33}\, \chi_1\,  \n_{R3}^T\,\mathcal{C}^{-1}\, \n_{R3} \\\\
&&-  Y^{2}_{\chi\, 11}\, \chi_2\,  \n_{R2}^T\,\mathcal{C}^{-1}\, \n_{R2} + Y^{2}_{\chi\, 12}\, \chi_2\,  \n_{R2}^T\,\mathcal{C}^{-1}\, \n_{R1} +  Y^{2}_{\chi\, 12}\, \chi_2\,  \n_{R1}^T\,\mathcal{C}^{-1}\, \n_{R2} - Y^{2}_{\chi\, 22}\, \chi_2\,  \n_{R1}^T\,\mathcal{C}^{-1}\, \n_{R1}\\\\
&& - Y^{3}_{\chi\, 11}\, \chi_3\,  \n_{R2}^T\,\mathcal{C}^{-1}\, \n_{R2} + Y^{3}_{\chi\, 12}\, \chi_3\,  \n_{R2}^T\,\mathcal{C}^{-1}\, \n_{R1} +  Y^{3}_{\chi\, 12}\, \chi_3\,  \n_{R1}^T\,\mathcal{C}^{-1}\, \n_{R2} -  Y^{3}_{\chi\, 22}\, \chi_3\,  \n_{R1}^T\,\mathcal{C}^{-1}\, \n_{R1}\\\\
&& +  Y^{1}_{D\, 11}\, \overline{D}_{L1}\, \tilde{\Phi}_2\,  \n_{R2} - Y^{1}_{D\, 12}\, \overline{D}_{L2}\, \tilde{\Phi}_1\,  \n_{R1} -
 Y^{1}_{D\, 21}\, \overline{D}_{L1}\, \tilde{\Phi}_1\,  \n_{R2} +  Y^{1}_{D\, 22}\, \overline{D}_{L1}\, \tilde{\Phi}_1\,  \n_{R1}  +  Y^{1}_{D\, 33}\, \overline{D}_{L3}\, \tilde{\Phi}_1\,  \n_{R3}\\\\
&& + i\, Y^{2}_{D\, 13}\, \overline{D}_{L2}\, \tilde{\Phi}_2\,  \n_{R3} -i\, Y^{2}_{D\, 23}\, \overline{D}_{L1}\, \tilde{\Phi}_2\,  \n_{R3} +i\,  Y^{3}_{D\, 31}\, \overline{D}_{L3}\, \tilde{\Phi}_3\,  \n_{R2} -i\, Y^{3}_{D\, 32}\, \overline{D}_{L3}\, \tilde{\Phi}_3\,  \n_{R1} \\\\
&&  +  Y^{4}_{\ell\, 11}\, \overline{D}_{L2}\, \Phi_5\,  \ell_{R1} +  Y^{4}_{\ell\, 12}\, \overline{D}_{L2}\, \Phi_5\,  \ell_{R2} -
 Y^{4}_{\ell\, 21}\, \overline{D}_{L1}\, \Phi_5\,  \ell_{R1} -  Y^{4}_{\ell\, 22}\, \overline{D}_{L1}\, \Phi_5\,  \ell_{R2}
 + i\, Y^{4}_{\ell\, 33}\, \overline{D}_{L3}\, \Phi_5\,  \ell_{R3}\\\\
&&  -  Y^{5}_{\ell\, 11}\, \overline{D}_{L2}\, \Phi_4\,  \ell_{R1} -  Y^{5}_{\ell\, 12}\, \overline{D}_{L2}\, \Phi_4\,  \ell_{R2} +  Y^{5}_{\ell\, 21}\, \overline{D}_{L1}\, \Phi_4\,  \ell_{R1}+  Y^{5}_{\ell\, 22}\, \overline{D}_{L1}\, \Phi_4\,  \ell_{R2}
-i\, Y^{5}_{\ell\, 33}\, \overline{D}_{L3}\, \Phi_4\,  \ell_{R3}.
\end{array}
\label{lagc33z2type1}
\ee
Thus, invariance under $Z_6 \times Z_2$ implies the following constraints on the Yukawa couplings:
\be
\begin{array}{lll}
&&Y^{1}_{\chi\, 33}= Y^{1}_{\chi\, 33},\; Y^{2}_{\chi\, 12}= Y^{2}_{\chi\, 12},\; Y^{3}_{\chi\, 12}= Y^{3}_{\chi\, 12},\;
 Y^{2}_{\chi\, 11}= -Y^{2}_{\chi\, 22}, \;Y^{3}_{\chi\, 11}= -Y^{3}_{\chi\, 22}, \\\\
&&Y^{1}_{D\, 33} =  Y^{1}_{D\, 33},\; Y^{1}_{D\, 11} =  Y^{1}_{D\, 22},\;  Y^{1}_{D\, 12} = - Y^{1}_{D\, 21},\; Y^{2}_{D\, 13} = -i\, Y^{2}_{D\, 23},\;
 Y^{3}_{D\, 31} =-i\,  Y^{3}_{D\, 32},\\\\
&&  Y^{4}_{\ell\, 11} =  Y^{5}_{\ell\, 21},\; Y^{4}_{\ell\, 12} =  Y^{5}_{\ell\, 22},\; Y^{4}_{\ell\, 21} =- Y^{5}_{\ell\, 11},\;
 Y^{4}_{\ell\, 22} = - Y^{5}_{\ell\, 12},\;  Y^{4}_{\ell\, 33} =  -i\,Y^{5}_{\ell\, 33},
\end{array}
\label{const_z6_z2_type1}
\ee
where all vanishing Yukawa couplings are omitted. In fact and as was done for the `rotated' symmetry (indirect realization), by brute force, one also could have used all the machinery encoded in the invariance equations, as given in Eqs.(\ref{constr1},\ref{constr2},\ref{constr3}), in order to obtain a system of linear equations involving Yukawa coupling constants. Solving this resulting system  of linear equations would have provided us then with the symmetry constraints (Eq. \ref{const_z6_z2_type1}).

Thus, the $Z_6 \times Z_2$  symmetry imposes some constraints on the Yukawa couplings which have to be taken into consideration when constructing mass terms after the relevant scalar fields acquire vevs. The emergent  $M_R$ and $M_D$ turn out to be,
\be
M_{R} = \left(
\begin {array}{ccc}
- Y^2_{\chi 22}\; v_{\chi_2} - Y^3_{\chi 22}\; v_{\chi_3}  &Y^2_{\chi 12} \;v_{\chi_2}+  Y^3_{\chi 12}\;v_{\chi_3}&  0 \\
\rule{0.7cm}{0.1mm}& Y^2_{\chi 22}\; v_{\chi_2} + Y^3_{\chi 22}\; v_{\chi_3}& 0\\
\rule{0.7cm}{0.1mm}&\rule{0.7cm}{0.1mm}& Y^1_{\chi 33}\; v_{\chi_1}
\end {array}
\right).
\label{M_R33_type1}
\ee
and,
\be
M_{D} = \left(
\begin {array}{ccc}
Y^1_{D 22}\; v_{\Phi_1} &- Y^1_{D 21}\; v_{\Phi_1}& -i\, Y^2_{D 23}\; v_{\Phi_2} \\
Y^1_{D 21}\; v_{\Phi_1} & Y^1_{D 22}\; v_{\Phi_1}&  Y^2_{D 23}\; v_{\Phi_2} \\
-i\,Y^3_{D 32}\; v_{\Phi_3} & Y^3_{D 32}\; v_{\Phi_3}&  Y^1_{D 33}\; v_{\Phi_1} \\
\end {array}
\right).
\label{M_D33_type1}
\ee
 One can check that the resulting $M_\n$ , through seesaw mechanism, satisfies the texture $\mathbf C_{33}$.

As to $M_\ell$ we get,
\be
M_{\ell } = \left(
\begin {array}{ccc}
Y^5_{\ell 21}\; v_{\Phi_4} +   Y^5_{\ell 11}\; v_{\Phi_5}&Y^5_{\ell 22}\; v_{\Phi_4} +   Y^5_{\ell 12}\; v_{\Phi_5}&0 \\
-Y^5_{\ell 11}\; v_{\Phi_4} +   Y^5_{\ell 21}\; v_{\Phi_5}&-Y^5_{\ell 12}\; v_{\Phi_4} +   Y^5_{\ell 22}\; v_{\Phi_5}&0 \\
0 & 0 &-i\,Y^5_{\ell 33}\; v_{\Phi_4}+
Y^5_{\ell 33}\; v_{\Phi_5}
\end {array}
\right).
\label{M_ell_33_type1}
\ee
Thus, and as an example, one can assume  $v \approx v_{\Phi_5} \gg v_{\Phi_4}$ so that to get
\be
M_{\ell} \approx v \left(
\begin {array}{ccc}
Y^5_{\ell 11} &  Y^5_{\ell 12}&0 \\
  Y^5_{\ell 21}& Y^5_{\ell 22} &0 \\
0 & 0 &   Y^5_{\ell 33}
\end {array}
\right) = v \left( \begin {array}{c}
{\bf a}^T\\{\bf b}^T\\{\bf c}^T
\end {array}
\right).
\label{M_ell_33_type1_approx}
\ee
Another time, one can by just imposing some reasonable assumptions on the ratios of the `free' vectors diagonalize $M_\ell M_\ell^\dagger$ by an infinitesimal rotation, which puts us thus to a good approximation in the flavor basis, as desired.

\subsection{Type I seesaw Direct realization of $\mathbf C_{11}$, $\mathbf C_{22}$ and  $\mathbf C_{13}$ }
Following the same method outlined in case $\mathbf C_{33}$, we state briefly the results of the cases $\mathbf C_{11}$, $\mathbf C_{22}$ and $\mathbf C_{13}$, in Tables~(\ref{11type1norotation},\ref{22type1norotation}) and (\ref{13type1norotation}) respectively.
\begin{table}[hbtp]
\begin{center}
\scalebox{0.8}{
\begin{tabular}{cccccc}
\toprule
\multicolumn{6}{c}{Matter Content and symmetry transformation  (Pattern $\mathbf C_{11}$) }\\
\midrule
\multicolumn{6}{c}{$ \mbox{Symmetry under}\; Z_6 $}\\
\midrule
$ \nu_{R1}\rightarrow \omega\, \nu_{R1} $ & $ \nu_{R2}\rightarrow \nu_{R2} $ & $ \nu_{R3}\rightarrow \nu_{R3}$ & $
\chi_1 \rightarrow  \omega^4\,\chi_1 $ & $  \chi_2 \rightarrow  \chi_2 $ & $ \chi_3 \rightarrow  \chi_3 $ \\
\midrule
 $D_{L1}\rightarrow \omega\, D_{L1} $ & $  D_{L2}\rightarrow D_{L2} $ & $  D_{L3}\rightarrow  D_{L3} $&
 $ \Phi_1 \rightarrow \Phi_1 $ & $ \Phi_2 \rightarrow \omega^5\,\Phi_2 $ & $  \Phi_3 \rightarrow \omega\,\Phi_3$ \\
\midrule
 $ \Phi_4 \rightarrow \omega^3\,\Phi_4 $ & $  \Phi_5 \rightarrow \omega^3\,\Phi_5 $ &
 $  \ell_{R1} \rightarrow \omega^4\, \ell_{R1} $ & $  \ell_{R2} \rightarrow \omega^3\, \ell_{R2} $ & $  \ell_{R3} \rightarrow \omega^3\, \ell_{R3}$ &
\\
\toprule
\multicolumn{6}{c}{$ \mbox{Symmetry under}\; Z_2 $}\\
\midrule
$\nu_{R1}\rightarrow \nu_{R1} $ & $ \nu_{R2}\rightarrow i\,\nu_{R3} $ & $ \nu_{R3}\rightarrow -i\,\nu_{R2}$ & $
\chi_1 \rightarrow  \chi_1 $ & $  \chi_2 \rightarrow  \chi_2 $ & $ \chi_3 \rightarrow  \chi_3 $ \\
\midrule
$  D_{L1}\rightarrow  D_{L1} $ & $  D_{L2}\rightarrow i\, D_{L3} $ & $  D_{L3}\rightarrow -i\, D_{L2} $ &
 $ \Phi_1 \rightarrow \Phi_1 $ & $ \Phi_2 \rightarrow  \Phi_2 $ & $   \Phi_3 \rightarrow -\Phi_3 $ \\
\midrule
 $  \Phi_4 \rightarrow i\,\Phi_5 $ & $  \Phi_5 \rightarrow -i\,\Phi_4 $ & $  \ell_{R1} \rightarrow  \ell_{R1} $ &
$  \ell_{R2} \rightarrow  \ell_{R2} $ & $  \ell_{R3} \rightarrow  \ell_{R3}$ &
\\
\toprule
\multicolumn{6}{c}{Mass matrices $M_R$, $M_D$, $M_\n$, and $M_\ell$}\\
\midrule
\multicolumn{6}{c}{$\small
 M_R =
\left(
\begin {array}{ccc}
x&0 & 0\\
0& y  & z \\
0 & z & -y
\end {array}
\right),
 M_D =
\left(
\begin {array}{ccc}
A&-i\,D & D\\
-I\,E & B  & - C \\
E & C & B
\end {array}
\right),
 M_\nu = \left(
\begin {array}{ccc}
\times &\times  &\times\\
\rule{0.3cm}{0.1mm} & \Delta  & \times\\
\rule{0.3cm}{0.1mm}& \rule{0.3cm}{0.1mm} & -\Delta
\end {array}
\right),
M_{\ell} \stackrel{v_{\Phi_5} \gg v_{\Phi_4}}{\approx} v_{\Phi_5} \left(
\begin {array}{ccc}
Y^5_{\ell 11} &  0&0 \\
  0 & Y^5_{\ell 22} &Y^5_{\ell 23} \\
0 &Y^5_{\ell 32}&   Y^5_{\ell 33}
\end {array}
\right) $}\\
\bottomrule
\end{tabular}}
\end{center}
 \caption{\small  The $Z_6 \times Z_2$ symmetry seesaw type I realization of the
  vanishing subtrace $\mathbf C_{11}$. $\Phi_a$ are five SM Higgs doublets ($a=1\cdots 5$), $D_L$ refers to the flavor three left handed lepton doublets, while the three right-handed charged lepton singlets are denoted by  $\ell_R$. $\omega$ denotes $e^{{i\,\pi/ 3}}$}
\label{11type1norotation}
 \end{table}
\begin{table}[H]
\begin{center}
\scalebox{0.8}{
\begin{tabular}{cccccc}
\toprule
\multicolumn{6}{c}{Matter Content and symmetry transformation  (Pattern $\mathbf C_{22}$) }\\
\midrule
\multicolumn{6}{c}{$ \mbox{Symmetry under}\; Z_6 $}\\
\midrule
$ \nu_{R1}\rightarrow  \nu_{R1} $ & $ \nu_{R2}\rightarrow \omega\, \nu_{R2} $ & $ \nu_{R3}\rightarrow \nu_{R3}$ & $
\chi_1 \rightarrow  \omega^4\,\chi_1 $ & $  \chi_2 \rightarrow  \chi_2 $ & $ \chi_3 \rightarrow  \chi_3 $ \\
\midrule
 $D_{L1}\rightarrow  D_{L1} $ & $  D_{L2}\rightarrow \omega\, D_{L2} $ & $  D_{L3}\rightarrow  D_{L3} $&
 $ \Phi_1 \rightarrow \Phi_1 $ & $ \Phi_2 \rightarrow \omega\,\Phi_2 $ & $  \Phi_3 \rightarrow \omega^5\,\Phi_3$ \\
\midrule
 $ \Phi_4 \rightarrow \omega^3\,\Phi_4 $ & $  \Phi_5 \rightarrow \omega^3\,\Phi_5 $ &
 $  \ell_{R1} \rightarrow \omega^3\, \ell_{R1} $ & $  \ell_{R2} \rightarrow \omega^4\, \ell_{R2} $ & $  \ell_{R3} \rightarrow \omega^3\, \ell_{R3}$ &
\\
\toprule
\multicolumn{6}{c}{$ \mbox{Symmetry under}\; Z_2 $}\\
\midrule
$\nu_{R1}\rightarrow i\, \nu_{R3} $ & $ \nu_{R2}\rightarrow \nu_{R2} $ & $ \nu_{R3}\rightarrow -i\,\nu_{R1}$ & $
\chi_1 \rightarrow  \chi_1 $ & $  \chi_2 \rightarrow  \chi_2 $ & $ \chi_3 \rightarrow  \chi_3 $ \\
\midrule
$  D_{L1}\rightarrow i\,  D_{L3} $ & $  D_{L2}\rightarrow  D_{L2} $ & $  D_{L3}\rightarrow -i\, D_{L1} $ &
 $ \Phi_1 \rightarrow \Phi_1 $ & $ \Phi_2 \rightarrow - \Phi_2 $ & $   \Phi_3 \rightarrow \Phi_3 $ \\
\midrule
 $  \Phi_4 \rightarrow i\,\Phi_5 $ & $  \Phi_5 \rightarrow -i\,\Phi_4 $ & $  \ell_{R1} \rightarrow  \ell_{R1} $ &
$  \ell_{R2} \rightarrow  \ell_{R2} $ & $  \ell_{R3} \rightarrow  \ell_{R3}$ &
\\
\toprule
\multicolumn{6}{c}{Mass matrices $M_R$, $M_D$, $M_\n$, and $M_\ell$}\\
\midrule
\multicolumn{6}{c}{$\small
 M_R =
\left(
\begin {array}{ccc}
x&0 & z\\
0& y  & 0 \\
z & 0 & -x
\end {array}
\right),
 M_D =
\left(
\begin {array}{ccc}
A&i\,B & - C\\
-I\,E & D  & E \\
C & - B & A
\end {array}
\right),
 M_\nu = \left(
\begin {array}{ccc}
\Delta &\times  &\times\\
\rule{0.3cm}{0.1mm} & \times  & \times\\
\rule{0.3cm}{0.1mm}& \rule{0.3cm}{0.1mm} & -\Delta
\end {array}
\right),
M_{\ell} \stackrel{v_{\Phi_5} \gg v_{\Phi_4}}{\approx} v_{\Phi_5} \left(
\begin {array}{ccc}
Y^5_{\ell 11} &  0& Y^5_{\ell 13} \\
  0 & Y^5_{\ell 22} & 0 \\
Y^5_{\ell 13 } & 0 &   Y^5_{\ell 33}
\end {array}
\right) $}\\
\bottomrule
\end{tabular}}
\end{center}
 \caption{\small  The $Z_6 \times Z_2$ symmetry seesaw type I realization of the
  vanishing subtrace $\mathbf C_{22}$. $\Phi_a$ are five SM Higgs doublets ($a=1\cdots 5$), $D_L$ refers to the flavor three left handed lepton doublets, while the three right-handed charged lepton singlets are denoted by  $\ell_R$. $\omega$ denotes $e^{{i\,\pi/ 3}}$}
\label{22type1norotation}
 \end{table}
\begin{table}[H]
\begin{center}
\scalebox{0.8}{
\begin{tabular}{cccccc}
\toprule
\multicolumn{6}{c}{Matter Content and symmetry transformation  (Pattern $\mathbf C_{13}$) }\\
\midrule
\multicolumn{6}{c}{$ \mbox{Symmetry under}\; Z_6 $}\\
\midrule
$ \nu_{R1}\rightarrow  \nu_{R1} $ & $ \nu_{R2}\rightarrow \omega\, \nu_{R2} $ & $ \nu_{R3}\rightarrow \nu_{R3}$ & $
\chi_1 \rightarrow  \omega^4\,\chi_1 $ & $  \chi_2 \rightarrow  \chi_2 $ & $ \chi_3 \rightarrow  \chi_3 $ \\
\midrule
 $D_{L1}\rightarrow  D_{L1} $ & $  D_{L2}\rightarrow \omega\, D_{L2} $ & $  D_{L3}\rightarrow  D_{L3} $&
 $ \Phi_1 \rightarrow \Phi_1 $ & $ \Phi_2 \rightarrow \omega\,\Phi_2 $ & $  \Phi_3 \rightarrow \omega^3\,\Phi_3$ \\
\midrule
 $ \Phi_4 \rightarrow \omega^3\,\Phi_4 $  &
 $  \ell_{R1} \rightarrow \omega^3\, \ell_{R1} $ & $  \ell_{R2} \rightarrow \omega^4\, \ell_{R2} $ & $  \ell_{R3} \rightarrow \omega^3\, \ell_{R3}$ & &
\\
\toprule
\multicolumn{6}{c}{$ \mbox{Symmetry under}\; Z_2 $}\\
\midrule
$\nu_{R1}\rightarrow i\, \nu_{R3} $ & $ \nu_{R2}\rightarrow \nu_{R2} $ & $ \nu_{R3}\rightarrow -i\,\nu_{R1}$ & $
\chi_1 \rightarrow  \chi_1 $ & $  \chi_2 \rightarrow - \chi_2 $ & $ \chi_3 \rightarrow  \chi_3 $ \\
\midrule
$  D_{L1}\rightarrow -\,  D_{L3} $ & $  D_{L2}\rightarrow  D_{L2} $ & $  D_{L3}\rightarrow -\, D_{L1} $ &
 $ \Phi_1 \rightarrow \Phi_1 $ & $ \Phi_2 \rightarrow  \Phi_2 $ & $   \Phi_3 \rightarrow i\, \Phi_4 $ \\
\midrule
 $  \Phi_4 \rightarrow - i\,\Phi_3 $ &  $  \ell_{R1} \rightarrow  \ell_{R1} $ &
$  \ell_{R2} \rightarrow  \ell_{R2} $ & $  \ell_{R3} \rightarrow  \ell_{R3}$ & &
\\
\toprule
\multicolumn{6}{c}{Mass matrices $M_R$, $M_D$, $M_\n$, and $M_\ell$}\\
\midrule
\multicolumn{6}{c}{$\small
 M_R =
\left(
\begin {array}{ccc}
x&0 & z\\
0& y  & 0 \\
z & 0 & u
\end {array}
\right),
 M_D =
\left(
\begin {array}{ccc}
i\, A& - D & -i\, C\\
0 & B  & 0 \\
C & D & A
\end {array}
\right),
 M_\nu = \left(
\begin {array}{ccc}
\times &\Delta  &\times\\
\rule{0.3cm}{0.1mm} & \times  & -\Delta\\
\rule{0.3cm}{0.1mm}& \rule{0.3cm}{0.1mm} & \times
\end {array}
\right),
M_{\ell} \stackrel{v_{\Phi_4} \gg v_{\Phi_3}}{\approx} v_{\Phi_4} \left(
\begin {array}{ccc}
Y^4_{\ell 11} &  0& Y^4_{\ell 13} \\
  0 & Y^4_{\ell 22} & 0 \\
Y^4_{\ell 31 } & 0 &   Y^4_{\ell 33}
\end {array}
\right) $}\\
\bottomrule
\end{tabular}}
\end{center}
 \caption{\small  The $Z_6 \times Z_2$ symmetry seesaw type I realization of the
 vanishing subtrace $\mathbf C_{13}$. $\Phi_a$ are four SM Higgs doublets ($a=1\cdots 4$), $D_L$ refers to the flavor three left handed lepton doublets, while the three right-handed charged lepton singlets are denoted by  $\ell_R$. $\omega$ denotes $e^{{i\,\pi/ 3}}$ and $u$ is a generic independent parameter present in $M_R$.}
\label{13type1norotation}
 \end{table}
\section{Direct realization of type II seesaw with $Z_2^\prime \times Z_2$-symmetry}
By the same token, we present now, within type II seesaw scenario, a ``direct" method which leads straightly to the vanishing subtrace texture without relating it to zero textures by `rotation'. It is applicable again only for the four textures ($\mathbf C_{33}, \mathbf C_{11}, \mathbf C_{22}$ and $\mathbf C_{31}$). Besides, the key idea behind this realization is having a  permutation performed through the group factor $Z_2^\prime$.

\subsection{Type II seesaw Direct realization of $\mathbf C_{33}$}
Within type II seesaw scenario, the term $Y^{\n a}_{ij}\; H^0_a\; \n^T_{Li}\; \mathcal{C}^{-1}\; \n_{Lj}$ in the Lagrangian of Eq.(\ref{type2Lagrangian}) is the term responsible for $M_\n$ where we introduced three Higgs triplets. We introduce two Higgs doublets $\Phi_b$ responsible for $M_\ell$ through the term $Y^{\ell a}_{ij}\; \overline{D}_{Li}\; \Phi_a \;\ell_{Rj}$ (see Eq.(\ref{type2LagrangianCharged})). We assume the field transformations defined
in Table~(\ref{33type2norotation}).
\begin{table}[hbtp]
\begin{center}
\begin{tabular}{cccc}
\toprule
\multicolumn{4}{c}{Matter Content (Pattern $\mathbf C_{33}$) }\\
\midrule
    $H$  & $D_L$ & $\ell_R$ & $\Phi$   \\
\midrule
\multicolumn{4}{c}{$ \mbox{Symmetry under}\; Z_2^\prime $}\\
\midrule
 $G^\prime_{H} H $ & $G^\prime_{D} D_L$ & $G^\prime_{\ell} \ell_R$ & $G^\prime_{\Phi} \Phi$
\\
 $G^\prime_{H}=\mbox{diag}\left( 1,1,-1\right)$& $G^\prime_{D}=\left(\begin {array}{ccc}
0 &-i&0 \\
+i &0&0\\
0 &0&1
\end {array} \right)$ & $G^\prime_{\ell} = \mbox{diag}\left( 1,1,1\right) $& $G^\prime_{\Phi} = \left(\begin {array}{cc}
0 &1 \\
1 &0
\end {array} \right)$ \\\\
\toprule
\multicolumn{4}{c}{$ \mbox{Symmetry under}\; Z_2 $}\\
\midrule
 $G_{H} H $ & $G_{D} D_L$ & $G_{\ell} \ell_R$ & $G_{\Phi} \Phi$
\\
$G_{H}= \mbox{diag}\left( 1,-1,-1\right)$ & $G_{D} = \mbox{diag}\left( -1,-1,1\right)$  & $G_{\ell} = \mbox{diag}\left( 1,1,-1\right)$ & $G_{\Phi} = \mbox{diag}\left( -1,-1\right)$ \\
\bottomrule
\end{tabular}
\end{center}
 \caption{\small  The $Z^\prime_2 \times Z_2$ symmetry seesaw type II realization of the
  vanishing subtrace $\mathbf C_{33}$. $H$ are three triplet scalars, $D_L$ refers to the flavor three left handed lepton doublets, while the three right-handed charged lepton singlets are denoted by  $\ell_R$. The $\Phi$ denotes two SM Higgs doublets.}
\label{33type2norotation}
 \end{table}

By forming bilinear terms in order to see the transformations under $Z_2$, we get for $\n^T_{Li}\;\n_{Lj}$,
 \be
\n^T_{Li}\;\n_{Lj} \stackrel{Z_2}{\cong}
\left(
\begin {array}{ccc}
1&1 & -1\\
1 & 1  & -1 \\
-1 & -1 & 1
\end {array}
\right) ,
\ee
and so,  when the bilinear  $\n^T_{L_i}\n_{L_j}$ is combined with the transformation of $H_a$ under $Z_2$, we get
\be
\begin{array}{lll}
 H_1^0\,\n^T_{Li}\;\n_{Lj} \stackrel{Z_2}{\cong}
\left(
\begin {array}{ccc}
1&1 & -1\\
1 & 1  & -1 \\
-1 & -1 & 1
\end {array}
\right), &    H_2^0\,\n^T_{Li}\;\n_{Lj} \stackrel{Z_2}{\cong}
\left(
\begin {array}{ccc}
-1&-1 & 1\\
-1 & -1  & 1 \\
1 & 1 & -1
\end {array}
\right), &
 H_3^0\,\n^T_{Li}\;\n_{Lj} \stackrel{Z_2}{\cong}
\left(
\begin {array}{ccc}
-1&-1 & 1\\
-1 & -1  & 1 \\
1 & 1 & -1
\end {array}
\right).
\end{array}
\label{z2norotationH}
\ee
Thus, the Lagrangian terms in Eq.(\ref{type2Lagrangian}), responsible for $M_\n$, which are due to the interaction with $H_1^0$ and are consistent with $Z_2$ symmetry are
\be
L^{Z_2}_{H_1\n} \propto H_1^0\, \left( Y^{\n 1}_{11}\,\mathcal{C}^{-1}\, \n_{L1}^T\, \n_{L1} +  Y^{\n 1}_{22} \,\n_{L2}^T\,\mathcal{C}^{-1}\, \n_{L2} + Y^{\n 1}_{33} \,\n_{L3}^T\,\mathcal{C}^{-1}\, \n_{L3} +  Y^{\n 1}_{12}\, \n_{L1}^T\, \mathcal{C}^{-1}\,\n_{L2}  + Y^{\n 1}_{12}\, \n_{L2}^T\, \mathcal{C}^{-1}\,\n_{L1} \right),
\ee
which transforms under $Z_2^\prime$ as
\be
L^{Z_2}_{H_1\n} \stackrel{Z^\prime_2}{\rightarrow} H_1^0\, \left( -Y^{\n 1}_{11}\, \n_{L2}^T\,\mathcal{C}^{-1}\, \n_{L2} -  Y^{\n 1}_{22}\, \n_{L1}^T \mathcal{C}^{-1}\,\n_{L1} + Y^{\n 1}_{33}\, \n_{L3}^T\,\mathcal{C}^{-1}\, \n_{L3} + Y^{\n 1}_{12}\, \n_{L1}^T\,\mathcal{C}^{-1}\, \n_{L2}
+  Y^{\n 1}_{12}\, \n_{L2}^T\,\mathcal{C}^{-1}\, \n_{L1}\right).
\ee
Thus, invariance under $Z_2 \times Z_2^\prime$ implies the constraint:
\be
Y^{\n 1}_{11} = - Y^{\n 1}_{22} ,\;\;  Y^{\n 1}_{13} = Y^{\n 1}_{23} =0.
\ee
By the same way, one can see the constraints on the Yukawa couplings due to interaction with $H_2^0$ and $H_3^0$, and we get
\be
\begin{array}{lll}
Y^{\n 2}_{13} = i\, Y^{\n 2}_{23} ,& & Y^{\n 2}_{11} = Y^{\n 2}_{22} = Y^{\n 2}_{12} = 0, \\
Y^{\n 3}_{13} = -i\, Y^{\n 3}_{23} ,&  & Y^{\n 3}_{11} = Y^{\n 3}_{22} = Y^{\n 3}_{12} = 0.
\end{array}
\ee
So when $H_a^0$ gets a vev $v^H_a$ we get $M_\n$ in the form
\be
\label{norotation-c33}
\begin{array}{lll}
M_{\n} &=& \left(
\begin {array}{ccc}
-Y^{\n1}_{22}\;v^H_{1}  &Y^{\n1}_{12}\; v^H_{1} & i\,\left(Y^{\n2}_{23}\; v^H_{2} - Y^{\n3}_{23}\; v^H_{3}\right) \\
\rule{0.7cm}{0.1mm}& Y^{\n1}_{22} \;v^H_{1} &  \left( Y^{\n2}_{23} \;v^H_{2} + Y^{\n3}_{23}\; v^H_{3} \right) \\
\rule{0.7cm}{0.1mm}&\rule{0.7cm}{0.1mm}&Y^{\n1}_{33}\; v^H_{1}
\end {array}
\right).
\end{array}
\ee
We see that the texture $\mathbf C_{33}$ is realized.

For the charged lepton mass matrix $M_\ell$, we follow the same procedure by forming bilinear terms in order to see the transformations under $Z_2$:
 \bea
\overline{D}_{Li} \,\ell_{Rj} \stackrel{Z_2}{\cong}
\left(
\begin {array}{ccc}
-1&-1 & 1\\
-1 & -1  & 1 \\
1 & 1 & -1
\end {array}
\right) &\Rightarrow& \overline{D}_{Li}\, \ell_{Rj}\, \Phi_1 \stackrel{Z_2}{\cong} \overline{D}_{Li} \, \ell_{Rj} \,\Phi_2 \stackrel{Z_2}{\cong}
\left(
\begin {array}{ccc}
1&1 & -1\\
1 & 1  & -1 \\
-1 & -1 & 1
\end {array}
\right),
\label{z2norotationPhi}
\eea
Thus, the Lagrangian terms in Eq. (\ref{type2LagrangianCharged}), responsible for $M_\ell$, which are due to the interaction with $\Phi_1, \Phi_2$ and are consistent with $Z_2$ symmetry are
\be
\begin{array}{lll}
L^{Z_2}_{\Phi\ell} &\propto& \Phi_1 \left( Y^{\ell 1}_{11}\; \overline{D}_{L1}\; \ell_{R1} +  Y^{\ell 1}_{12} \; \overline{D}_{L1}\; \ell_{R2} + Y^{\ell 1}_{21} \;\overline{D}_{L2}\; \ell_{R1}
+  Y^{\ell 1}_{22}\; \overline{D}_{L2} \;\ell_{R2} + Y^{\ell 1}_{33}\; \overline{D}_{L3}\; \ell_{R3} \right) +  \\\\
&& \Phi_2 \left( Y^{\ell 2}_{11}\; \overline{D}_{L1}\; \ell_{R1} +  Y^{\ell 2}_{12} \; \overline{D}_{L1}\; \ell_{R2} + Y^{\ell 2}_{21} \;\overline{D}_{L2}\; \ell_{R1}
+  Y^{\ell 2}_{22}\; \overline{D}_{L2} \;\ell_{R2} + Y^{\ell 2}_{33}\; \overline{D}_{L3}\; \ell_{R3} \right),
\end{array}
\ee
which transforms under $Z_2^\prime$ as
\be
\begin{array}{lll}
L^{Z_2}_{\Phi\ell} &\stackrel{Z^\prime_2}{\rightarrow}& \Phi_2 \left( -i \,Y^{\ell 1}_{11}\; \overline{D}_{L2}\; \ell_{R1} -i  Y^{\ell 1}_{12}\; \overline{D}_{L2}\; \ell_{R2} +i\, Y^{\ell 1}_{21}\; \overline{D}_{L1}\; \ell_{R1}
+ i\, Y^{\ell 1}_{22} \overline{D}_{L1}\; \ell_{R2} + Y^{\ell 1}_{33} \overline{D}_{L3}\; \ell_{R3} \right) + \\\\
&&\Phi_1 \left( -i \,Y^{\ell 2}_{11}\; \overline{D}_{L2}\; \ell_{R1} -i\,  Y^{\ell 2}_{12}\; \overline{D}_{L2}\; \ell_{R2} +i\,  Y^{\ell 2}_{21}\; \overline{D}_{L1}\; \ell_{R1}
+ i\, Y^{\ell 2}_{22}\; \overline{D}_{L1} \;\ell_{R2} + Y^{\ell 2}_{33}\; \overline{D}_{L3}\; \ell_{R3} \right).
\end{array}
\ee
Thus, invariance under $Z_2 \times Z_2^\prime$ implies the constraint:
\be
\begin{array}{lll}
&&Y^{\ell 1}_{11} = i \,Y^{\ell 2}_{21},\; Y^{\ell 1}_{12} = i\, Y^{\ell 2}_{22},\; Y^{\ell 1}_{21} =- i\, Y^{\ell 2}_{11},\;
Y^{\ell 1}_{22} = -i\, Y^{\ell 2}_{12},\; Y^{\ell 1}_{33} =  Y^{\ell 2}_{33},\\
&&  Y^{\ell 1(2)}_{13} = Y^{\ell 1(2)}_{23} =Y^{\ell 1(2)}_{31} = Y^{\ell 1(2)}_{32}=0
\end{array}
\ee
So when $\Phi_a^0$ gets a vev $v^\Phi_a$ we get $M_\ell$ in the form
\bea
\label{norotation-c33}
M_{\ell} &=& \left(
\begin {array}{ccc}
Y^{\ell 1}_{11}\; v^\Phi_{1} + i \,Y^{\ell 1}_{21}\; v^\Phi_{2}   &Y^{\ell 1}_{12} \;v^\Phi_{1} + i\, Y^{\ell 1}_{22}\; v^\Phi_{2}  & 0 \\
Y^{\ell 1}_{21}\; v^\Phi_{1} - i\, Y^{\ell 1}_{11}\; v^\Phi_{2}  &  Y^{\ell 1}_{22}\; v^\Phi_{1} - i\, Y^{\ell 1}_{12} \;v^\Phi_{2} &0 \\
0&0&Y^{\ell 1}_{33}\; (v^\Phi_{1}+v^\Phi_{2})
\end {array}
\right).
\eea
Thus, when $v \approx v^\Phi_{2} \gg v^\Phi_{1}$ we get
\bea
\label{norotation-c33-v}
M_{\ell} = v \left(
\begin {array}{ccc}
 i Y^{\ell 1}_{21}    & i Y^{\ell 1}_{22}   & 0 \\
-i Y^{\ell 1}_{11}  &   - i Y^{\ell 1}_{12}  &0 \\
0&0&Y^{\ell 1}_{33}
\end {array}
\right) &=& v \left( \begin {array}{c}
{\bf a}^T\\{\bf b}^T\\{\bf c}^T
\end {array}
\right)
\eea
One can by just imposing some reasonable assumptions on the ratios of the `free' vectors diagonalize $M_\ell M_\ell^\dagger$ by an infinitesimal rotation, which puts us thus to a good approximation in the flavor basis, as desired.

\subsection{Type II seesaw Direct realization of $\mathbf C_{11}$, $\mathbf C_{22}$ and  $\mathbf C_{13}$ }
Following the same method outlined in case $\mathbf C_{33}$, we state briefly the results of the cases $\mathbf C_{11}$, $\mathbf C_{22}$ and $\mathbf C_{13}$, in Tables~(\ref{11type2norotation},\ref{22type2norotation}) and (\ref{13type2norotation}) respectively.
\begin{table}[H]
\begin{center}
\begin{tabular}{cccc}
\toprule
\multicolumn{4}{c}{Matter content (Pattern $\mathbf C_{11}$)}\\
\midrule
    $H$  & $D_L$ & $\ell_R$ & $\Phi$   \\
\midrule
\multicolumn{4}{c}{$ \mbox{Symmetry under}\; Z_2^\prime $}\\
\midrule
 $G^\prime_{H} H $ & $G^\prime_{D} D_L$ & $G^\prime_{\ell} \ell_R$ & $G^\prime_{\Phi} \Phi$
\\
 $G^\prime_{H}=\mbox{diag}\left( 1,1,-1\right)$& $G^\prime_{D}=\left(\begin {array}{ccc}
1 &0&0 \\
0 &0&-i\\
0 &i&0
\end {array} \right)$ & $G^\prime_{\ell} = \mbox{diag}\left( 1,1,1\right) $& $G^\prime_{\Phi} = \left(\begin {array}{cc}
0 &1 \\
1 &0
\end {array} \right)$ \\
\toprule
\multicolumn{4}{c}{$ \mbox{Symmetry under}\; Z_2 $}\\
\midrule
 $G_{H} H $ & $G_{D} D_L$ & $G_{\ell} \ell_R$ & $G_{\Phi} \Phi$
\\
$G_{H}= \mbox{diag}\left( 1,-1,-1\right)$ & $G_{D} = \mbox{diag}\left( 1,-1,-1\right)$  & $G_{\ell} = \mbox{diag}\left( -1,1,1\right)$ & $G_{\Phi} = \mbox{diag}\left( -1,-1\right)$ \\
\midrule
\multicolumn{4}{c}{Mass matrices}\\
\midrule
\multicolumn{4}{c}{$
M_{\n} = \left(
\begin {array}{ccc}
Y^{\n1}_{11}\; v^H_{1}  &i\,\left(Y^{\n 2}_{13}\; v^H_{2} - Y^{\n 3}_{13}\;v^H_{3}\right) &  Y^{\n2}_{13}\; v^H_{2} + Y^{\n3}_{13} \;v^H_{3} \\
\rule{0.7cm}{0.1mm} &-Y^{\n1}_{33} \;v^H_{2} &   Y^{\n1}_{23}\; v^H_{1} \\
\rule{0.7cm}{0.1mm}&\rule{0.7cm}{0.1mm}&- Y^{\n1}_{33}\; v^H_{1}
\end {array}
\right), M_{\ell} \stackrel{v^{\Phi}_2 \gg v^{\Phi}_1}{\approx} v^\Phi_{2} \left(
\begin {array}{ccc}
Y^{\ell 2}_{11}   &0  & 0 \\
0  &  Y^{\ell 2}_{22}  & Y^{\ell 2}_{23} \\
0& Y^{\ell 2}_{32}& Y^{\ell 2}_{33}
\end {array}
\right) $}\\
\bottomrule
\end{tabular}
\end{center}
 \caption{\small  The $Z^\prime_2 \times Z_2$ symmetry seesaw type II realization of the
  vanishing subtrace $\mathbf C_{11}$. $H$ are three triplet scalars, $D_L$ refers to the flavor three left-handed lepton doublets, while the three right-handed charged lepton singlets are denoted by  $\ell_R$. The $\Phi$ denotes two SM Higgs doublets.}
\label{11type2norotation}
 \end{table}
\begin{table}[H]
\begin{center}
\scalebox{0.95}{
\begin{tabular}{cccc}
\toprule
\multicolumn{4}{c}{Matter content (Pattern $\mathbf C_{22}$)}\\
\midrule
    $H$  & $D_L$ & $\ell_R$ & $\Phi$   \\
\midrule
\multicolumn{4}{c}{$ \mbox{Symmetry under}\; Z_2^\prime $}\\
\midrule
 $G^\prime_{H} H $ & $G^\prime_{D} D_L$ & $G^\prime_{\ell} \ell_R$ & $G^\prime_{\Phi} \Phi$
\\
 $G^\prime_{H}=\mbox{diag}\left( 1,1,-1\right)$& $G^\prime_{D}=\left(\begin {array}{ccc}
0 &0&-i \\
0 &1&0\\
i &0&0
\end {array} \right)$ & $G^\prime_{\ell} = \mbox{diag}\left( 1,1,1\right) $& $G^\prime_{\Phi} = \left(\begin {array}{cc}
0 &1 \\
1 &0
\end {array} \right)$ \\
\toprule
\multicolumn{4}{c}{$ \mbox{Symmetry under}\; Z_2 $}\\
\hline
 $G_{H} H $ & $G_{D} D_L$ & $G_{\ell} \ell_R$ & $G_{\Phi} \Phi$
\\
$G_{H}= \mbox{diag}\left( 1,-1,-1\right)$ & $G_{D} = \mbox{diag}\left( -1,1,-1\right)$  & $G_{\ell} = \mbox{diag}\left( 1,-1,1\right)$ & $G_{\Phi} = \mbox{diag}\left( -1,-1\right)$ \\
\midrule
\multicolumn{4}{c}{Mass matrices}\\
\midrule
\multicolumn{4}{c}{$
M_{\n} = \left(
\begin {array}{ccc}
-Y^{\n1}_{33}\; v^H_{1}  &i\,\left(Y^{\n 2}_{23}\; v^H_{2} - Y^{\n 3}_{23}\; v^H_{3}\right) & Y^{\n 1}_{13}\; v^H_{1}   \\
\rule{0.7cm}{0.1mm} &Y^{\n 1}_{22}\; v^H_{1} &  \left( Y^{\n 2}_{23}\; v^H_{2} + Y^{\n 3}_{23}\; v^H_{3} \right) \\
\rule{0.7cm}{0.1mm}&\rule{0.7cm}{0.1mm}& Y^{\n1}_{33}\; v^H_{1}
\end {array}
\right), M_{\ell} \stackrel{v^{\Phi}_2 \gg v^{\Phi}_1}{\approx} v^\Phi_{2} \left(
\begin {array}{ccc}
 Y^{\ell 2}_{11}   &0  & Y^{\ell 2}_{13}  \\
0  &  Y^{\ell 2}_{22}  & 0 \\
-i \,Y^{\ell 2}_{31} &0&  Y^{\ell 2}_{33}
\end {array}
\right) $}\\
\bottomrule
\end{tabular}}
\end{center}
 \caption{\small  The $Z^\prime_2 \times Z_2$ symmetry seesaw type II realization of the
  vanishing subtrace $\mathbf C_{22}$. $H$ are three triplet scalars, $D_L$ refers to the flavor three left-handed lepton doublets, while the three right-handed charged lepton singlets are denoted by  $\ell_R$. The $\Phi$ denotes two SM Higgs doublets.}
\label{22type2norotation}
 \end{table}
\begin{table}[H]
\begin{center}
\begin{tabular}{cccc}
\toprule
\multicolumn{4}{c}{Matter content (Pattern $\mathbf C_{13}$)}\\
\midrule
    $H$  & $D_L$ & $\ell_R$ & $\Phi$   \\
\midrule
\multicolumn{4}{c}{$ \mbox{Symmetry under}\; Z_2^\prime $}\\
\midrule
 $G^\prime_{H} H $ & $G^\prime_{D} D_L$ & $G^\prime_{\ell} \ell_R$ & $G^\prime_{\Phi} \Phi$
\\
 $G^\prime_{H}=\mbox{diag}\left( 1,-1,-1\right)$& $G^\prime_{D}=\left(\begin {array}{ccc}
0 &0&1 \\
0 &1&0\\
1 &0&0
\end {array} \right)$ & $G^\prime_{\ell} = \mbox{diag}\left( 1,1,1\right) $& $G^\prime_{\Phi} = \left(\begin {array}{cc}
0 &1 \\
1 &0
\end {array} \right)$ \\
\midrule
\multicolumn{4}{c}{$ \mbox{Symmetry under}\; Z_2 $}\\
\midrule
 $G_{H} H $ & $G_{D} D_L$ & $G_{\ell} \ell_R$ & $G_{\Phi} \Phi$
\\
$G_{H}= \mbox{diag}\left( 1,1,-1\right)$ & $G_{D} = \mbox{diag}\left( -1,1,-1\right)$  & $G_{\ell} = \mbox{diag}\left( 1,-1,1\right)$ & $G_{\Phi} = \mbox{diag}\left( -1,-1\right)$ \\
\midrule
\multicolumn{4}{c}{Mass matrices}\\
\midrule
\multicolumn{4}{c}{$
M_{\n} = \left(
\begin {array}{ccc}
Y^{\n1}_{33}\; v^H_{1} - Y^{\n2}_{33}\; v^H_{2}  &-Y^{\n 3}_{32}\; v^H_{3} & Y^{\n 1}_{13}\; v^H_{1}   \\
\rule{0.7cm}{0.1mm}&Y^{\n 1}_{22}\; v^H_{1} &    Y^{\n 3}_{23}\; v^H_{3}  \\
\rule{0.7cm}{0.1mm}&\rule{0.7cm}{0.1mm}& Y^{\n1}_{33}\; v^H_{1} + Y^{\n 2}_{33}\; v^H_{2}
\end {array}
\right), M_{\ell} \stackrel{v^{\Phi}_2 \gg v^{\Phi}_1}{\approx} v^\Phi_{2} \left(
\begin {array}{ccc}
 Y^{\ell 2}_{11}   &0  &  Y^{\ell 2}_{13}  \\
0  &  Y^{\ell 2}_{22}  & 0 \\
 Y^{\ell 2}_{31} &0&  Y^{\ell 2}_{33}
\end {array}
\right) $}\\
\bottomrule
\end{tabular}
\end{center}
 \caption{\small  The $Z^\prime_2 \times Z_2$ symmetry seesaw type II realization of the
  vanishing subtrace $\mathbf C_{13}$. $H$ are three triplet scalars, $D_L$ refers to the flavor three left-handed lepton doublets, while the three right-handed charged lepton singlets are denoted by  $\ell_R$. The $\Phi$ denotes two SM Higgs doublets.}
\label{13type2norotation}
 \end{table}
\section{Summary-Discussion and Conclusion}

We have studied a specific texture characterized by one vanishing subtrace of the neutrino mass matrix. We found that all textures, whether they be of  inverted or normal  type, can accommodate the recent experimental bounds. Moreover, four textures of inverted type can accommodate data in case one neutrino mass is zero. We have carried out a complete phenomenological and analytic analysis, but did not state the analytic expressions, as they are too cumbersome, even the first terms in a Taylor expansion in powers of $s_{13}$. Finally, for the model building of the texture, we first proposed a `generic' strategy to justify such a specific texture form based on finding a corresponding symmetry implying certain zeros at $M_{\n 0}$, which when `rotated' to a new `rotated symmetry' leads to the desired form of vanishing subtrace in $M_\n$. We applied this strategy for both types of seesaw scenarios and in both invertible and singular neutrino mass matrices. We also presented a direct method  to realize the textures without `rotation' for both types of seesaw scenarios  based on discrete symmetry.

{  In all these theoretical models, the SSB of a discrete symmetry $Z_n \times Z_m$, triggered by some fields ($\Phi, \chi$) -some of which are very heavy- taking a vev, led by construction to  a texture of vanishing subtrace, and this presumably happens at high scale. However, the question arises as to whether or not the running of the Yukawa couplings from high scale to weak scale spoils the form of the texture. In our work we assumed this change is slight, and that the texture would be kept when running to weak scale, in line with \cite{cherif}. However, it was argued in \cite{Liu} that the ``entries-equality'' condition is not stable against radiative corrections, and surely this question is worthy of a thorough analysis in its own right.

Moreover, we have not discussed the scalar and Higgs potential. In the appendix we stated the general form of the renormalizable scalar-Higgs potential in one case ($C_{33}$) respecting the discrete symmetry ($Z_2 \times Z_6$), and therein put in constraints on the coupling constants, and stated the tadpole equations and the corresponding mass matrices.

 The questions of stability and perturbative unitarity, in that the potential is co-positive and bounded from below or at least it accepts local minima with sufficiently large decaying time-scales through tunneling, and that the coupling constants are small enough for perturbative expansion, are all assumed to be met through tuning of the potential parameters. Surely, a term like $\Phi^\dagger \Phi \chi^\dagger \chi$ would break the discrete symmetry when $\chi$ takes a vev, and here we wonder whether or not the $\Phi$ can behave as an SM Higgs field, as we assumed in our realization models, in which case we ask how its mass is kept small whereas such a term  gives a contribution of the form $Y v^2_\chi$ where $v_\chi$ is expected to be high. Again, we assume a sort of fine tuning imposing the corresponding $Y$ to be too small, and this is reminiscent of the Higgs hierarchy problem, where even in the SM and at tree levels, one needs fine tuning in order to keep the Higgs mass at electroweak scale assuming it is coupled to heavy scalars.

Finally, the existence in our realization models of many electroweak doublets $\Phi$'s which get a vev at the electroweak scale could lead to an interesting phenomenology of the extra states at the Large Hadron Collider {(\bf LHC)}. We did not discuss this, but rather assumed again a fine tuning position. First, the parameters of the Higgs-Scalar potential are assumed to be fine tuned so that several Higgs fields $\Phi$ get a vev $v_\Phi$ at the electroweak scale, while other fields $\chi$ get a vev  $v_\chi$ at high scale. Second, the corresponding `quartic' Yukawas for some of these electroweak doublets ($Y_\Phi \Phi^4$) are tuned to be high, so that the `low scale' contribution to their masses ($Y_\Phi v^2_\Phi$), when added to the `high scale' contribution to their masses ($Y_{\Phi \chi} v_\chi^2$) originating from the coupling term ($Y_{\Phi \chi} \Phi^2 \chi^2$), is in such a way that the resulting mass for all the electroweak doublets $\Phi$'s, except the SM one, are beyond the reach of current accelerators. Actually, this is a `common' assumption because fine tuning of parameters is required whenever there are two different
scales in the theory, which are generated by the Higgs vevs \cite{sarkar}. In \cite{grimus}, a similar enriched scalar sector with three
Higgs doublets and two scalar gauge singlets was studied, and again a high level of fine tuning
was required in order to eliminate the large radiative corrections, originated from the existence of a high scale in the model, through renormalization to the mass of light scalars.

We hope to address some of the above mentioned points in the future.
}

\section*{{\large \bf Acknowledgements}}
M.A. thanks the staff members of the  CPPM-Marseille and LPTM-Cergy-Pontoise Laboratories. N.C. acknowledges support from ICTP-Associate program, ITP-PIFI program and Humboldt Foundation.

\appendix
{

\section{Scalar potential for the case $C_{33}$ under $Z_2 \times Z_6$ symmetry }

Looking at Table~(\ref{33type1norotation}) as a representative case of direct realization\footnote{We could equally choose the case $C_{33}$ of indirect realization as presented in Table~(\ref{newtableC33}) but we chose the case of direct realization since it is simpler.}, we see that for the sake of constructing the scalar potential, we can drop the two gauge singlets so that we have five doublets , ${\bf \phi}= \left( \phi_1 , \phi_2, \phi_3, \phi_4, \phi_5 \right)^T$,  and   one singlet $\chi$.  The scalar fields transform under $Z_2 \times Z_6$ as follows ( $\omega = e^{\pi i/3}$).
\bea
{\bf \phi}\stackrel{Z_6}{\longrightarrow}
\mbox{\small diag} \left(1,\omega,\omega^5,\omega^3,\omega^3\right)
 {\bf \phi} =  {\bf U_{6} \phi}
&\!\!\!,&\!\!\!\!\!
{\bf \phi}  \stackrel{Z_2}{\longrightarrow}
\left(\begin{array}{ccccc} 1&0&0&0&0 \\ 0&-1&0&0&0 \\0&0& 1 &0&0 \\0&0&0&0&i \\0&0&0&-i&0 \end{array}\right)
{\bf \phi} =  {\bf U_{2} \phi},\nn\\
\chi \stackrel{Z_6}{\longrightarrow} \omega^4 \chi &,& \chi \stackrel{Z_2}{\longrightarrow}  \chi.
\eea
Dictated by gauge symmetry, we see directly that the most general renormalizable potential expressing the coupling amidst the  ${\bf \phi}$ and $\chi$ fields can be restricted to the following form:
\bea
V &=& V_\phi + V_\chi + V_{\phi \chi},\nn \\
V_\phi &=& \mu_{a\,b}\, \phi_a^{\dagger}\, \phi_b + \lambda_{a\,b\,c\,d}\, \phi_a^{ \dagger}\, \phi_b\, \phi_c^{\dagger}\, \phi_d,\nn \\
 V_\chi &=&  m_{\chi}^2 \, \chi^\dagger\, \chi + \lam_{\chi}\, (\chi^\dagger\, \chi)^2 ,\nn \\
V_{\phi \chi} &=&  \lam_{\chi a\,b}\, (\chi^\dagger \, \chi)(\phi_a^{\dagger}\, \phi_b ),
\eea
where $ \mu_{ab}$ and  $m_{\chi}^2$ are mass square parameters while $ \lambda_{a\,b\,c\,d},   \lam_{\chi ab}$ and $\lam_{\chi}$ are Yukawa coupling coefficients of fields having the corresponding indices. The sum convention over repeated indices is implied. We see directly from the definition of quartic couplings that $\lambda_{a\,b\,c\,d}= \lambda_{c\,d\,a\,b}$. Together with hermiticity, the coupling coefficients satisfy:
\bea
\label{hermiticity}
 \mu_{a\,b} =  \mu_{b\,a}^*, \; \lam_{\chi a\,b} = \lam_{\chi b\,a}^* \;,\left( m_{\chi}^2\;\mbox{and}\;  \lam_{\chi}\; \mbox{are real}\right), \;
 \lambda_{a\,b\,c\,d}= \lambda_{c\,d\,a\,b}= \lambda_{b\,a\,d\,c}^*.
\eea
In constructing the scalar potential we follow closely the notation of \cite{silva,branco}.

Let us find the independent couplings from the constraints of Eq.(\ref{hermiticity}) alone. Since this is a common problem, we consider $n$ doublets $\phi_a$ (in our case $n=5$) with one gauge singlet scalar $\chi$. In order to ease the counting and also for notational simplicity, we arrange the indices of the  mass parameters and Yukawa couplings in 2-tuples and 4-tuples as
$(a,b)$ and $(a,b,c,d)$.  Thus,  it is easy to count the number of  independent couplings for each of the quadratic $(\mu_{a\,b})$ and cubic $(\lam_{\chi a\,b})$ couplings as:
 \bea
 (a,b): a\leq b, & \mbox{counting} & 1 + 2 + 3 + \cdots + n =\sum_{k=1}^n  k = { n(n+1)\over 2}.
 \eea
For the quartic coefficients, one can, using the constraints of Eq. \ref{hermiticity}, span all the couplings once and only once, by restricting to:
\bea
 (a,b,c,d):\left\{  a=\mbox{Min}\{a,b,c,d\} \wedge a \notin \{b,c,d\} \right\} &\mbox{counting}& \sum_{k=1}^n (n-k)^3  =
{n^2\,(n-1)^2\over 4}, \nn \\
\begin{array}{l}
(a,a,c,d):\left\{ c \leq d, a=\mbox{Min}\{a,c,d\} \wedge a \notin \{c,d\} \right\}  \\
(a,b,a,d):\left\{ b \leq d, a=\mbox{Min}\{a,b,d\} \wedge a \notin \{b,d\} \right\}    \\
(a,b,c,a) : \left\{ b \leq c, a=\mbox{Min}\{a,b,c\} \wedge a \notin \{b,c\} \right\}    \\
\end{array}
  & \mbox{counting} & 3\sum_{k=1}^n \sum_{j=k+1}^n (n-j+1) = {n\,(n^2-1)\over 2},   \nonumber \\
(a,a,a,d): \left\{ a < d \right\}& \mbox{counting} & \sum_{k=1}^n (n-k) = {n\,(n-1)\over 2},\nonumber \\
(a,a,a,a)  & \mbox{counting} & n.
\eea
By adding the number of independent parameters, we get $\frac{n^4+3n^2}{4}$, giving $175$ independent quartic Yukawa coefficient. Imposing the invariance under $Z_2 \times Z_6$ symmetry we find the following constraints
\be
\left.
\begin{array}{lll}
\mu_{c\,d} &=& {\bf U^*_\a}_{a\,c} \,{\bf U_\a}_{b\,d}\, \mu_{a\,b}\\
\lam_{\chi c\,d} &=& {\bf U^*_\a}_{a\,c} \,{\bf U_\a}_{b\,d}\, \lam_{\chi a\,b}\\
\lam_{e\,f\,g\,h} &=& {\bf U^*_\a}_{a\,e}\, {\bf U_\a}_{b\,f} {\bf U^*_\a}_{c\,g}\, {\bf U_\a}_{d\,h}\, \lam_{a\,b\,c\,d}
\end{array}
\right\}
\a \in \{\bf 2,6\}.
\ee
We end up with the following  independent Yukawa couplings  ($5$ quadratic, $5$ cubic and $33$ quartic coefficients) shown in Table~(\ref{indep_coeff}).
\begin{table}[H]
\begin{center}
\begin{tabular}{cc}
\toprule
Independent Yukawa Couplings & $\in\;$ which number field\\
\toprule
$\mu_{1\,1}, \mu_{2\,2}, \mu_{3\,3}, \mu_{4\,4}=\mu_{5,5};\;\;\;\;\;\lam_{\chi\, 1\,1}, \lam_{\chi\, 2\,2},\lam_{\chi\, 3\,3},\lam_{\chi\, 4\,4}=\lam_{\chi\, 5,5}$ &
 ${\mathds R}$ \\
\midrule
$\mu_{4\,5};\;\;\;\;\;  \lam_{\chi\, 4\,5} $ & $  i{\mathds R}$ \\
\toprule
$
\lam_{1\,1\,1\,1},\;\lam_{1\,1\,2\,2},\; \lam_{1\,1\,3\,3},\; \lam_{1\,1\,4\,4}=\lam_{1\,1\,5\,5},
\lam_{1\,2\,2\,1},\; \lam_{1\,3\,3\,1},\;  \lam_{1\,4\,4\,1}=\lam_{1\,5\,5\,1}$ &  \\
$\lam_{2\,2\,2\,2},\;\lam_{3\,3\,3\,3},\;  \lam_{4\,4\,4\,4}=\lam_{5\,5\,5\,5},\;
\lam_{2\,2\,3\,3},\; \lam_{3\,3\,4\,4}=\lam_{3\,3\,5\,5},\; \lam_{2\,2\,4\,4}=\lam_{2\,2\,5\,5}$ &  ${\mathds R}$ \\
$\lam_{2\,3\,3\,2},\; \lam_{3\,4\,4\,3}=\lam_{3\,5\,5\,3},\;  \lam_{2\,4\,4\,2}=\lam_{2\,5\,5\,2},\;
\lam_{4\,4\,5\,5},\; \lam_{4\,5\,4\,5},\;  \lam_{4\,5\,5\,4}  $ &  \\
\midrule
$\lam_{1\,1\,4\,5},\;  \lam_{1\,4\,5\,1},\; \lam_{2\,2\,4\,5},\; \lam_{2\,4\,5\,2},\;
\lam_{3\,3\,4\,5},\; \lam_{3\,4\,5\,3} $ & $  i {\mathds R}$ \\
\midrule
$\lam_{1\,2\,1\,3},\;  \lam_{1\,4\,1\,5},\;  \lam_{1\,4\,1\,4}=-\lam_{1\,5\,1\,5},\;
\lam_{2\,3\,2\,4}= -i\lam_{2\,3\,2\,5},\;  \lam_{2\,3\,4\,3}=-i\,\lam_{2\,3\,5\,3}$ &  $   {\mathds C}$ \\
 $ \lam_{2\,4\,3\,4}=\lam_{2\,5\,3\,5},\;
\lam_{2\,4\,3\,5}=\lam_{2\,5\,3\,4},\;
\lam_{4\,4\,4\,5}=-\lam^*_{4\,5\,5\,5}
 $ & $   {\mathds C}$\\
\bottomrule
\end{tabular}
\end{center}
 \caption{\small  Independent Yukawa coupling in the case $C_{33}$ under $Z_2 \times Z_6$ symmetry}
\label{indep_coeff}
 \end{table}
We assume that the Yukawa coupling  are such that  the potential gets local minima, around one of which the fields are expanded as follows.
\bea \label{expansion}
\phi_{k}=\left(\begin{array}{c}
\varphi_{k}^{+} \\
\frac{1}{\sqrt{2}}  (v_{k}+\varphi^o_{k})
\end{array}\right)&,& \chi = \frac{v_{\chi}+\chi^o}{\sqrt{2}}.
\eea
In terms of the fields ($\varphi_a^+, \varphi_b^o, \chi^o$), the potential can be decomposed into constant, linear, quadratic, cubic and quartic terms.
Being in a local minimum means that the linear term is vanishing, whence we get the tadpole conditions:
\bea \label{tadpole}
\left[ \mu_{a\,b} + \lam_{a\,b\,c\,d}\, v_c^*\, v_d \right]  v_b + \frac{1}{2}\,\lam_{\chi\, a\,b}\, v^*_\chi \,v_\chi v_b=0, \\
\left(  m_{\chi}^2  + \frac{1}{2}\, \lam_{\chi\, a\, b}\, v^*_a\,  v_b\right) v_\chi =0
\eea

As to the quadratic term, it gives the mass matrices:
\be
V_2 = {\varphi^-_a} ({\bf M}^2_\pm)_{ab} \varphi^+_b + {1\over 2}\,
\mathcal{N}^T \, {\bf M}^2_o \, \mathcal{N},
\label{quadratic}
\ee
where  $\mathcal{N}$ represents the neutral fields organized as
\be
\mathcal{N }\equiv \left( \mbox{Re} [\varphi^o_1], \ldots ,\mbox{Re}[\varphi^o_5],\mbox{Im}[\varphi^o_1], \ldots ,\mbox{Im}[\varphi^o_5],
 \mbox{Re}[\chi^o] , \mbox{ Im}[\chi^o]  \right)^T.
\ee
The  hermitian $5 \times 5$ matrix ${\bf M}^2_\pm$, in Eq.(\ref{quadratic}),  represents the mass matrix of  charged fields and is given as
\be
\label{m2+-}
({\bf M}^2_\pm)_{ab} = \mu_{a\,b} + \lam_{a\,b\,c\,d}\, v_c^* v_d + \frac{1}{2}\,\lam_{\chi\, a\,b}\, v^*_\chi\, v_\chi,
\ee
while the $12 \times 12$  matrix ${\bf M}^2_o$ for neutral fields is expressed in terms of   block matrices as
\bea
\label{m2o}
{\bf M}^2_o &=& \left( \begin{array}{cc}{\bf M^2_\varphi} & {\bf M^2_{\chi \varphi}}\\
{\bf M^2_{\chi \varphi}}^T & {\bf M^2_\chi}
 \end{array}\right)
\eea
where the $10 \times 10$ matrix ${\bf M^2_\varphi}$ is in turn  given in terms of block matrices as
\bea
\label{m2phi}
{\bf M}^2_\varphi = \left( \begin{array}{cc}{\bf M^2_R} & {\bf M^2_{RI}}\\
{\bf M^2_{RI}}^T & {\bf M^2_{I}}
 \end{array}\right) &\;\mbox{with}\;&
  \begin{array}{l}
 ({\bf M^2_R})_{ab}= \mbox{Re}[({\bf M}^2_\pm)_{ab} +\lam_{a\,c\,d\,b}\,  v_c\, v_d^* + \lam_{a\,c\,b\,d}\, v_c\, v_d], \\
 ({\bf M^2_I})_{ab}=\mbox{Re}[({\bf M}^2_\pm)_{ab} +\lam_{a\,c\,d\,b}\,  v_c\, v_d^* - \lam_{a\,c\,b\,d}\, v_c\, v_d],  \\
 ({\bf M^2_{RI}})_{ab}= -\mbox{Im}[({\bf M}^2_\pm)_{ab} +\lam_{a\,c\,d\,b}\,  v_c\, v_d^* - \lam_{a\,c\,b\,d}\, v_c\, v_d],
\end{array}
  \eea
and the $10 \times 2$ matrix ${\bf M^2_{\chi \varphi}}$ is given in terms of four $5 \times 1$ matrices as follows
\bea
\label{m2chiphi}
{\bf M}^2_{\chi\varphi} = \left( \begin{array}{cc}{\bf M^2_{\chi \varphi R}} & {\bf M^2_{\chi \varphi RI}}\\
{\bf M^2_{\chi \varphi IR}} & {\bf M^2_{\chi \varphi  I}}
 \end{array}\right)  & \mbox{with}&
 \begin{array}{l}
 ({\bf M^2_{\chi \varphi R}})_{a1}= \mbox{Re}[\lam_{\chi\,a\,b} (v_\chi^*\, v_b + v_\chi\, v_b )],  \\
 ({\bf M^2_{\chi \varphi I}})_{a1}= \mbox{Re}[\lam_{\chi\,a\,b} (v_\chi^*\, v_b - v_\chi\, v_b )], \\
  ({\bf M^2_{\chi \varphi RI}})_{a1}=  \mbox{Im}[\lam_{\chi\,a\,b} (v_\chi^*\, v_b + v_\chi\, v_b )],   \\
 ({\bf M^2_{\chi \varphi IR}})_{a1}=  -\mbox{Im}[\lam_{\chi\,a\,b} (v_\chi^*\, v_b - v_\chi\, v_b )].
  \end{array}
 \eea
Finally, the $2 \times 2$ matrix ${\bf M^2_\chi}$ is given by
\bea
\label{m2chi}
{\bf M^2_\chi} &=& (m_{\chi}^2 + {1\over 2}\, \lam_{\chi\, a\,b}\, v_a^*\, v_b)\,\mbox{diag}\left(1, 1\right).
\eea

Note that ${\bf M}^2_\pm$ is hermitian, whereas ${\bf M^2_R}, {\bf M^2_I}, {\bf M^2_\chi}$ are real symmetric. The matrices ${\bf M^2_{RI}}, {\bf M^2_{\chi\varphi R}}$, ${\bf M^2_{\chi\varphi I}}, {\bf M^2_{\chi\varphi RI}}, {\bf M^2_{\chi\varphi IR}}$ are real, so we get ${\bf M}^2_o$ as real symmetric.

 One should diagonalize the mass matrices in order to get the physical masses, but we anticipate at least three vanishing masses which would correspond to the would be Goldstone bosons. With such a large number of free parameters, we do not carry out this task, but just assume that the Yukawa couplings are chosen in their parameters space so that the mass spectrum is such that one doublet would play the role of the SM Higgs whereas the others would be outside the  {\bf LHC} reach.
}


\begin{thebibliography}{99}

\bibitem{neutrinoMass} see e.g. K. Nakamura et al. (Particle Data Group), J. Phys. G 37,
075021 (2010).

\bibitem{oscillation} see e.g. Y. Fukuda et al., Phys. Lett. B 436, 33 (1998); Phys. Rev.
Lett. 81, 1562 (1998);  C. K. Jung, C.
McGrew, T. Kajita, and T. Mann, Annu. Rev. Nucl. Part.
Sci. 51, 451 (2001), and references therein.

\bibitem{xing} P. H. Frampton, S. L. Glashow, and D. Marfatia, Phys.
Lett. B 536, 79 (2002)., \\
Z. Z. Xing, Phys. Lett. B 530, 159 (2002). \\
H. Fritzsch, Z. Z. Xing, and S. Zhou, J. High Energy Phys.
09 (2011) 083.

\bibitem{M0texture} A. Merle and W. Rodejohann, Phys. Rev. D 73, 073012
(2006).
\bibitem{0texture} E. I. Lashin and N. Chamoun, Phys. Rev.  D 85, 113011 (2012)


\bibitem{minor} L. Lavoura, Phys. Lett. B 609, 317 (2005). \\
E. I. Lashin and N. Chamoun, Phys. Rev. D 78, 073002
(2008) and Phys. Rev. D 80, 093004
(2009).

\bibitem{pmt} E. I. Lashin and N. Chamoun, C. Hamzaoui and S. Nasri, Phys. Rev. D 89, 093004 (2014), Phys. Rev. D 91, 113014 (2015) and
Phys. Rev. D 96, 015003 (2017)

\bibitem{Lashin} H. A. Alhendi, E. I. Lashin, and A. A. Mudlej, Phys. Rev.
D 77, 013009 (2008); arXiv:0708.2007.

\bibitem{mtas} Kitabayashi T and Yasue M, Phys. Lett. B621 (2005), 133-138 (arXiv:hep-ph/0504212). \\
Grimus W, Kaneko S, Lavoura L, Sawanaka H and Tanimoto M, JHEP 0601 (2006) 110 (arXiv:hep-ph/0510326).


\bibitem{Liu} Ji-Yuan Liu, Shun Zhou, Phys. Rev. D 87 (2013) 093010, arXiv: 1304.2334
\bibitem{Gautam} S. Dev, R.R. Gautam, and L. Singh, Phys. Rev. D 87, 073011 (2013)


\bibitem{Lisi} F. Capozzi, E. D. Valentino, E. Lisi,
A. Marrone, A. Melchiorri and A. Palazzo, Phys. Rev. D 95, 096014 (2017)

\bibitem{daya} F. P. An et al. (DAYA-BAY Collaboration), Phys. Rev.
Lett. 108, 171803 (2012).

\bibitem{cherif} W. Grimus and L. Lavoura, Eur. Phys. J. C 39, 219 (2005);
A. Dighe, S. Goswami, and P. Roy, Phys. Rev. D 76, 096005
(2007); S. Luo and Z.-z. Xing, Phys. Rev. D 86, 073003
(2012).


\bibitem{heinrich} Heinrich P\"{a}s and Werner Rodejohann, (2015) New J. Phys. {\bf 17} 115010

\bibitem{Cuoricino} E. Andreotti et al., Astropart. Phys. 34, 822 (2011).


\bibitem{ckm} M. Bargiotti, {\it et al.}, Riv.Nuovo Cim.23N3:1,2000, arXiv:hep-ph/0001293
\bibitem{seesaw2} T. P. Cheng and L. F. Li, Phys. Rev. D 22, 2860 (1980);
R. N. Mohapatra and G. Senjanovic, Phys. Rev. D 23, 165
(1981).

\bibitem{grimus} W. Grimus, L. Lavoura, Phys.Lett. B572, 189( 2003).

\bibitem{ma2004}  E. Ma,  Phys. Lett. B 583, 157 (2004).

\bibitem{sarkar} Particle and
Astroparticle Physics, Series in High Energy Physics, Cosmology, and Gravitation by Utpal Sarkar
, Taylor and Francis group (2008).

\bibitem{silva} F. J. Botella and J. P. Silva,  Phys. Rev. D 51, 3870 (1995).

\bibitem{branco} G. C. Branco, L. Lavoura and J. P. Silva, CP violation, Oxford University Press, Oxford U.K., (1999).

\end{thebibliography}
\end{document}